\def\tooee{LaTeX2e}
\ifx\fmtname\tooee
   \documentclass[11pt]{article}\usepackage{pb-diagram}
\else
   \documentstyle[11pt,pb-diagram]{article}
\fi

\usepackage{epsfig}
\usepackage{amsmath}
\usepackage{amssymb}
\usepackage{xypic}

\newcommand{\mbf}[1]{{\boldsymbol {#1} }}

\renewcommand{\thefootnote}{\fnsymbol{footnote}}
\newcommand{\newsection}{\setcounter{equation}{0}\section}

\def\appendix#1{\addtocounter{section}{1}\setcounter{equation}{0}
\renewcommand{\thesection}{\Alph{section}}
\section*{Appendix \thesection\protect\indent \parbox[t]{11.715cm} {#1}}
\addcontentsline{toc}{section}{Appendix \thesection\ \ \ #1} }
\newcommand{\complex}{{\mathbb C}} %% complex numbers
\newcommand{\zed}{{\mathbb Z}} %% integers
\newcommand{\nat}{{\mathbb N}} %% naturals
\newcommand{\real}{{\mathbb R}} %% real numbers
\newcommand{\reals}{{\mathbb R}} %% small real numbers
\newcommand{\zeds}{{\mathbb Z}} %% small integers
\newcommand{\rat}{{\mathbb Q}} %% rational numbers
\newcommand{\torus}{{\mathbb T}}
\newcommand{\sphere}{{\mathbb S}}

\newcommand{\halfplane}{{\mathbb U}}
\newcommand{\NO}{\,\mbox{$\circ\atop\circ$}\,} % Normal ordering
\newcommand{\prym}{{\rm pr}}

\newif\ifold             \oldtrue

\newcommand{\tr}[1]{\:{\rm tr}\,#1}
\newcommand{\Tr}{\:{\rm Tr}\,}

\def\e{{\,\rm e}\,}
\newcommand{\ii}{\,{\rm i}\,}
\newcommand{\dd}{{\rm d}}

\def\benum{\begin{enumerate}}
\def\eenum{\end{enumerate}}
\def\be{\begin{equation}}
\def\ee{\end{equation}}
\def\bea{\begin{eqnarray}}
\def\eea{\end{eqnarray}}
\def\bd{\begin{displaymath}}
\def\ed{\end{displaymath}}

\def\={\ =\ }

\newcommand{\beq}{\begin{equation}}
\newcommand{\eeq}{\end{equation}}

\newcommand{\g}{\gamma}

\newcommand{\pr}{\partial}

\renewcommand{\a}{\alpha}
\renewcommand{\b}{\beta}
\renewcommand{\th}{\theta}

\renewcommand{\t}{\tau}
\newcommand{\tb}{\theta^\bullet}
\newcommand{\ts}{\theta^\#}
\newcommand{\ttb}{\tilde{\theta}^\bullet}
\newcommand{\tts}{\tilde{\theta}^\#}

\begin{document}
\begin{titlepage}
\begin{flushright}

\baselineskip=12pt

HWM--07--17\\
EMPG--07--13\\
\hfill{ }\\
June 2007
\end{flushright}

\begin{center}

\baselineskip=24pt

\vspace{2cm}

{\Large\bf DLCQ Strings, Twist Fields\\ and One-Loop Correlators on a
  Permutation Orbifold}

\baselineskip=14pt

\vspace{2cm}

{\bf Henry C.D. Cove}, {\bf Zolt\'an K\'ad\'ar} and {\bf Richard J. Szabo}
\\[4mm]
{\it Department of Mathematics}\\ and\\ {\it Maxwell Institute for
  Mathematical Sciences\\Heriot-Watt University\\ Colin Maclaurin
  Building, Riccarton, Edinburgh EH14 4AS, U.K.}
\\{\tt H.Cove , Z.Kadar , R.J.Szabo @ma.hw.ac.uk}
\\[50mm]

\end{center}

\begin{abstract}

We investigate some aspects of the relationship between matrix
string theory and light-cone string field theory by analysing the
correspondence between the two-loop thermal partition function of DLCQ
strings in flat space and the integrated two-point correlator of twist
fields in a symmetric product orbifold conformal field theory at
one-loop order. This is carried out by deriving combinatorial
expressions for generic twist field correlation functions in
permutation orbifolds using the covering surface method, by deriving
the one-loop modification of the twist field interaction vertex,
and by relating the two-loop finite temperature DLCQ string theory to
the theory of Prym varieties for genus two covers of an elliptic
curve. The case of bosonic $\zed_2$~orbifolds is worked out explicitly
and precise agreement between both amplitudes is found. We use these
techniques to derive explicit expressions for $\zed_2$~orbifold spin
twist field correlation functions in the Type~II and heterotic string
theories.

\end{abstract}

\end{titlepage}

\setcounter{page}{2}
\renewcommand{\thefootnote}{\arabic{footnote}}
\setcounter{footnote}{0}

\newsection{Introduction and Summary\label{Intro}}

Large $N$ matrix field theories obtained as dimensional reductions of
maximally supersymmetric $U(N)$ Yang-Mills theory in ten spacetime
dimensions provide nonperturbative descriptions of M-theory and string
theory in various backgrounds, and associated superconformal field
theories (see~\cite{Taylor:2001vb} for a review). The best understood
example is matrix string
theory~\cite{Motl:1997th}--\cite{Dijkgraaf:1997vv} which takes the
form of maximally supersymmetric $U(N)$ Yang-Mills theory in two
dimensions. In this case the gauge coupling is inversely proportional
to the string coupling, so that the free string limit corresponds to
the infrared limit and the first order interaction term to the least
irrelevant operator in the gauge theory. In this strong coupling limit
the supersymmetric Yang-Mills theory approaches a superconformal fixed
point which is conjectured to be the supersymmetric sigma model on the
symmetric product orbifold $(\real^{8})^N/S_N$. The spectrum of this
orbifold superconformal field theory can be canonically identified
with that of the free second quantized Type~IIA
string~\cite{Dijkgraaf:1997vv,DMVV1}. 

This equivalence is demonstrated qualitatively in discrete light-cone
quantization (DLCQ) by matching configurations obtained by gluing
different copies of the strings winding around a light-like circle to
twisted sectors of the symmetric product. The quantitative
demonstration is given by matching the torus partition function
of the superconformal field theory~\cite{DMVV1} to the
thermodynamic free energy of the free Type~IIA
superstring~\cite{GS1}. It is conjectured~\cite{Dijkgraaf:1997vv} that
the equivalence holds generally in the interacting string theory as
well. Strings interact by means of splitting and joining, and the
interaction points correspond to insertions of twist field operators
in the orbifold superconformal field theory. It has been recently
argued~\cite{Dijkgraaf:2003nw}--\cite{Kishimoto:2006xc} that the
structure of the contact interactions in Green-Schwarz light-cone
superstring field theory simplifies within the twist field formulation
of matrix string theory. Unlike the light-cone string field theory,
however, the matrix model provides a full nonperturbative definition
of the string dynamics in the large $N$ limit.

In this paper we will investigate this conjectural perturbative
correspondence further by
examining the relationship between the thermodynamic free energy of
Type~II superstring theory in DLCQ and correlation functions of the
leading irrelevant twist field operators in the symmetric product
orbifold conformal field theory. The Polyakov path integral for the
former quantity is known~\cite{Grignani:2000zm} to truncate the sum
over contributing string worldsheets to those which are branched
covers of the spacetime torus arising from the null compactification
at finite temperature. The free energy at the leading non-vanishing
order in the string coupling constant is the two-loop string
amplitude which has been calculated in~\cite{csz}. In order to check
the conjecture one needs to compute the corresponding amplitude in the
orbifold conformal field theory, which is given by the one-loop two-point
function of appropriate twist fields. These operators create twisted
sectors out of the vacuum state, in that the local fields of the sigma
model acquire non-trivial monodromy about the twist field insertion
points. Computing their correlation functions is thus not
straightforward, and a good portion of our analysis will centre around
the technicalities involved in these calculations. 

There are several strategies presented in the literature for computing
twist field correlation functions. The stress tensor method was
originally introduced in~\cite{Dixon:1986qv} and used to compute
$\zeds_2$~orbifold~\cite{Dixon:1986qv}--\cite{Saleur}, and more
generally $\zeds_N$~orbifold~\cite{Atick,LawSever}, correlation
functions on worldsheets of arbitrary topology, and $S_N$ orbifold correlation
functions on the sphere~\cite{ArFro1,Arutyunov:1997gi}. In this method
one first determines the twisted Green's function (the $n$-point
function of the stress energy tensor in the twisted sector) by
demanding the correct short distance behaviour and monodromy about the
twist field insertion points. A closely related but more general
technique is the covering space method. It makes direct use of the
fact that a monodromy is associated to a covering surface. If a field
is multi-valued when transported around a closed curve, then it is
well-defined as a single-valued function on the appropriate cover of
the worldsheet without any special points. In this way the twist field
correlation functions can be expressed as vacuum amplitudes of the
free conformal field theory on the covering surfaces. This method was
exploited in~\cite{Hamidi:1986vh}--\cite{Lunin:2000yv}. It is also the
main principle behind computing essentially all quantities in
permutation orbifolds as shown in~\cite{bantaym} where, in particular,
the partition function was given for arbitrary orbifold twist group.

In this paper we use the covering space method for the definition and
computation of twist field correlation functions in symmetric products
defined on worldsheets of non-trivial topology. The vacuum amplitudes
of these conformal field theories are known in complete generality,
\emph{i.e.}, for worldsheets of arbitrary genus and arbitrary finite
twist group~\cite{bantayhg}. We generalize these results to the
$n$-point correlation functions of twist field operators. When the
worldsheet has non-trivial fundamental group and the twist group is
nonabelian, the definition of the corresponding twisted Green's
functions is problematic and the covering space technique is the only
possible way to define the amplitudes. To make these formulae
completely explicit, one needs to determine the dependence of the
complex structure of the covering space on that of the worldsheet and
the location of the twist field operator insertions. This 
is a very difficult problem in the general case when the covering
surface does not admit any conformal automorphisms. We have not been
able to solve this problem in full generality and are not aware of any
solution to it for any specific cases of such a cover. All known
computations of twist field correlation functions are done with
respect to covers with automorphisms (this is the case, in particular,
for the $\zeds_N$~orbifolds), or to worldsheets of trivial topology
when the covering space can be parametrized explicitly in terms of the
complex coordinate $z$ of the sphere. Nevertheless, using our
technique we are able to determine the bosonic two-point twist field
correlation function of the orbifold
$\real^{24}\wr\zeds_2:=(\real^{24}\times\real^{24})/\zeds_2$ and
compare it to the appropriate power of the $\zeds_2$ orbifold twist
field correlation function of the one-dimensional free boson computed
in~\cite{dvvc1}, yielding a highly non-trivial check of our
methods. Although throughout we deal only with orbifolds of flat space
$\real^d$, most of our considerations and results apply to more
general symmetric products as well.

When writing down generating functions of amplitudes in symmetric
products, one has to sum over all covers of the worldsheet in such
a way that only the connected covering surfaces contribute. This fact
lies behind the conjecture that these amplitudes naturally arise in
physical string theories. We generalize the resummation procedure
which was done originally for the torus partition function
in~\cite{DMVV1} and for the Klein bottle amplitude
in~\cite{bantaysymm} for the case of closed strings, and then for the
annulus and M\"obius diagrams in~\cite{Fuji:2000fa,Fuji:2001kt} for
the case of open strings. The generalization to the twist field
$n$-point function is possible due to a general combinatorial
formula~\cite{bantaysymm} which is the crux of all of these
calculations.

The main technical achievement of the two-loop calculation
of~\cite{csz} was a modification of the Weierstrass-Poincar\'e theory
of reduction. Reduction may be described entirely in terms of the Riemann
matrix of periods of a curve, and it has the effect of expressing
theta functions at a given genus in terms of lower dimensional theta
functions. This happens exactly when the curve in question covers a
surface of lower genus (but it may also occur without there being a
covering map). The remarkable feature of this reduction is the simple
universal form that the genus two DLCQ free energy takes in terms of
Jacobi elliptic functions on the base torus. For the
contributions from double covers of the torus to the two-loop free
energy of the critical bosonic string, we find perfect agreement
between the string free energy and the correlator of twist fields
computed as the appropriate power of the $\zeds_2$~orbifold twist
field two-point function of~\cite{dvvc1}. We will find generally that
the original genus zero interaction vertex proposed
in~\cite{Dijkgraaf:1997vv} must be modified at one-loop order to
ensure equivariance under the action of the non-trivial modular group
in this instance. Since the structure of the result depends only on
the orbifold twist group and not on the data of the specific string
theory, we use this equivalence and the known formulae from~\cite{csz}
for the two-loop DLCQ free energy of the Type~II and heterotic
strings to derive the two-point functions of the appropriate spin
twist fields in the corresponding $\zed_2$~orbifold superconformal
field theories. To the best of our knowledge, these correlation
functions have not been previously computed, and our explicit formulae
should be useful for further clarifying the role of the twist field
interaction vertex in light-cone string field theory.

The difficulty in establishing the correspondence is writing down the
period matrix of the covering surface explicitly in terms of the
modulus of the worldsheet torus and the branch point loci. This is
achieved in part by elucidating the geometric meaning of the reduced
genus two period matrix. In~\cite{csz} it was shown that this period
depends on two elliptic moduli, one of which lies in a modular orbit
of an unramified (one-loop) cover of the base torus. Here we show that
the second elliptic modulus determines the complex structure of a Prym
variety. Prym varieties arise in special instances of covering
surfaces. A theorem due to Mumford asserts that there are only three
types of branched covers which give rise to Prym varieties, namely
unramified double covers, ramified double covers with two branch
points, and precisely our instance of genus two covers over an
elliptic curve. When this is in addition a double cover of the torus,
we use the canonical involution of the genus two surface to explicitly
construct the dependence of the periods on the branch points (which
are the images of the fixed points of the involution). This procedure
unfortunately doesn't generalize to higher degree covering surfaces
(although the identification with a Prym variety always holds).

The organisation of the rest of this paper is as follows. In
Section~\ref{CorrsPermOrbs} we give a general introduction to the
theory of bosonic permutation orbifolds, and use the one-loop
sigma model to illustrate the typical combinatorial structure of
amplitudes therein. We apply the combinatorial resummation formula for
symmetric products to compute a large class of correlation functions
which are invariant under the action of the twist group. We show how
to generalize these formulae to correlation functions of twist field
operators, and briefly review the structure of the DLCQ string
partition function. In Section~\ref{Z2Orbifolds} we present detailed
and explicit calculations for $\zeds_2$ orbifolds. In the course of
this analysis, we make the generic connection between DLCQ string
theory and the theory of Prym varieties, and also derive the explicit
modification of the twist field interaction vertex for toroidal
worldsheets in the symmetric product sigma model. In
Section~\ref{NonabOrbs} we discuss the technical issues surrounding
the generalizations of these results to $S_N$~orbifolds with
$N>2$. We examine the uniformization construction, which is used to
build vacuum amplitudes, in the context of a generic twist field
$n$-point function, and the problem of determining the period of the
genus two covering surface in terms of the branch point data. We also
study the combinatorial expansion in more detail and indicate that,
while computable in principle, the combinatorics become very
non-trivial for $N>2$. Finally, in Section~\ref{FermOrb} we describe
the modifications of permutation orbifolds required in the presence of
fermionic degrees of freedom, and of twist field correlation functions
therein. We then apply these and previous considerations to derive
explicit formulae for the one-loop spin twist field correlation
functions in the $\zed_2$ orbifold supersymmetric and heterotic string
theories.

\newsection{Correlation Functions on Permutation
  Orbifolds\label{CorrsPermOrbs}}

In this section we will discuss some general aspects of permutation
orbifolds of conformal field theories, and in particular the case of
two-dimensional sigma models on symmetric product orbifolds of flat
space. We first describe the general structure of the partition
functions of these models, and then explain the construction of
various classes of correlation functions including those of twist
field operators. For the moment we treat only bosonic sigma models
explicitly in order to highlight the essential details, defering a
more detailed analysis of the supersymmetric and heterotic cases to
Section~\ref{FermOrb}. We also explain how these orbifold theories can
be interpreted as string field theories.

\subsection{Permutation Orbifolds}

When a two-dimensional conformal field theory has a discrete symmetry,
one can consider the orbifold theory arising from quotienting with
respect to the symmetry. The simplest example is the free boson on
the circle $\sphere^1$. Its action
$\frac1{4\pi\,\alpha'}\,\int_\Sigma\,\dd^2z~\|\pr X\|^2$ is
invariant under the reflection $X\to -X$. The quotient of the target
space is the well-known geometric orbifold $\sphere^1/\zeds_2$, and
the coordinate field $X$ can have non-trivial monodromy when
encircling a non-contractible cycle of the worldsheet $\Sigma$. If the
radius of the circle is equal to the fundamental string length
$\ell_s=\sqrt{\alpha'}$, then the resulting orbifold conformal field
theory is the ``square'' of the critical Ising model~\cite{ginsparg}.
       
Permutation orbifolds represent a large class of orbifolds where the
parent conformal field theory (whose quotient is taken) has physical Hilbert
space $\cal H$ with a discrete symmetry. This concept was first introduced
in~\cite{klemmschmidt}, and used for the construction of a $\zeds_2$
orbifold of the $E_8\times E_8$ heterotic string in~\cite{fhpv}. One
of their main applications is to the second quantization of string
theory~\cite{DMVV1}, and they have recently been
argued~\cite{Halpern:2007vj} to describe new physical string theories
at multiples of the critical dimension. A permutation orbifold of an
arbitrary conformal field theory ${\cal C}$, by any finite symmetry
group $G$ regarded as a subgroup of a symmetric group of some degree,
is a consistent conformal field theory. All of its important
quantities (central charge, conformal weights, genus one characters,
modular $S$ and $T$ matrices, genus one partition function,
\emph{etc.}) were worked out originally for cyclic groups
in~\cite{bhs}, and then generalized to arbitrary finite groups
in~\cite{bantaym}. These formulae express a given quantity as a
combinatorial expansion, depending on the twist group ${G}$, of the
same quantity in the parent theory.

Highest weight states in a permutation orbifold ${\cal C}\wr
{G}:=({\cal C})^{\otimes N}/G$ correspond to orbits of a subgroup
${G}<S_N$ of the symmetric group of degree $N$ acting on the $N$-fold
tensor product of states in the parent theory ${\cal
  C}$.\footnote{There is an additional label corresponding to the
  irreducible character of the double of the stabilizer of the
  orbit. See~\cite{bantaym} for the precise definition.}  In the case
that ${\cal C}$ admits a sigma model description with embedding
coordinate field $X\in M$, there is a corresponding sigma model
description of ${\cal C}\wr {G}$ on the geometric orbifold
$M^N/G$~\cite{Fuji:2000fa}. One introduces $N$ identical coordinate
fields $X^a=X$, $a=1,\dots,N$ on the worldsheet $\Sigma$ and allows
for ${G}$-twisting of them along non-trivial cycles. For
example, on the torus $\Sigma=\torus$ with modulus $\tau$, the
boundary conditions of the $N$ coordinate fields are labelled by two
commuting permutations $P,Q\in {G}<S_N$ such that
 \beq X^a(z+1)\=X^{P(a)}(z) \qquad \mbox{and}
\qquad X^a(z+\tau)\=X^{Q(a)}(z)
 \label{mbond} \ , \eeq
where in general $g(a)$ denotes the image of the label $a$ under the
permutation $g\in{G}$. For a non-trivial pair $(P,Q)$, these boundary
conditions are called twisted sectors of the theory. Two pairs $(P,Q)$
and $(g\,P\,g^{-1},g\,Q\,g^{-1})$ with $g\in S_N$ correspond to the
same twisted sector, since we can get from one to the other by
relabelling the coordinate fields $a\to g(a)$.

In general, a twisted sector is given by an equivalence class of
homomorphisms from the fundamental group $\pi_1(\Sigma)$ of the
worldsheet to the twist group $G$. (Since
$\pi_1(\torus)=\zeds\oplus\zeds$, on the torus one specifies a
homomorphism by choosing the image in $G$ of the two commuting
generators.) Two homomorphisms $\Phi,\Phi'$ define the same twisted
sector, and are said to be equivalent, if they are related by conjugation as
$\Phi'(-)=g\,\Phi(-)\,g^{-1}$ for some $g\in S_N$. The geometric
interpretation is provided by the fact that every equivalence class
$[\Phi]$ of homomorphisms $\Phi:\pi_1(\Sigma)\to S_N$ determines an
unramified cover $\hat\Sigma$ of degree $N$ over the Riemann surface
$\Sigma$. The coordinate label $a$ corresponds to the label of a sheet
and $\Phi$ is called the monodromy homomorphism of the
covering. Conjugation of homomorphisms corresponds to relabelling of
the sheets. In the case of the torus $\Sigma=\torus$ with the boundary
conditions (\ref{mbond}), and with the subgroup generated by the pair of
permutations $P,Q$ acting transitively on the set of coordinate labels
$a=1,\dots,N$, one can define a single new field ${\cal X}(z)$
which generates all of the fields $X^a(z)$ through the identifications
\beq
{\cal X}(z+m+n\,\tau)=X^{P^m\,Q^n(a)}(z)
\label{calXdef}\eeq
with $n,m\in\zed$ and a fixed choice of $a$. This field is
single-valued on a torus which is a cover the original torus $\torus$,
whose modular parameter can be determined from the doubly periodic
function $\cal X$ on $\torus$.

The modular invariant partition function of the permutation orbifold
is determined entirely by the above data. It is given
by~\cite{bantaym}~\footnote{It is also expressible as a sesquilinear
  expansion in the Virasoro characters $\Tr_{\cal H}(q^{L_0-c/24})$,
  whose form is known in permutation
  orbifolds~\cite{bantaym,kadar}.}
\beq Z^{G}(\tau)=\frac{1}{|{G}|}\,\sum_{\Phi:\pi_1(\Sigma)\to {G}}\;
\Big(\,\prod_{\xi\in {\cal O}(\Phi)}\,Z\big(\tau^{\xi}\big)\,\Big)
\label{partf}\eeq 
where the product runs over the orbits $\xi$ of the image
$\Phi(\pi_1(\Sigma))$ in $G$ and $Z(\tau^{\xi})$ is the modular
invariant partition function of the parent conformal field theory on
the connected component, corresponding to $\xi$, of the cover of
$\Sigma$ given by the homomorphism $\Phi$. (The covering space
$\hat\Sigma$ is connected if and only if $\Phi(\pi_1(\Sigma))$ acts
transitively in $S_N$). The summation over $\Phi$ defines the
projection onto $G$-invariant states and ensures modular invariance of
the partition function. We will now explain how to
determine (\ref{partf}) in practice.

The complex structure $\tau$ of the worldsheet $\Sigma=\Sigma_\tau$ is
encoded by a monomorphism $u:\pi_1(\Sigma)\to I$, where $I$ is the
isometry group of the universal cover $U$ of $\Sigma$. For genus $g>1$
the latter space is a two-dimensional hyperbolic space, say the upper
half plane $U=\halfplane$, and $I=PSL(2,\real)$. The surface $\Sigma$
equipped with a complex structure can be presented as the quotient
$\Sigma_\tau=\halfplane/u(\pi_1(\Sigma))$ and its complex structure
inherited from $\halfplane$ is encoded by the uniformizing group 
$u(\pi_1(\Sigma))$. Given a monodromy homomorphism $\Phi$, the fundamental 
group of the corresponding cover $\hat{\Sigma}$ is isomorphic to the 
stabilizer subgroup
$H_a=\pi_1(\Sigma)_\xi:=\{\gamma\in \pi_1(\Sigma)\;|\;\Phi(\gamma)(a)=a\}$
with fixed $a\in \xi$ (represented by closed loops based at sheet
$a$). Its index is equal to the length of the orbit
$[\pi_1(\Sigma):H_a]=|\xi|$, which is the number of sheets of the
corresponding connected component of $\hat\Sigma$. Thus the monodromy
homomorphism determines the topology of the covering space
$\hat\Sigma$. We can now define the uniformizing group (and hence the
complex structure $\tau^\xi$) of the cover $\hat\Sigma$ to be given by
$u(H_a)$ (\emph{i.e.}, $\hat{\Sigma}_{\tau^{\xi}}=\halfplane/u(H_a)$),
which is a subgroup of $u(\pi_1(\Sigma))$ in accordance with the
expected property $\pi_1(\hat{\Sigma})<\pi_1(\Sigma)$. Note that the
representative of the orbit $a\in\xi$ can be arbitrarily chosen. This
is because $H_a=\gamma\,H_{a'}\,\gamma^{-1}$ with
$\gamma\in\pi_1(\Sigma)$ for any $a,a'\in\xi$ and conjugate subgroups
of $\pi_1(\Sigma)$ give rise to isometric quotients, hence determining
equivalent surfaces. The homomorphism $u$ is not unique, as it can be
composed with a modular transformation, but the partition function is
modular invariant which makes the formula (\ref{partf}) well defined.

The expression $(\ref{partf})$ is an example of the typical structure
of a quantity defined on a Riemann surface $\Sigma$ in a permutation
orbifold. It is given by a combinatorial expansion (depending only on
${G}$) over the same quantity in the parent theory $\cal C$ defined on
all of those surfaces which cover $\Sigma$ whose monodromy group is a
subgroup of ${G}$. Its direct applicability is limited somewhat by the
Riemann-Hurwitz formula for the genus $\hat g$ of the unramified cover
$\hat\Sigma$ given by
\beq \hat{g}=N\,(g-1)+1 \ . \label{rhu} \eeq 
This implies that, unless $g=1$, we would need to
know the partition functions of the parent theory on surfaces of
genera higher than $g$ in order to write down the genus $g$ partition
function of the orbifold.

The case $g=1$ is, however, much simpler. The universal cover of the
torus $\torus$ is $U=\complex$ and $I=\{T_c~|~c\in\complex\}$ is the group
of translations $T_c:z\mapsto z+c$ of the complex plane. We saw above
that specifying a homomorphism $\Phi$ amounts to assigning commuting
elements $P,Q\in {G}$ for the generators $(\alpha,\beta)$ of
$\pi_1(\torus)={\mathbb Z}\oplus{\mathbb Z}$. The stabilizer subgroup
$H_a$ of any representative of an orbit $a\in\xi$ can be characterized
by three positive integers $s,m,r$ such that $r$ is the smallest positive
integer satisfying $P^r(a)=a$, $0\leq s<r$ and $H_a$ is generated by
$\alpha^r,\alpha^s\,\beta^m$. Then the index of this subgroup is given
by $|\xi|=r\,m$. The image of $H_a$ under the isomorphism
$u:(\alpha,\beta)\mapsto (T_1,T_\tau)$ determines a subgroup
$<T_r\,,\,T_{s+m\,\tau}>$ and the corresponding quotient of $\complex$
is the torus with Teichm\"uller parameter given by
\beq \tau^{\xi}=\frac{s+m\,\tau}{r} \ . \eeq
The fact that the finite index subgroups of the group ${\mathbb
  Z}\oplus{\mathbb Z}$ are all isomorphic to the group itself implies
that all unramified covers of the torus are tori. In sigma model
language, the path integral over the multi-valued fields $X^a$ on the
torus $\torus$ is constructed by calculating the path integral over
the single-valued field $\cal X$ on the covering torus and summing
over every possible $\cal X$ constructed by different choices of the
commuting pair $P,Q\in G$. For example, the genus one partition
function of the $S_3$ orbifold is given by~\cite{bantaypo}
\begin{eqnarray}
Z^{S_{3}}( \tau) &=&\mbox{$\frac{1}{6}$}\,Z( \tau)^{3} +
\mbox{$\frac{1}{2}$}\,Z( \tau ) \,\Big( Z( 2\tau ) +Z\big( 
\mbox{$\frac{\tau }{2}$}\big)
+Z\big( \mbox{$\frac{\tau +1}{2}$}\big) \Big) \nonumber \\ && +\,
\mbox{$\frac{1}{3}$}\,\Big( Z( 3\tau) +Z\big( \mbox{$\frac{\tau }{3}$}
\big) +Z\big( \mbox{$\frac{\tau +1}{3}$}\big) +Z\big( 
\mbox{$\frac{\tau +2}{3}$}\big) \Big) \ .
\label{S3orbpart}\end{eqnarray}
Note that the individual terms in (\ref{S3orbpart}) are not modular
invariant, but their sum is.

\subsection{Symmetric Products\label{SymProds}}

Permutation orbifolds whose twist group $G$ is the full symmetric
group $S_N$ are called symmetric products ${\rm Sym}^N({\cal
  C}):=({\cal C})^{\otimes N}/S_N$. In this case the formula
(\ref{partf}) takes into account all $N$-sheeted coverings. Starting
from a fixed parent theory $\cal C$ and a given worldsheet genus $g$,
the generating function of partition functions for all $N$ can be
written in a closed form thanks to a combinatorial identity due to
B\'antay~\cite{bantaysymm}. This identity translates the sum over
homomorphisms in (\ref{partf}) to a sum over finite index subgroups of
the group $\Gamma=\pi_1(\Sigma)$ and is given by
\beq 1+\sum_{N=1}^{\infty}\,\frac{1}{N!}~
\sum_{\Phi:\Gamma\to S_N}\Big(\,\prod_{\xi\in{\cal O}(\Phi)}\,
{\cal Z}(\Gamma_{\xi})\,\Big)=\exp\Big(\,\sum_{H<\Gamma}\,
\frac{{\cal Z}(H)}{[\Gamma:H]}\,\Big) \ ,
\label{combi} \eeq
where $\Gamma_\xi$ is the stabilizer of the orbit $\xi$ and
$[\Gamma:H]$ denotes the index of the subgroup $H$ in $\Gamma$. The
formula (\ref{combi}) holds generally for any finitely generated group
$\Gamma$ and any conjugation invariant function $\cal Z$ (\emph{i.e.},
${\cal Z}(\gamma\,H\,\gamma^{-1})={\cal Z}(H)$ for all
$\gamma\in\Gamma$) from the set of finite index subgroups of $\Gamma$
to a commutative ring $R$.

The proof of (\ref{combi}) is instructive. A given
term $\prod_{\xi}\,{\cal Z}(\Gamma_\xi)$ in the sum
on the left-hand side of (\ref{combi}) depends only on the equivalence
class of the homomorphism $\Phi$. An equivalence class can be written
as
\beq
[\Phi]=\bigoplus_{k=1}^N\, n_k\,\phi_k \ ,
\label{equivclassdecomp}\eeq
where $\phi_k$ is a transitive equivalence class whose orbits all have
length $k$ and $n_k\geq0$ is its
integer multiplicity with $\sum_k\,n_k=N$. One can then rewrite the
product $\prod_{\xi}\,{\cal Z}(\Gamma_\xi)=\prod_k\,{\cal
  Z}(\Gamma_k)^{n_k}$, where $\Gamma_k$ is the stabilizer subgroup of
an arbitrary representative of the image of $\phi_k$ in $S_N$. The cardinality
of the equivalence class $[\Phi]$ can be determined as follows. The
total number of possible elements to conjugate with is $|S_N|=N!$, but
not all of these give inequivalent homomorphisms $\Phi$. The permutations
which exchange the orbits that have the same $S_N$-action do not
change $[\Phi]$, so we have to divide by their number which is
$n_k!$. Finally, we have to divide out the number of cosets
$\gamma\,\Gamma_k$ with $\gamma\,\Gamma_k\,\gamma^{-1}=\Gamma_k$,
which is the index $\gamma_k=[N_\Gamma(\Gamma_k):\Gamma_k]$ of the
stabilizer $\Gamma_k$ in its normalizer subgroup
$N_\Gamma(\Gamma_k)$. Thus $|[\Phi]|=N!/\prod_k\,
n_k!\,\gamma_k^{n_k}$.

One can now rewrite the left-hand side of (\ref{combi}) as
\bea
1+\sum_{N=1}^{\infty}\, \frac{1}{N!}
   \sum_{\stackrel{\scriptstyle\{n_k\}}{\scriptstyle n_1+
\cdots+n_N=N}}~ \frac{N!}{\prod\limits_{k=1}^N\, n_k!\,\gamma_k^{n_k}}~
   \Big(\,\prod_{k=1}^N\, {\cal Z}(\Gamma_k)^{n_k}\,\Big)&=&
   \prod_{k=1}^\infty\,\Big(\,\sum_{n_k=0}^{\infty}\,
\frac{{\cal Z}(\Gamma_k)^{n_k}} {n_k!\,\gamma_k^{n_k}}\,\Big)
\nonumber\\[4pt] &=&
\prod_{k=1}^\infty\,\exp\Big(\frac{{\cal Z}(\Gamma_k)}{\gamma_k}\Big) \ .
\eea
Note that here a summation over conjugacy classes of index $k$
subgroups is implicitly assumed. The final step consists in rewriting
the product of exponentials as the exponential of a sum over $k$, and
then translating the latter summation into a sum over index $k$
subgroups. There are $[\Gamma:N_\Gamma(\Gamma_k)]$ distinct subgroups
in the conjugacy class of $\Gamma_k$ (as
$\gamma\,\Gamma_k\,\gamma^{-1}\neq\Gamma_k$ if $\gamma\notin
N_\Gamma(\Gamma_k)$), so we need to divide by this number if we wish
to sum over all index $k$ subgroups. Then the resulting factor in the
denominator
\beq \gamma_k\,\big[\Gamma:N_\Gamma(\Gamma_k)\big]
\=\big[N_\Gamma(\Gamma_k):\Gamma_k\big]\,
\big[\Gamma:N_\Gamma(\Gamma_k)\big]\=[\Gamma:\Gamma_k]\=k \eeq
is precisely the index of $\Gamma_k$ in $\Gamma$ and we have arrived
at (\ref{combi}).

Let us now apply the identity (\ref{combi}) to the uniformizing group
$\Gamma=u(\pi_1(\Sigma))$ of a compact Riemann surface
$\Sigma=\Sigma_\tau$ with the definition
\beq {\cal Z}(H):=Z\big(\tau^H\big)~\kappa^{[\Gamma:H]} \label{gcpf}
\eeq
where $Z(\tau^H)$ is the modular invariant partition function of $\cal
C$ defined on the surface $\Sigma_{\tau^H}=U/H$, with $U$ the universal 
cover of $\Sigma_{\tau}$, and $\kappa$ is a formal variable which is
determined by physical constants in applications. The result is the
grand canonical partition function
\beq Z^{\rm Sym}(\tau,\kappa)~:=~1+\sum_{N=1}^\infty\,\kappa^N~
Z^{S_N}(\tau)\=\exp\Big(\,\sum_{N=1}^\infty\,\kappa^N~{\cal H}_N
Z(\tau) \,\Big)  \ ,
\label{combim} \eeq 
where $Z^{S_N}(\tau)$ is the partition function for the $S_N$ orbifold
given by the formula (\ref{partf}) and the operator ${\cal H}_N$ is
defined on modular invariant functions by
\beq {\cal H}_NZ(\tau)=\frac{1}{N}\,\sum_{[\Gamma:H]=N}\, Z\big(\tau^H
\big) \ . \label{hecke} \eeq
Note that the product over the orbits $\xi$ in (\ref{combi}) gives a
sum for the power of $\kappa$ equal to
$\sum_\xi\,[\Gamma:\Gamma_\xi]=\sum_\xi\,|\xi|=N$. This generating
function is a sum over all possible (finite-sheeted) covers of the
surface $\Sigma$ that the parent conformal field theory $\cal C$ is
defined on, and its logarithm gives the restricted sum over connected
covers. The operator defined by (\ref{hecke}) yields a sum over
subgroups $H<\pi_1(\Sigma)$ of index $N$, and in the case of the
torus $\Sigma=\torus$ it coincides with the Hecke operator acting on
the partition function of the parent theory by
\begin{equation}
{\cal H}_NZ(\tau)=\frac{1}{N}\,\sum_{r\,m=N}~
\sum_{s\in\zed/r\,\zed}\,Z\big(\mbox{$\frac{s+m\,\tau}{r}$}\big) \ .
\label{hecke1loop}\end{equation}

\subsection{Sigma Models at One-Loop}

Our primary example of a permutation orbifold in this paper will be
that of sigma models on symmetric products of flat space
$\real^d$ at one-loop order in string perturbation theory. Let us
describe this example explicitly in the case of a single boson $X$ in
$\real$. The path integral of the sigma model conformal field theory
on a symmetric product is gotten by considering the grand canonical
partition function
\begin{equation}
  Z^{\rm Sym}(\tau,\kappa)=1+\sum_{N=1}^\infty\,\kappa^N~
\sum_{\stackrel{\scriptstyle P,Q\in S_N}{\scriptstyle P\,Q=Q\,P}}\,\frac{1}{N!}~
\int_{(P,Q)}\,{\cal D}X^1\cdots
  {\cal D}X^N~\exp{\Big(-\sum_{a=1}^N\,I(X^a)\Big)} \ ,
\label{sigmaZSym}\end{equation}
where 
\beq
I(X)=\frac{1}{4\pi\,\alpha'}\,\int_\torus\,
\dd^2z~\frac{1}{2\ii\tau_2}\,\partial X
(z)\,\overline{\partial}X(z)
\label{bosPolyakov}\eeq
is the bosonic Polyakov action and $z=\sigma^1-\tau\,\sigma^2$,
$\sigma^1,\sigma^2\in [0,1]$ are complex coordinates on the torus with
respect to the complex structure $\tau=\tau_1+\ii\tau_2$,
$\tau_1\in\real,\tau_2>0$. The sum over commuting pairs of
permutations, specifying monodronomy homomorphisms
$\Phi:\pi_1(\torus)\to S_N$, is taken over worldsheet instantons of
the field theory labelled by the boundary conditions
(\ref{mbond}). Note that any metric on the torus can be written as
\beq
\dd s^2=\e^{2\phi(z)}~|\dd z|^2
\label{torusmetric}\eeq
where the scalar field $\phi(z)$ on $\torus$ is an arbitrary conformal
factor.

From the general formulas (\ref{combim}) and (\ref{hecke1loop}) above
it follows that the partition function (\ref{sigmaZSym}) is given by
the combinatorial formula
\begin{equation}
  \label{eq:firsthecke}
  Z^{\rm Sym}(\tau,\kappa)
=\exp{\Big(\,\sum_{N=1}^\infty\,\kappa^N~\sum_{r\,m=N}~
\sum_{s\in\zed/r\,\zed}\,\frac{1}{N}~\mathfrak{z}\big(
\mbox{$\frac{s+m\,\tau}{r}$}\big)\,\Big)} \ ,
\end{equation}
where
\beq
\mathfrak{z}(\tau)=\int\,{\cal D}X~\e^{-I(X)}
\label{sigmazsingle}\eeq
is the sigma model partition function on the torus with target space
$\reals$. This gives a sum of the partition function on a particular
torus $\torus$ over the discrete set of covering tori. The Gaussian
integral (\ref{sigmazsingle}) can be evaluated in terms of a Quillen
norm as
\beq
\mathfrak{z}(\tau)=\left(\frac{\mbox{vol}(\torus)\,\det'\Delta}
{4\pi^2\,\alpha'}\right)^{-{1}/{2}}
\label{sigmazdets}\eeq
where $\Delta$ is the scalar Laplacian operator on $\torus$ with
respect to the torus metric (\ref{torusmetric}), $\mbox{vol}(\torus)$
is the volume of the surface $\torus$ in (\ref{torusmetric}), and
$\det'\Delta$ denotes the determinant of $\Delta$ with zero modes
excluded. At genus one, this determinant has a natural holomorphic
splitting and $\mathfrak{z}(\tau)$ is a section of the determinant
line bundle
$\underline{\det}\,(\,\overline{\partial}\,)^{-{1}/{2}}\otimes\,
\underline{\det}\,(\partial)^{-{1}/{2}}$ over the moduli space of
complex structures on $\torus$. The determinant of the Dolbeault
operator $\overline{\partial}$ is the automorphic form on
Teichm\"uller space given by
\beq
{\det}'~\overline{\partial}=\e^{S_{\rm L}(\phi)/{24\pi}}~\eta(\tau)^2
\ ,
\label{detdelbar}\eeq
where $S_{\rm L}(\phi)$ is the Liouville action and
$\eta(\tau)=\e^{\pi\ii\tau/12}\,\prod_{n\in\nat}\,(1-\e^{2\pi\ii n\,\tau})$ is
the Dedekind function. The partition function (\ref{sigmazsingle}) is
thus given explicitly by
\begin{equation}
  \label{eq:evaldet}
  \mathfrak{z}(\tau)=\e^{-S_{\rm L}(\phi)/{24\pi}}~\Big(\,
\frac1{4\pi^2\,\alpha'}\,\int_\torus\, \dd^2z~\e^{\phi(z)}\,
\Big)^{-{1}/{2}}~\frac{1}{\big|\eta(\tau)\big|^2} \ .
\end{equation}

By replacing $\mathfrak{z}(\tau)$ with $\mathfrak{z}(\tau)^d$ in
(\ref{eq:firsthecke}) we get the corresponding result for the parent
conformal field theory of a free boson on the target space
$\real^d$. Moreover, the combinatorial formula (\ref{eq:firsthecke})
is completely generic and holds for any sigma model partition function
on the torus. For example, we may simply replace $\mathfrak{z}(\tau)$
by the appropriate superstring or heterotic string partition functions
at one-loop (with some modifications that we discuss in
Section~\ref{FermOrb}).

The formula (\ref{combim}) can also be used to compute any correlation
function of fields which are unaffected by the orbifolding. These are the
operators which are symmetric under permutations of the indices of the
scalar field $X$. Given any function $f$, we use the notation $\Tr
f(X):=\sum_a\,f(X^a)$ for such an operator refering to a diagonal
matrix of the $N$ independent fields $X^a$. The (normalized)
correlation function is defined by
\bea
  \label{eq:defcorrelation}
&& \big\langle\Tr f(X)\,\big\rangle^{\rm Sym}(\tau,\kappa)\\ &&
\qquad ~:=~\frac1{Z^{\rm Sym}(\tau,\kappa)}\,\Big(1+
\sum_{N=1}^\infty\,\frac{\kappa^N}{N!}~\sum_{\stackrel{\scriptstyle
    P,Q\in S_N}{\scriptstyle P\,Q=Q\,P}}~
\int_{(P,Q)}\,{\cal D}X^1\cdots{\cal D}X^N ~\Tr f(X)~\e^{-\Tr I(X)}
\Big) \ . \nonumber
\eea
Rather than trying to determine the combinatorics of this amplitude
directly, we will calculate instead the generating function
\begin{equation}
  \label{eq:generatingfunc}
Z^{\rm Sym}_\zeta(\tau,\kappa)~:=~
\big\langle\e^{\zeta\Tr f(X)}\,\big\rangle^{\rm
  Sym}(\tau,\kappa)\= \sum_{n=0}^\infty\,\big\langle\big(\Tr f(X)\,
\big)^n\big\rangle^{\rm Sym}(\tau,\kappa)~\frac{\zeta^n}{n!} \ .
\end{equation}
Then we can get the correlation function (\ref{eq:defcorrelation}) by
differentiation as
\beq
\big\langle\Tr f(X)\,\big\rangle^{\rm Sym}(\tau,\kappa)=
\left.\frac{\partial
Z^{\rm Sym}_\zeta(\tau,\kappa)}{\partial\zeta}\right|_{\zeta=0} \ .
\label{symcorrdiff}\eeq

The generating function (\ref{eq:generatingfunc}) is just the
symmetric product partition function of the sigma model conformal
field theory with a shifted action
\beq
I_\zeta(X)=I(X)-\zeta\,f(X)
\label{shiftedaction}\eeq
and the normalization $Z^{\rm Sym}_{\zeta=0}(\tau,\kappa)=1$. It can
thus be calculated by using the combinatorial formulae (\ref{combim})
and (\ref{hecke1loop}) as above, with the result
\begin{equation}
  \label{eq:modifiedpartionon}
  Z^{\rm Sym}_\zeta(\tau,\kappa)=\frac1{Z^{\rm
        Sym}(\tau,\kappa)}\,\exp\Big(\,\sum_{N=1}^\infty\,
\frac{\kappa^N}{N}~\sum_{r\,m=N}~\sum_{s\in\zed/r\,\zed}\,
\mathfrak{z}_\zeta\big(\mbox{$\frac{s+m\,\tau}{r}$}\big)\,\Big)
\end{equation}
where
\beq
\mathfrak{z}_\zeta(\tau)=\int\,{\cal D}X~\e^{-I_\zeta(X)}
\label{partsinglezeta}\eeq
is the sigma model partition function on the torus with respect to the
modified action (\ref{shiftedaction}). To carry out the
differentiation in (\ref{symcorrdiff}), we first calculate
\beq
\left.\frac{\partial\,\mathfrak{z}_\zeta(\tau)}{\partial\zeta}
\right|_{\zeta=0}=\big\langle f(X)\big\rangle(\tau)
\label{diffpartsingle}\eeq
where the (unnormalized) expectation values are calculated as Gaussian
moments with respect to the original action
(\ref{bosPolyakov}). Combining these results along with the elementary
identity $\frac{\dd}{\dd\zeta}\e^{F(\zeta)}=F'(\zeta)~\e^{F(\zeta)}$
gives finally
\beq
\big\langle\Tr f(X)\,\big\rangle^{\rm Sym}(\tau,\kappa)=
\sum_{N=1}^\infty\,\frac{\kappa^N}N~
\sum_{r\,m=N}~\sum_{s\in\zed/r\,\zed}\,
\big\langle f(X)\big\rangle\big(\mbox{$\frac{s+m\,\tau}{r}$}\big) \ .
\label{symcorrfinal}\eeq
The correlation function of the symmetric operator $\Tr f(X)\,$ in the
symmetric product is thus likewise expressed in terms of the correlation
function of the operator $f(X)$ on all unramified covering spaces over
the base torus $\torus$. These formulae have natural extensions to
higher loops, but in those instances they require knowledge of the
correlation functions of $f(X)$ on all higher genus Riemann surfaces.

\subsection{Twist Fields\label{TwistFields}}

A twist field $\sigma_P(w)$ in a generic permutation orbifold ${\cal
  C}\wr {G}$ is a primary field that creates the vacuum state
of a twisted sector at a point $w\in\Sigma$. In a sigma model
conformal field theory, its insertion results in non-trivial local
monodromy
\beq X^a\big((z-w)~\e^{2\pi\ii}\big)\,\sigma_P
(w)=X^{P(a)}(z)\,\sigma_P(w) \label{locmon} \eeq
where the permutation $P$ is an element of the twist group $G<S_N$.
Its effect is to thus make the local field $X$ multi-valued about
the insertion point $w\in\Sigma$. If $P=(n)$ consists of a single
cycle of length $n>1$, then the corresponding twist field
$\sigma_{(n)}(w)$ permutes $n$ copies of $\cal C$ in a
$\zed_n$-twisted sector and is a primary field with conformal
weight~\cite{ArFro1}
\beq \Delta_{(n)}=\mbox{$\frac{d}{24}\,\big(n-\frac{1}{n}\big)$}
\label{Deltand}\eeq
for a $d$-dimensional boson. The corresponding fields $X^{a_i}(z)$,
$i=1,\dots,n$ can then be glued together into one field ${\cal X}(z)$
which is identified with a long string of length $n$.

In the general case, we have seen that twisted sectors are
in one-to-one correspondence with conjugacy classes of $G$. The
conjugacy class $[P]$ of an element $P\in S_N$ can be decomposed into
combinations of cyclic permutations as $[P]=\prod_n\,(n)^{N_n}$ with
$N_n\geq0$ and $\sum_n\,n\,N_n=N$. For a bosonic sigma model in $d$
dimensions, the corresponding twist field has conformal dimension
\beq
\Delta_P\=\sum_{n=1}^N\,N_n\,\Delta_{(n)}\=
\frac d{24}\,\Big(N-\sum_{n=1}^N\,\frac{N_n}n\Big) \ .
\label{DeltaPgen}\eeq
An $S_N$-invariant twist field creating the twisted sector $[P]$ of
the permutation orbifold is defined by averaging over all twist fields
in the conjugacy class of $P$ to get
\beq
\sigma_{[P]}(w)=\frac1{N!}\,\sum_{g\in S_N}\,\sigma_{g\,P\,g^{-1}}(w)
\ .
\label{invtwistfield}\eeq

In this paper we will be primarily interested in correlation functions
$\langle\sigma_{[P_1]}(w_1)\cdots\sigma_{[P_k]}(w_k)\rangle^G$ of twist
field operators in the permutation orbifold ${\cal C}\wr G$. These
averages are difficult to calculate directly within a
path integral formalism, because the twist fields are non-local
operators. However, since these correlation functions are the vacuum
functionals with twisted boundary conditions due to (\ref{locmon}),
it is natural to extend the covering surface principle as
in~\cite{Hamidi:1986vh}--\cite{Lunin:2000yv} and compute them via a
generalization of the permutation orbifold partition function
(\ref{partf}) on a Riemann surface $\Sigma$ of genus $g>0$. Whenever
we have twist fields inserted at $k$ distinct points
$\underline{w}:=\{w_1,\dots,w_k\}$ of the worldsheet, a twisted sector
is given by a conjugacy class of homomorphisms
$\Phi:\pi_1(\Sigma_{\,\underline{w}\,})\to G<S_N$ where
$\Sigma_{\,\underline{w}\,}:=\Sigma\setminus\,\underline{w}$ is the
marked Riemann surface with the $k$ twist field insertion points
deleted. It is restricted by admissibility criteria which require that
the images of the generators $\gamma_i$ of
$\pi_1(\Sigma_{\,\underline{w}\,})$ which are contractible to $w_i$
must be simple cycles of length $\nu_i>1$ if a ${\mathbb
  Z}_{\nu_i}$ twist field $\sigma_{(\nu_i)}(w_i)$ is inserted at
$w_i$. Each such homomorphism $\Phi$ determines a cover of the
worldsheet $\Sigma$ on which a single new field ${\cal X}(z)$, defined
by a formula analogous to (\ref{calXdef}), is single-valued. Namely,
after going around a curve $\gamma$ which is closed on the marked
worldsheet $\Sigma_{\,\underline{w}\,}$, one sews the fields
$X^{\Phi(\gamma)(a)}(z)$ into ${\cal X}(z)$. Thus, the contribution to
the correlation function from the worldsheet instanton sector determined by
the homomorphism $\Phi$ is the free partition function on the cover of
$\Sigma$ determined by $\Phi$.

While the sum arising in the orbifold partition function (\ref{partf})
is only over unramified covers $\hat\Sigma$ of $\Sigma$, the twist
field correlation functions involve sums over branched covers
$\hat\Sigma_{\,\underline{\hat w}\,}$ where
$\underline{\hat w}:=f^{-1}(\,\underline{w}\,)$ is the set of pre-images of
the set $\underline{w}$ under the covering map
$f:\hat\Sigma\to\Sigma$. The Riemann-Hurwitz formula
for the genus $\hat g$ of the covering space with the given monodromy
homomorphism is the general one for covers with ramification given by
\beq \hat g\=N\,(g-1)+1+\mbox{$\frac{B}{2}$} \qquad\mbox{with}\quad
B\= \sum_{i=1}^k\, (\nu_i-1) \ , \label{rh} \eeq
where $\nu_i$ is the ramification index given by the length of the
cycle of the $i$-th primary twist field. As before, we have to take
into account those homomorphisms $\Phi$ whose image does not act
transitively on the coordinate labels $a=1,\dots,N$. In this case the
simple cycle condition for fixed length $\nu_i$ has to hold for each
orbit $\xi$. This ensures that the genus of the connected component of
the cover determined by the action of
$\Phi(\pi_1(\Sigma_{\,\underline{w}\,}))$ on each orbit $\xi$ is equal
to $\hat g$. We may now write down a formula analogous to
(\ref{partf}) for the normalized $k$-point correlation function of
twist field operators given by
\beq
\Big\langle\,\prod_{i=1}^k\,
\sigma_{[P_i]}(w_i)\,\Big\rangle^G=\frac{1}{|G|}\,
\sum_{\Phi:\pi_1(\Sigma_{\,\underline{w}\,})\to G}~
\frac1{Z^G(\tau)}~\Big(\,
\prod_{\xi\in {\cal O}(\Phi)}\,Z\big(\tau^{\xi,\,\underline{w}\,}
\big)\,\Big) \ , \label{correlator} \eeq
where $\tau^{\xi,\,\underline{w}\,}$ is the complex structure of the
covering surface determined by the worldsheet modulus $\tau$, the
stabilizer $\pi_1(\Sigma_{\,\underline{w}\,})_\xi$, and the branch
point loci $\underline{w}$.

There are three crucial differences between the formulae
(\ref{correlator}) and (\ref{partf}). Firstly, the twist field
correlation functions are not expressed in terms of correlation functions but
instead in terms of partition functions. Secondly, there is a
restriction on the admissible homomorphisms $\Phi$ to ensure that they
have the prescribed monodromy around the punctures,
\emph{i.e.}, $\Phi(\gamma_i)$ has to be a simple cycle of length
$\nu_i$ in each orbit. Thirdly, while the uniformization theorem
provided us with a computational recipe for obtaining the
Teichm\"uller coordinate $\tau^{\xi}$ in terms of $\tau$ via knowledge
of $\Phi$, it does not apply to the twist field $k$-point
functions. The reason is that $\tau$ parametrizes the uniformizing
group of the compact Riemann surface $\Sigma$, which is isomorphic to
$\pi_1(\Sigma)$, while the domain of the monodromy homomorphism $\Phi$
is $\pi_1(\Sigma_{\,\underline{w}\,})$ which differs from the domain
of the isomorphism from the abstract group $\pi_1(\Sigma)$ to the
uniformizing group $u(\pi_1(\Sigma))$. Therefore, the complex
structure of the ramified cover $\hat\Sigma_{\,\underline{\hat w}\,}$ is a
function of that of the base space $\Sigma$, the locations
$\underline{w}$ of the branch points, and the monodromy homomorphism
$\Phi$.

We are also interested in twist field correlation functions on symmetric
products. In order to apply a version of (\ref{combi}) we need to pass
the constraint, which is imposed on the admissible homomorphisms
$\Phi$ in (\ref{correlator}), to the definition of the function ${\cal
  Z}(H)$. Let us specialize the discussion to the torus
$\Sigma=\torus$ for definiteness. In this case, the genus of the
covering surface $\hat\Sigma$ is $\hat g$ whenever its branching
number is $B=2(\hat g-1)$. A standard presentation of the 
fundamental group of the marked torus is given by
\beq \Gamma~:=~\pi_1({\mathbb T}_{\,\underline{w}\,})\=\,
<\alpha,\beta,\gamma_1,\dots,\gamma_k~\big|~
[\alpha,\beta]\,\gamma_1\cdots\gamma_k=1> \ . 
\label{fundgpmarked}\eeq
To each $N$-sheeted cover of $\torus$ there corresponds a conjugacy
class of subgroups of $\Gamma$ of index~$N$~\cite{ezell}, which is the
stabilizer of the monodromy homomorphism $\Phi$ acting in $S_N$. Note
that the group (\ref{fundgpmarked}) is isomorphic to the free group on
$k+1$ generators $\alpha,\beta,\gamma_1,\dots,\gamma_{k-1}$, and any
subgroup of a free group is also free. This is consistent with the
fact~\cite{ezell} that the stabilizer subgroup is isomorphic to
$\pi_1(\hat\Sigma_{\,\underline{\hat w}\,})<\pi_1({\mathbb
  T}_{\,\underline{w}\,})$. To decide when a given finite index
subgroup $H<\Gamma$ corresponds to a stabilizer subgroup of an
admissible homomorphism $\Phi$ in (\ref{correlator}), we proceed as
follows. Let $\hat\imath:\hat\Sigma_{\,\underline{\hat w}\,}\hookrightarrow
\hat\Sigma$ be the natural inclusion of surfaces. The induced
homomorphism $\hat\imath_*:\pi_1(\hat\Sigma_{\,\underline{\hat w}\,})\rightarrow
\pi_1(\hat\Sigma)$ is then the natural forgetful map. Since
$H\cong\pi_1(\hat\Sigma_{\,\underline{\hat w}\,})$, a formal criterion for
the admissibility of a finite index subgroup
$H<\pi_1(\torus_{\,\underline{w}\,})$ is given by
\beq
H/\ker\big(\hat\imath_*\big)\cong \pi_1\big(\hat\Sigma\big) \ .
\label{gcov}\eeq
We can use (\ref{gcov}) to check whether a given subgroup $H$ is
admissible. If the quotient is defined and it yields a group
isomorphic to $\pi_1(\hat\Sigma)$, then $H$ is admissible. This
property does not depend on the conjugacy class of $H$ in
$\pi_1(\torus_{\,\underline{w}\,})$. We can thus give an implicit
definition for the function appearing in (\ref{combi}) as
\beq
{\cal Z}(H):=
\left\{\begin{array}{ll} \displaystyle
\frac{Z\big(\tau^{H,\,\underline{w}\,}\big)}
{Z^{S_N}(\tau)}~\kappa^{[\Gamma:H]} & \mbox{if}\;
H\;\mbox{satisfies}\;(\ref{gcov}) \ , 
\\[4pt] 0 & \mbox{otherwise}\ . \end{array}\right.
\label{gcpftwist}\eeq
We may then apply the formula (\ref{combi}) to get the generating
function of twist field correlation functions.

In the following we will apply this formalism to study the
perturbation of the sigma model conformal field theory, on the
symmetric product of $\real^d$, by an irrelevant operator of conformal
dimension $\frac32$. For this, we introduce the bosonic
Dijkgraaf-Verlinde-Verlinde (DVV) interaction
vertex~\cite{Dijkgraaf:1997vv,Rey:1997hj} which is defined with
respect to the $\zed_2$ twist field $\sigma_{ab}(w)$ corresponding to
the transposition in $S_N$ that interchanges the fields
$X^a$ and $X^b$ while leaving all others invariant. These twist fields
generate the elementary joining and splitting of strings in the
symmetric product, and they can be built out of standard
$\zed_2$ orbifold twist
operators~\cite{Dixon:1986qv,Hamidi:1986vh}. Then the translationally
invariant vertex operator is defined by
\beq
V_{\rm bos}=-\frac{\lambda\,N}{{\rm vol}(\torus)}\,\int_\torus\,
\dd\mu(z)~\sum_{1\leq a<b\leq N}\,\sigma_{ab}(z) \ ,
\label{DVVint}\eeq
where $\lambda$ is a coupling constant proportional to the string
coupling $g_s$. In contrast to the originally proposed genus zero
case~\cite{Dijkgraaf:1997vv,ArFro1,Rey:1997hj}, we will find that the
DVV vertex operator at genus one needs to be defined using a
non-constant measure $\dd\mu(z)=\dd^2z/\mu(z)$ on the torus
$\torus$. It will be determined explicitly in the ensuing sections (as
will the coupling constant $\lambda$) by modular invariance
requirements. When $d=24$, the twist field $\sigma_{ab}(w)$ is a
primary field of conformal weight $\frac32$. Starting from the
one-loop action (\ref{bosPolyakov}), the interacting symmetric product
sigma model is defined by the action
\beq
I_{\rm int}^{S_N}(X)=\Tr I(X)+V_{\rm bos}
\label{intsigmamodel}\eeq
with $\Tr I(X)=\sum_a\,I(X^a)$.

In this paper we will compute the leading order effect of this
perturbation. Using translational invariance of the sigma model path
integral to move one of the branch points to the origin $z=0$, we are
thus interested in computing the translationally invariant correlator
\beq
\big\langle\NO V_{\rm bos}\,V_{\rm bos}\NO\big\rangle^{S_N}=
\frac{\lambda^2\,N^2}{{\rm vol}(\torus)\,\mu(0)}~\sum_{a_i<b_i}~
\int_{\torus}\,\dd\mu(z)~\big\langle\sigma_{a_1b_1}(z)\,
\sigma_{a_2b_2}(0)\big\rangle^{S_N} \ .
\label{transinvcorr}\eeq
The computation of the two-point functions in (\ref{transinvcorr})
specializes the above discussion to the case $g=1$, $\hat g=2$, and
$k=2$. There are two simple branch points with ramification indices
$\nu_1=\nu_2=2$ and $\Gamma=\pi_1({\mathbb T}\setminus\{z,0\})$. Then
the logarithm of the generating function (\ref{combi}) with the
definition (\ref{gcpftwist}) is given by a sum over the modular
invariant vacuum amplitudes on all connected $N$-sheeted genus two covers
$\hat\Sigma$ with two fixed simple branch points. In this case the
first quantized modular invariant partition function for the parent
theory is the two-loop version of (\ref{sigmazdets}) on $\hat\Sigma$
(with vanishing Liouville field $\phi=0$ for simplicity) given
by~\cite{Belavin:1986tv,Moore1}
\beq
\mathfrak{z}^{(2)}(\tau)=\frac{\big(\det({\rm Im}\,\tau)\big)^{3-d/2}}
{\big(4\pi^2\,\alpha'\,\big)^{-d/2}\,\big|\Psi_{10}(\tau)\big|^2} \ ,
\label{Z2bos}\eeq
where $d$ is the spacetime dimension ($d=26$ for the critical bosonic
string). Here $\Psi_{10}(\tau)$ is the genus two parabolic modular
form of weight ten with no zeroes or singularities (the Igusa cusp
form), defined on the Siegel half-space $\mathbb{U}^2=\{\tau~|~{\rm
  Im}(\tau_{11})>0~,~{\rm Im}(\tau_{22})>0~,~\det({\rm Im}\,\tau)>0\}$
of $2\times2$ Riemann period matrices $\tau$ with the boundary
component $\halfplane\times\halfplane$ consisting of diagonal matrices
removed. It can be expressed in terms of the ten genus two
theta-constants $\Theta({}^{\mbf a}_{\mbf b})(\tau):= \Theta({}^{\mbf
  a}_{\mbf b})(0,0|\tau)$ with even binary characteristics $\mbf
a=(a_1,a_2),\mbf b=(b_1,b_2)\in\zed^2/2\zed^2$ as
\beq
\Psi_{10}(\tau)=2^{-12}\,\prod_{\mbf a\cdot\mbf b\equiv0\,{\rm
    mod}\,2}\,\Theta\big({}^{\mbf a}_{\mbf b}
\big)(\tau)^2 \ .
\label{Psi10Omega}\eeq

\subsection{Thermodynamics of DLCQ Strings\label{DLCQ}}

In the genus one case $\Sigma=\torus$, the logarithm of the right-hand
side of (\ref{combim}) coincides with the free energy of second
quantized string theory on the target space
$M\times\sphere^1\times\real$ when the parent theory is the
corresponding conformal field theory on the spacetime $M$ in the free
string limit $g_s\to0$~\cite{DMVV1,GS1}. The matching is
provided by identifying the modulus of the worldsheet and that of the
spacetime torus, where the second compact direction is timelike and
is generated by the trace taken in computing the free energy
amplitude. Its radius is identified with the inverse temperature
$\beta$. In discrete light cone quantization (DLCQ), the light cone
Hamiltonian and momentum are given by
\beq
H\=P^+ \qquad \mbox{and} \qquad P^-\=N/R
\label{DLCQHP}\eeq
where $R$ is the radius of the compactified light-like direction
$x^+\in\sphere^1$ and $N\in\nat_0$. The thermodynamic free energy
$F^{(1)}_{\rm DLCQ}$ is then defined by
\beq
\e^{-\beta\,F^{(1)}_{\rm DLCQ}}\=
\Tr\e^{-\frac{\beta}{\sqrt2}\,(P^++P^-)}\=
\sum_{N=0}^\infty\,\e^{-\beta\,N/\sqrt2\,R}~\Tr_{{\cal H}_N}
\e^{-\beta\,P^+/\sqrt2} \ ,
\label{FDLCQ1loop}\eeq
where ${\cal H}_N$ denotes the sector of the physical Hilbert space
with definite total light cone momentum $P^-=N/R$. The trace over this
subspace can be computed by using the mass-shell relation
$P^+=H^\perp/P^-$, where $H^\perp$ is the Hamiltonian for the
transverse degrees of freedom along $M$.

In this way one arrives at the expression (\ref{combim}) with the
definition (\ref{hecke1loop}) and $\kappa:=\e^{-\beta/\sqrt2\,R}$. The
Teichm\"uller parameter of the base torus $\torus$ on which the string
bits live is
\beq
\tau^\bullet:=\frac{4\pi\ii\alpha'}{\sqrt2\,\beta\,R} \ .
\label{taubulletdef}\eeq
The qualitative reason for the equivalence is that the second quantized
vacuum amplitude is given by the integral of the conformal field theory
partition function over the moduli space of complex structures, but
the only contributing surfaces at one-loop order are those which arise
by winding the string around the compact directions. In other words,
only the discretized moduli space of unramified covers of the torus
$\torus$ is summed over and taking the logarithm eliminates the
disconnected covers.

When $M=\real^{24}$ one finds that the DLCQ partition function for
bosonic string theory coincides exactly with the partition function of
the symmetric product in the limit $N\to\infty$, with the length $n_i$
of a long string identified with the light cone momentum $P_i^-=n_i/R$
for $i=1,\dots,24$. Checking the equivalence of
perturbative bosonic string dynamics and the corresponding interacting
symmetric product of $\real^{24}$ beyond the free string limit $g_s\to0$
requires computing the thermal free energy in DLCQ at higher genus and
the appropriate amplitudes in the permutation orbifold perturbed by
the DVV interaction vertex (\ref{DVVint}). The former amplitudes
truncate to sums over branched covers of the spacetime torus $\torus$
arising in the null compactification at finite
temperature~\cite{Grignani:2000zm}, while the local structure of the
operator $V_{\rm bos}$ matches nicely with the cubic string interaction
vertices in light cone Green-Schwarz string field
theory~\cite{Dijkgraaf:2003nw}--\cite{Kishimoto:2006xc}. In this
setting the string interactions are generated by sewing together torus
worldsheets along branch cuts.

On the DLCQ side, the next-to-leading order contribution is the
two-loop free energy which was computed in~\cite{csz} with the result
\bea
F_{\rm DLCQ}^{(2)}\big(\tau^\bullet\,,\,\kappa\big)&=&
-g_s^2\,\left|\frac{\tau^\bullet}{32\pi^2\,\alpha'}
\right|^{12}\,\sum_{N=2}^\infty\,\frac{\kappa^N}{N^2}~
\sum_{r\,m=N}\,\left(\frac rm\right)^{10} \nonumber\\ && \times\,
\sum_{\stackrel{\scriptstyle s,t\in\zed/r\,\zed}{\scriptstyle
    t\neq0}}~\int_\triangle\,\frac{\dd^2\tau^\#}{\big(\tau^\#_2
\big)^{12}}~\big|\Psi_{10}\big(\tau_{r,m,s,t}(\tau^\bullet,\tau^\#)
\big)\big|^{-2} \ .
\label{FDLCQ2}\eea
This thermal string amplitude is just the weighted integral over a
fundamental modular domain of the
genus two bosonic string partition function (\ref{Z2bos}) with respect
to the modular invariant integration measure on the space of
$2\times2$ Riemann period matrices with diagonal matrices excluded,
but with integration domain restricted to the partially discretized
moduli space of genus two simple branched covers $\hat\Sigma$ of the
torus $\torus$ with modulus $\tau^\bullet$. The integers appearing in
(\ref{FDLCQ2}) can be assembled into the $2\times4$ matrix
\beq
{\sf M}=\begin{pmatrix}0&0&-m&0\\ r&0&-s&-t\end{pmatrix}
\label{sfMints}\eeq
which determines a homology basis for the cover in which the
push-forward $f_*:H_1(\hat\Sigma,\zed)\to H_1(\torus,\zed)$, induced
by the holomorphic covering map $f:\hat\Sigma\to\torus$, is given on a
basis of canonical homology cycles $\hat\alpha_i,\hat\beta_i$,
$i=1,2$ for $\hat\Sigma$ by
\beq
f_*\big(\hat\alpha_1\,,\,\hat\alpha_2\,,\,\hat\beta_1\,,\,
\hat\beta_2\big)=(\alpha,\beta)~{\sf M}
\label{fpushM}\eeq
with respect to a canonical homology basis $(\alpha,\beta)$ of the
base torus. It specifies the way in which the cycles of the cover
$\hat\Sigma$ wind around the cycles of $\torus$. The period matrix
$\tau\in\mathbb{U}^2\setminus(\halfplane\times\halfplane)$ of the
cover in this basis is given by the normal form
\beq
\tau_{r,m,s,t}\big(\tau^\bullet\,,\,\tau^\#\big)=
\begin{pmatrix}-\frac{s+m/\tau^\bullet}r&-\frac tr\\[4pt]
-\frac tr&\tau^\#\end{pmatrix}
\label{Omegacover}\eeq
with $\tau^\#\in\halfplane$, and the integration in (\ref{FDLCQ2}) is
taken over the standard fundamental domain
$\triangle\subset\halfplane$ for the action of the genus one modular
group $SL(2,\zed)$ on $\tau^\#$.

The diagonal elements of the period matrix (\ref{Omegacover})
naturally capture the modulus of the degree $N=r\,m$ unramified cover
of the base torus $\torus$ of modulus $(\tau^\bullet)^{-1}$, along with a
second torus of modulus $\tau^\#$. The key feature of the homology
basis in which we have expressed the genus two amplitude
(\ref{FDLCQ2}) is that the genus two theta functions appearing in
(\ref{Psi10Omega}) admit reduction to genus one theta functions on
these two tori, due to the rational-valued off-diagonal entries of
(\ref{Omegacover}). Hence the $\tau^\bullet$-dependence of the
two-loop free energy is expressible in terms of elliptic functions,
analogously to the one-loop case. Recall that the elliptic Jacobi
theta function with characteristics $a,b\in\zed/2\zed$ is defined by
\beq \theta\big({}^a_b\big)(z|\tau)=\sum_{n\in {\mathbb Z}}\,
\exp\Big(\pi \ii\tau \big(n+\mbox{$\frac a2$}\big)^2+2\pi\ii\big(n+
\mbox{$\frac a2$}\big)\,\big(z+\mbox{$\frac b2$}\big)\Big) \eeq   
along with the Erd\'elyi notation
\bea
\theta_1(z|\tau)&=&\theta\big({}^1_1\big)(z|\tau) \qquad \mbox{and}
\qquad
\theta_2(z|\tau)\=\theta\big({}^1_0\big)(z|\tau) \ , \nonumber \\[4pt]
\theta_3(z|\tau)&=&\theta\big({}^0_0\big)(z|\tau) \qquad \mbox{and}
\qquad
\theta_4(z|\tau)\=\theta\big({}^0_1\big)(z|\tau) \ .
\eea
Then one has the decompositions~\cite{csz}
\bea
\Theta\big({}^{\mbf a}_{\mbf b}\big)\big(
\tau_{r,m,s,t}(\tau^\bullet,\tau^\#)\big)&=&
\frac{\e^{\pi \ii a_2 \,b_2/2}}{N\,\sqrt{-\ii\tau^\#}}~
\sum_{n=0}^{N-1}\, (-1)^{b_2\, n}~
\theta\big({}^{a_1}_{b_1}\big)\Big(\mbox{$\big(n+\frac{a_2}{2}
\big)\,\frac{m\,t}{N}\,\Big|\,
\frac{m\,s+{m^2}/{\tau^\bullet}}{N}$}\Big)\nonumber\\ &&
\qquad\qquad\qquad\qquad \times~\theta_j\big(\mbox{$
\frac{n+{a_2}/{2}}{N}\,\big|\,
-\frac{1}{N^2\,\tau^\#}$}\big) \label{poi} \eea
where $j=2$ (resp.~$j=3$) when the integer
$a_1\,m\,t+b_2\,N$ is odd (resp.~even). For notational ease, this
formula is written after performing a projective rotation
$\tau_{r,m,s,t}(\tau^\bullet,\tau^\#)\to-\tau_{r,m,s,t}(\tau^\bullet,-\tau^\#)$
along with a reflection in the modulus $\tau^\#$.

In this paper we shall present a detailed comparison between the free
energy (\ref{FDLCQ2}) and the integrated (with respect to the 
branch point loci) two-point
correlation function (\ref{transinvcorr}) of twist fields
corresponding to transpositions, which requires the generalization of
the combinatorial identity (\ref{combim}) to coverings with two simple
branch points as explained in Section~\ref{TwistFields} above. While
the auxilliary genus one surface of modulus $\tau^\#$ above is
anticipated {\it a posteriori} on general grounds from the
Weierstrass-Poincar\'e reduction theory for branched
covers~\cite{csz}, its geometrical significance has been hithereto
unclear. In the following we will identify this torus
explicitly, which among other things will provide the transformation
from the branch point loci to the modulus $\tau^\#$ required to match
the expressions (\ref{transinvcorr}) and (\ref{FDLCQ2}), as well as
the measure $\dd\mu(z)$ and coupling constant $\lambda$ required to
define the DVV vertex operator (\ref{DVVint}) on an elliptic curve.

\newsection{$\zed_2$ Orbifolds\label{Z2Orbifolds}}

The purpose of this section is to establish the equivalence of the
two-point function for the DVV vertex operator in the symmetric product
$\real^{24}\wr\zeds_2$ with the $N=2$ contribution to the genus two free
energy (\ref{FDLCQ2}) of the bosonic DLCQ string. For the former
calculation we will exploit the known formulae~\cite{dvvc1} for the
multi-loop partition functions and twist field correlation functions on the
geometric orbifold $\sphere^1/\zed_2$. For the latter computation we
connect the form of the total reduced free energy (\ref{FDLCQ2}) to
the theory of Prym varieties for generic genus two covers of the
torus $\torus$ of modulus $\tau^\bullet$. By a theorem due to
Mumford~\cite{birkenhake}, the only coverings that generate Prym
varieties are double covers with at most two branch points, and our
case of genus two covers over an elliptic curve. Our proof puts the
covering surface principle sketched in Section~\ref{TwistFields} on
more solid ground, and provides a non-trivial explicit check for the
computation of twist field correlation functions through two rather
distinct methods.

\subsection{Target Space vs. Permutation Orbifold\label{TSvsPO}}

For later use, we begin by elucidating the correspondence between the
sigma model conformal field theories on the geometric orbifold
$\real^{24}/\zed_2$ and on the permutation orbifold
$\real^{24}\wr\zed_2$. For this, let us consider the
$\sphere^1/\zeds_2$ target space orbifold of a free boson $X$
compactified on a circle $\sphere^1$ of radius $R$, where the group
action is the reflection involution $X\mapsto-X$. On the other hand,
the permutation orbifold $\sphere^1\wr\zeds_2$ is defined on the
tensor product of the $\sphere^1$ conformal field theory with
itself. Labelling the two copies of the boson $X$ by $X^a$, $a=1,2$,
the group action of the permutation orbifold is given by $X^1\mapsto
X^2$, $X^2\mapsto X^1$. This can be compared to the geometric orbifold
group action by introducing new coordinate fields $X^\pm=X^1\pm X^2$,
so that the $\zed_2$ permutation group now acts as
$X^\pm\mapsto\pm\,X^\pm$. It follows that the permutation orbifold is
equivalent to the target space orbifold plus an independent free boson
$X^+$ on $\sphere^1$. The partition functions of the two theories are
thus related by
\beq Z^{\zeds_2}(\tau,R)={\mathfrak z}(\tau,R)~
Z_{\rm orb} (\tau,R) \ , \label{egy} \eeq
where ${\mathfrak z}(\tau,R)$ denotes the partition function of the
compactified scalar field $X^+$ and $Z_{\rm orb} (\tau,R)$ that of the
$\sphere^1/\zed_2$ theory.

It is instructive to check the identity (\ref{egy}) explicitly at
one-loop order in the decompactified circle theory. The amplitude for
the boson $X^+$ on $\sphere^1$ is given by the worldsheet instanton
sum
\beq
{\mathfrak z}(\tau,R)\={\mathfrak z}(\tau)~
{\mathfrak z}^{\rm cl}(\tau,R)~:=~
\frac{\sqrt{4\pi^2\,\alpha'}}{\sqrt{\tau_2}\,\big|\eta(\tau)
\big|^2}~\sum_{m,m'\in\zed}\,\frac{R}{\sqrt{\alpha'}}\,\exp\Big(-
\frac{\pi\, R^2\,\big|m\,\tau-m'\,\big|^2}{\alpha'\,\tau_2}\Big) \ ,
\label{ycft}\eeq
where ${\mathfrak z}(\tau)$ is the modular invariant amplitude
(\ref{eq:evaldet}) for the free boson on the real line (so that
${\mathfrak z}^{\rm cl}(\tau,R=\infty)=1$) and henceforth we set the
Liouville field $\phi=0$. The sum in (\ref{ycft}) runs over classical
solutions with the given winding numbers around the generating cycles
of a canonical homology basis. For the partition function
of the target space orbifold, we note that the oscillator part
$\mathfrak{z}(\tau)$ of the partition function (\ref{ycft}) is
independent of the radius $R$. A monodromy homomorphism $\Phi$ for an
unramified double cover of a genus one surface is characterized by a
binary pair $(\varepsilon,\delta)\in(\zed/2\zed)^2$, where $0$
(resp.~$1$) labels periodic (resp.~antiperiodic) global monodromy
around the canonical homology cycles $(\alpha,\beta)$ of the base. In
the twisted sectors, the $\zed_2$ action $X\mapsto-X$ kills
non-trivial instantons at one-loop (as a consequence of the
Riemann-Roch theorem), while the quantum parts may be computed by
equating the $\zed_2$-twisted partition function at $R=\sqrt{\alpha'}$
with that of the untwisted $\sphere^1$ theory at the self-dual radius
$R=1/\sqrt{\alpha'}$ which coincides with the multi-critical
Ashkin-Teller model. The result is~\cite{dvvc1}
\beq Z_{\rm orb}(\tau,R)=\mbox{$\frac{1}{2}$}\,
\mathfrak{z}(\tau,R)+
\left|\frac{\eta(\tau)}{\theta_2(\tau)}\right|+
\left|\frac{\eta(\tau)}{\theta_3(\tau)}\right|+
\left|\frac{\eta(\tau)}{\theta_4(\tau)}\right| \ ,
\eeq
where we have denoted the Jacobi-Erd\'elyi theta constants by
$\theta_i(\tau):=\theta_i(0|\tau)$. Finally, the vacuum
amplitude of the $\zeds_2$ permutation orbifold can be determined from
the formula (\ref{partf}) as
\beq
Z^{\zeds_2}(\tau,R)=\mbox{$\frac{1}{2}$}\,\Big(\mathfrak{z}(\tau,R)^2+
\mathfrak{z}(2\tau,R)+\mathfrak{z}\big(\mbox{$\frac{\tau}{2}$}\,,\,R
\big)+\mathfrak{z}\big(\mbox{$\frac{\tau+1}{2}$}\,,\,R\big)\Big) \
. \eeq

Clearly the contributions to both sides of the formula (\ref{egy})
from the untwisted sector match. For the contributions from the
twisted sectors, we use the identities
$\theta_3(\tau+1)=\theta_4(\tau)$ and
\beq
\theta_2(\tau)\,\theta_3(\tau)\,\theta_4(\tau)=2\eta(\tau)^3
\label{thetaeta3id}\eeq
to derive the elliptic function relation
\bea
&&\frac{1}{\big|\theta_2(\tau)\,\eta(\tau)\big|}+\frac{1}
{\big|\theta_3(\tau)\,
\eta(\tau)\big|}+\frac{1}{\big|\theta_4(\tau)\,\eta(\tau)\big|}
\nonumber\\[4pt] && \qquad\qquad\qquad\qquad\qquad\qquad\qquad \=
\left|\frac{\theta_3(\tau)\,\theta_3(\tau+1)}{2\eta(\tau)^4}\right|+
\frac{1}{\big|\theta_3(\tau)\,\eta(\tau)\big|}+
\frac{1}{\big|\theta_3(\tau+1)\,\eta(\tau)\big|}\nonumber\\[4pt]
&& \qquad\qquad\qquad\qquad\qquad\qquad\qquad \=
\frac{1}{2\big|\eta(2\tau)\big|^2}+\frac{1}
{\big|\eta(\frac{\tau}{2})\big|^2}+\frac{1}
{\big|\eta(\frac{\tau+1}{2})\big|^2} \ ,
\label{ellfnid}\eea
where in the last line we substituted the identity
$\theta_3(\tau)=\eta(\frac{\tau+1}{2})^2/\eta(\tau+1)$ and used
$|\eta(\tau+1)|=|\eta(\tau)|$. This equation establishes the
$R\to\infty$ limit of the formula (\ref{egy}), for each twisted
sector, which easily generalizes to $\zed_2$ orbifolds of $\real^d$ by
taking appropriate powers.

\subsection{DLCQ Strings on Double Covers\label{DLCQDouble}}

We now turn to the explicit form of the $N=2$ part of the genus two
bosonic DLCQ free energy (\ref{FDLCQ2}) which is given explicitly by
\beq
{\cal F}_2\big(\tau^\bullet\big)=-\frac{g_s^2}{16}\,
\left|\frac{\tau^\bullet}{16\pi^2\,\alpha'}\right|^{12}\,\sum_{s=0,1}~
\int_\triangle\,\frac{\dd^2\tau^\#}{\big(\tau^\#_2
\big)^{12}}~\big|\Psi_{10}\big(\tau_{s}(\tau^\bullet,\tau^\#)
\big)\big|^{-2} \ ,
\label{FDLCQ2N2}\eeq
where the corresponding period matrices read
\beq
\tau_s\big(\tau^\bullet\,,\,\tau^\#\big)~:=~
\tau_{r=2,m=1,s,t=1}\big(\tau^\bullet\,,\,\tau^\#\big)\=
\begin{pmatrix}-\frac{1}{2\tau^\bullet}-\frac{s}{2}&-\frac{1}{2}\\[4pt]
                   -\frac{1}{2}&\tau^\#\end{pmatrix} \ .
\label{x0}\eeq
By modular invariance it suffices to restrict to the $s=0$
contribution. To see this, we define the $SL(2,\zed)$ modular
transformation $\tilde\tau^\#=\tau^\#/(2\tau^\#+1)$. Then the period
matrices $\tau_1(\tau^\bullet,\tau^\#)$ and $\tau_0(\tau^\bullet,\tilde\tau^\#)$
are related by the $Sp(4,\zeds)$ modular transformation
\beq
\tau_0\big(\tau^\bullet\,,\,\tilde\tau^\#\big)=\big(A\,\tau_1(\tau^\bullet,
\tau^\#)+B\big)\,\big(C\,\tau_1(\tau^\bullet,\tau^\#)+D\big)^{-1}
\eeq
given by the matrix 
\beq
g\=\begin{pmatrix}1&-1&0&0\\0&1&0&0\\0&0&1&0\\
0&2&1&1\end{pmatrix}~=:~\begin{pmatrix}A&B\\C&D\end{pmatrix} \ . \eeq
Since the integration over $\tau^\#$ in (\ref{FDLCQ2N2}) runs over
a fundamental domain $\triangle$ for $SL(2,\zed)$, we can compensate
the omission of the $s=1$ term by simply doubling the $s=0$
contribution.

Let us now simplify the integrand of (\ref{FDLCQ2N2}) by working out
explicitly the product of theta constants appearing in the genus two
modular form (\ref{Psi10Omega}). Starting from the reduction
(\ref{poi}) with $N=2$, one has $j=2$ when $a_1=1$ and $j=3$ when
$a_1=0$, and hence
\bea
\Theta\big({}^{\mbf a}_{\mbf b}\big)\big(
\tau_0(\tau^\bullet,\tau^\#)\big)&=&\frac{
\e^{\pi\ii a_2\,b_2/2}}{2\,\sqrt{-\ii
\tau^\#}}\,\Big(\theta\big({}^{a_1}_{b_1}\big)\big(\mbox{$\frac{a_2}4\,
\big|\,\frac1{2\tau^\bullet}$}\big)\,\theta\big({}^{a_1}_0\big)
\big(\mbox{$\frac{a_2}{4}\,\big|\,-\frac1{4\tau^\#}$}\big)
\label{Theta2decomp}\\ && \qquad\qquad\qquad
+\,(-1)^{b_2}\,\theta\big({}^{a_1}_{b_1}\big)\big(\mbox{$
\frac{a_2}{4}+\frac{1}{2}\,\big|\,\frac1{2\tau^\bullet}$}\big)\,
\theta\big({}^{a_1}_0\big)\big(\mbox{$\frac{a_2}{4}+\frac{1}{2}\,
\big|\,-\frac1{4\tau^\#}$}\big)\Big) \ . \nonumber
\eea
Using the property
\beq 
\theta\big({}^a_b\big)\big(z+\mbox{$\frac12$}\,\big|\,\tau
\big)=(-1)^{a\,b}~
\theta\big({}^{~a}_{b+1}\big)(z|\tau)
\label{tricky}\eeq
where $b+1$ is understood modulo~$2$, one can now write down the
product of the even genus two theta constants in
(\ref{Psi10Omega}). To simplify the formulae somewhat, in the ensuing
calculations we will use the shorthand notations
$\theta_i^\bullet:=\theta_i(0|\frac1{2\tau^\bullet})$,
$\tilde\theta^\bullet_i:=\theta_i(\frac14|\frac1{2\tau^\bullet})$,
$\theta^\#_i:=\theta_i(0|-\frac1{4\tau^\#})$ and
$\tilde\theta^\#_i:=\theta_i(\frac14|-\frac1{4\tau^\#})$.

Then the modular form (\ref{Psi10Omega}) can be expressed as
\beq
\Psi_{10}\big(\tau_0(\tau^\bullet,\tau^\#)\big)=\frac{{\cal A}^2~{\cal B}^2}
{2^{32}\,\big(\tau^\#\big)^{10}}
\label{Psi10AB}\eeq
where
\bea
{\cal A}&=&
\big(\theta^\bullet_3\,\theta^\#_3+\theta^\bullet_4\,\theta^\#_4\big)\,
\big(\theta^\bullet_2\,\theta^\#_2+\theta^\bullet_1\,\theta^\#_1\big)\,
\big(\theta^\bullet_4\,\theta^\#_3+\theta^\bullet_3\,\theta^\#_4\big)
\nonumber\\
&& \times~\big(\theta^\bullet_3\,\theta^\#_3-\theta^\bullet_4\,
\theta^\#_4\big)\,
\big(\theta^\bullet_4\,\theta^\#_3-\theta^\bullet_3\,\theta^\#_4\big)\,
\big(\theta^\bullet_2\,\theta^\#_2-\theta^\bullet_1\,\theta^\#_1\big) \ ,
\label{calAPsi}\\[4pt]
{\cal B}&=&\big(\tilde\theta^\bullet_3\,\tilde\theta^\#_3+
\tilde\theta^\bullet_4\,\tilde\theta^\#_4\big)\,
\big(\tilde\theta^\bullet_2\,\tilde\theta^\#_2+\tilde\theta^\bullet_1\,
\tilde\theta^\#_1\big)\,\big(\tilde\theta^\bullet_4\,
\tilde\theta^\#_3+\tilde\theta^\bullet_3\,\tilde\theta^\#_4\big)\,
\big(\tilde\theta^\bullet_1\,\tilde\theta^\#_2+\tilde\theta^\bullet_2\,
\tilde\theta^\#_1\big) \ .
\label{calBPsi}\eea
The products (\ref{calAPsi}) can be immediately simplified by noticing
that $\theta^\bullet_1=\theta_1(0|\frac1{2\tau^\bullet})=0$ (and similarly
$\theta^\#_1=0$). One finds
\bea
{\cal A}&=&{\theta^\bullet_2}\,^2\,{\theta^\#_2}\,^2\,
\big({\theta^\bullet_3}\,^2\,
{\theta^\bullet_4}\,^2\,({\theta^\#_3}\,^4+{\theta^\#_4}\,^4)-
{\theta^\#_3}\,^2\,
{\theta^\#_4}\,^2\,({\theta^\bullet_3}\,^4+{\theta^\bullet_4}\,^4)\big)
\ ,
\label{calAsimpl1}\\[4pt]
{\cal B}&=&\tilde\theta^\bullet_1\,\tilde\theta^\bullet_2\,
\tilde\theta^\bullet_3\,
\tilde\theta^\bullet_4\,\big({\tilde\theta^\#_1}\,^2+
{\tilde\theta^\#_2}\,^2
\big)\,\big({\tilde\theta^\#_3}\,^2+{\tilde\theta^\#_4}\,^2\big)+
\tilde\theta^\#_1\,\tilde\theta^\#_2\,\tilde\theta^\#_3\,
\tilde\theta^\#_4\,\big({\tilde\theta^\bullet_1}\,^2+
{\tilde\theta^\bullet_2}\,^2
\big)\,\big({\tilde\theta^\bullet_3}\,^2+{\tilde\theta^\bullet_4}\,^2
\big) \label{calBsimpl1}\\[4pt]
&& +\,\tilde\theta^\bullet_1\,\tilde\theta^\bullet_2\,
\tilde\theta^\#_3\,
\tilde\theta^\#_4\,\big({\tilde\theta^\#_1}\,^2+
{\tilde\theta^\#_2}\,^2
\big)\,\big({\tilde\theta^\bullet_3}\,^2+{\tilde\theta^\bullet_4}\,^2
\big)+\tilde\theta^\#_1\,\tilde\theta^\#_2\,\tilde\theta^\bullet_3\,
\tilde\theta^\bullet_4\,\big({\tilde\theta^\bullet_1}\,^2+
{\tilde\theta^\bullet_2}\,^2
\big)\,\big({\tilde\theta^\#_3}\,^2+{\tilde\theta^\#_4}\,^2\big) \ .
\nonumber\eea
Using (\ref{tricky}) and the parity properties of the theta functions,
one notices that $\tilde\theta^\bullet_1=-\tilde\theta^\bullet_2$ and
$\tilde\theta^\bullet_3=\tilde\theta^\bullet_4$. We may thus simplify
(\ref{calBsimpl1}) further to
\beq
{\cal B}\=-16\,\tilde\theta^\bullet_1\,\tilde\theta^\bullet_2\,
\tilde\theta^\bullet_3\,\tilde\theta^\bullet_4\,\tilde\theta^\#_1\,
\tilde\theta^\#_2\,\tilde\theta^\#_3\,\tilde\theta^\#_4\=
-4\,{\theta^\bullet_2}\,^2\,\theta^\bullet_3\,\theta^\bullet_4\,
{\theta^\#_2}\,^2\,\theta^\#_3\,\theta^\#_4
\eeq
where the second equality is a consequence of the identity for
products of theta functions with identical modulus given by
\beq \theta_1(2z|\tau)\,\theta_2(0|\tau)\,\theta_3(0|\tau)\,
\theta_4(0|\tau)=2\,\theta_1(z|\tau)\,\theta_2(z|\tau)\,
\theta_3(z|\tau)\,\theta_4(z|\tau) \ ,
\label{id1}\eeq
applied with $z=\frac14$.

The next step consists in using the modulus doubling identities
\bea 
\theta_2(0|\tau)^2&=&2\,\theta_2(0|2\tau)\,\theta_3(0|2\tau) \ ,
\nonumber\\[4pt]
\theta_3(0|\tau)\,\theta_4(0|\tau)&=&\theta_4(0|2\tau)^2 \ ,
\nonumber\\[4pt]
\theta_3(0|\tau)^2+\theta_4(0|\tau)^2&=&2\,\theta_3(0|2\tau)^2
\label{doublingids}\eea
along with the Jacobi abstruse identity
\beq
\theta_3(0|\tau)^4-\theta_4(0|\tau)^4=\theta_2(0|\tau)^4
\label{id2}\eeq
on both $\theta_i^\bullet$ and $\theta_i^\#$. After introducing the
notations  $\bar{\theta}^\bullet_i:=\theta_i(0|\frac1{\tau^\bullet})$ and 
$\bar{\theta}^\#_i:=\theta_i(0|-\frac1{2\tau^\#})$ we find
\beq {\cal A}~{\cal B}=-128\,{\bar{\theta}^\bullet_2}\,^2\,
{\bar{\theta}^\bullet_3}\,^2\,{\bar{\theta}^\bullet_4}\,^2\,
{\bar{\theta}^\#_2}\,^2\,{\bar{\theta}^\#_3}\,^2\,
{\bar{\theta}^\#_4}\,^2\,\big({\bar{\theta}^\bullet_4}\,^4\,(
{\bar{\theta}^\#_2}\,^4+{\bar{\theta}^\#_3}\,^4)-
{\bar{\theta}^\#_4}\,^4\,
({\bar{\theta}^\bullet_2}\,^4+{\bar{\theta}^\bullet_3}\,^4)\big) \ .
\label{10th1}\eeq
We now undo the projective rotation $\tau_0\to-\tau_0$ and the
reflection $\tau^\#\to-\tau^\#$ that were used to write (\ref{poi}),
in order to use theta functions which are convergent on the standard
domain of genus one moduli $\tau_2>0$. This affects 
only $\bar{\theta}^\bullet_i$, because its modulus changes as 
$\theta_i(0|\frac1{\tau^\bullet})\to\theta_i(0|-\frac1{\tau^\bullet})$.
The reflection of the off-diagonal elements of the period matrix
(\ref{Omegacover}) which flips the sign of the argument of $\theta_i$
via (\ref{poi}) is easily checked to have no effect on the product
(\ref{10th1}).

The final transformation we perform on the product (\ref{10th1}) is a
modular $S$ transformation on both $\bar\theta^\bullet_i$ and
$\bar\theta^\#_i$ given by
\bea
\theta_2\big(0\,\big|\,\mbox{$-\frac{1}{\tau}$}\big)&=&
\sqrt{-\ii\tau}~\theta_4(0|\tau) \ , \nonumber\\[4pt]
\theta_3\big(0\,\big|\,\mbox{$-\frac{1}{\tau}$}\big)&=&
\sqrt{-\ii\tau}~\theta_3(0|\tau) \ , \nonumber\\[4pt]
\theta_4\big(0\,\big|\,\mbox{$-\frac{1}{\tau}$}\big)&=&
\sqrt{-\ii\tau}~\theta_2(0|\tau) \ .
\label{Stransfs}\eea
Then we can write the modular form (\ref{Psi10AB}) as
\bea
\Psi_{10}\big(\tau_0(\tau^\bullet,\tau^\#)\big)&=&
\big(\tau^\bullet\big)^{10}\,\eta\big(\tau^\bullet\big)^{12}\,
\eta\big(2\tau^\#\big)^{12}\label{Psi10final}\\ && \times~\Big(
{\theta}_2(2\tau^\#)^4\,\big({\theta}_4(\tau^\bullet)^4+
{\theta}_3(\tau^\bullet)^4
\big)-{\theta}_{2}(\tau^\bullet)^4\,\big({\theta}_4(2\tau^\#)^4+
{\theta}_3(2\tau^\#)^4\big)\Big)^2 \nonumber
\eea
where we have used (\ref{thetaeta3id}). Substituting into
(\ref{FDLCQ2N2}) and using (\ref{id2}) we arrive at our final form for
the two-loop DLCQ free energy given by
\beq
{\cal F}_2\big(\tau^\bullet\big)=-\frac{g_s^2}{8\,\big(16\pi^2\,
\alpha'\,\big)^{12}}\,\frac{\big|\eta(\tau^\bullet)\big|^{-24}}
{\big|\tau^\bullet\big|^8}\,
\int_\triangle\,\frac{\dd^2\tau^\#}{\big(\tau^\#_2
\big)^{12}}~\left|\frac{\eta\big(2\tau^\#\big)^{-6}}
{\theta_3\big(\tau^\bullet\big)^4\,\theta_4\big(2\tau^\#\big)^4-
\theta_4\big(\tau^\bullet\big)^4\,\theta_3\big(2\tau^\#\big)^4}
\right|^4 \ .
\label{10th2}\eeq

\subsection{Prym Varieties\label{Prym}}

Our next goal is to determine the genus one modulus $\tau^\#$
explicitly in terms of the branch point loci on the base torus
$\torus$. This modulus arose generically from the algebraic
Weierstrass-Poincar\'e reduction of the period matrix $\tau$ of the
covering surface $\hat\Sigma$ to the normal form (\ref{Omegacover}),
which is a consequence of the fact that the genus two Riemann period
matrix in this instance satisfies a Hopf condition~\cite{csz}. We will
now elucidate the geometrical significance of this modulus for a
generic genus two cover over $\torus$ of degree $N=r\,m$, and then
show how in the case of double covers this geometrical realization
determines it explicitly as a function of branch points on the
worldsheet $\torus$.

Let $f:\hat\Sigma\to\torus$ be a holomorphic map. Let $\omega_i$,
$i=1,2$ be the canonical, normalized abelian holomorphic differentials
on $\hat\Sigma$ with the periods
\beq
\oint_{\hat\alpha_i}\,\omega_j\=\delta_{ij} \qquad \mbox{and} \qquad
\oint_{\hat\beta_i}\,\omega_j\=\tau_{ij} \ .
\label{omegaperiods}\eeq
On the base elliptic curve $\torus$ the holomorphic one-form is $\dd
z$ with the periods $\oint_{\alpha}\,\dd z=1$ and $\oint_{\beta}\,\dd
z=\tau^\bullet$. The two sets of differentials are related by the
pull-back homomorphism $f^*:H^{1,0}(\torus,\complex)\to
H^{1,0}(\hat\Sigma,\complex)$ through
\beq
f^*(\dd z)=h_1\,\omega_1+h_2\,\omega_2
\label{holdiffsrel}\eeq
for some complex numbers $h_i$. These numbers can be determined by
integrating the relation (\ref{holdiffsrel}) over a canonical homology
basis of $H_1(\hat\Sigma,\zed)$ using (\ref{fpushM}), and with respect
to the basis specified by (\ref{sfMints}) they are given by
\beq
h_1\=r~\tau^\bullet \qquad \mbox{and} \qquad h_2\=0 \ .
\label{hred}\eeq

Let ${\rm Jac}(\hat\Sigma):=H^{1,0}(\hat\Sigma,\complex)/
H^{1,0}(\hat\Sigma,\Lambda_\tau)$ be the principally polarized
Jacobian variety of $\hat\Sigma$, where
$\Lambda_\tau=\zed^2\oplus\tau\,\zed^2$ is the lattice of rank four
induced by the period matrix $\tau$ of $\hat\Sigma$. It can be
identified with the Picard group ${\rm Pic}^0(\hat\Sigma)$ of
isomorphism classes of flat line bundles over $\hat\Sigma$, in
correspondence with degree zero divisors, and it is isomorphic to the
complex two-dimensional torus
$\complex^2/\Lambda_\tau$. There is an embedding of $\hat\Sigma$ into
${\rm Jac}(\hat\Sigma)$ provided by the Abel map $\mathfrak{A}:\hat
z\mapsto\int^{\hat z}\,(\omega_1,\omega_2)$, which also provides the
mapping from divisors to the Jacobian variety. The theta divisor is
the analytic subvariety of the Jacobian defined by the equation
$\Theta\big({}^{\mbf0}_{\mbf0}\big)(z_1,z_2|\tau)=0$. On the base, the
Jacobian torus can instead be identified with the elliptic curve
$\torus$ itself and one has ${\rm Jac}(\torus)\cong\torus$.

It follows from a general property of finite morphisms between smooth
projective curves~\cite{birkenhake} that the holomorphic map
$f:\hat\Sigma\to\torus$ can be factorized by means of a commutative
triangle
\beq
\xymatrix{ \hat\Sigma\ar[r]^{g} ~ \ar[rd]_f& ~
\Sigma_1 \ar[d]^{f_1}\\ & ~\torus }
\label{coverfact}\eeq
where $f_1:\Sigma_1\to\torus$ is an unramified cover. The induced
pullback morphisms on the Jacobian tori have the properties that
$\ker(f^*)\cong\ker(f_1^*)$ and $g^*:\Sigma_1\to{\rm Jac}(\hat\Sigma)$
is injective. This accounts for the first diagonal entry in the period
matrix (\ref{Omegacover}). The complimentary subvariety to ${\rm
  im}(f^*)\cong\torus$ in the Jacobian torus $\complex^2/\Lambda_\tau$
is gotten from the norm morphism
\beq
\Omega_f\,:\,{\rm Jac}\big(\hat\Sigma\big)~\longrightarrow~
\torus \qquad \mbox{with}\quad \Omega_f(z_1,z_2)~:=~
h_1\,z_1+h_2\,z_2
\label{normmorph}\eeq
which takes the divisor class $D$ of degree zero by applying
$f$ to each point of the divisor. The kernel of this morphism is
a principally polarized subvariety of ${\rm Jac}(\hat\Sigma)$ called
the Prym variety of the cover and in the present case it is a
complex one-dimensional torus $\complex/(\zed\oplus\Pi\,\zed)$ whose
period $\Pi$ is called the Prym modulus. In the basis defined by
(\ref{sfMints}), from (\ref{hred}) it follows that the kernel of
(\ref{normmorph}) in $\complex^2$ consists of all points of the form
$(z_1,z_2)=(\frac mr,z)$ with $m\in\zed$ and $z\in\complex$. Passing
to the quotient $\complex^2/\Lambda_\tau$ using (\ref{Omegacover})
truncates to points $(0,z)$ with the identifications $z\sim
z+\frac{m_1}r+\tau^\#\,m_2$ for any $m_1,m_2\in\zed$. It follows that
the Prym modulus in this basis is given by
\beq
\Pi=r~\tau^\#
\label{Pitaurel}\eeq
and we have explicitly identified the second elliptic modulus in
(\ref{Omegacover}). Using the factorization (\ref{coverfact}) one
shows~\cite{birkenhake} that the induced theta divisor on
$\ker(\Omega_f)$ is $r$ times the theta divisor defining its principal
polarization, and hence that $\ker(\Omega_f)$ is a Prym-Tyurin
variety.

So far everything we have said holds generally for any $N$-sheeted
genus two cover of the torus $\torus$. When $N=2$, wherein only the
$r=2$ term contributes in (\ref{FDLCQ2}), the Prym variety possesses a
special characterization~\cite{fay} which enables one to make this
construction much more explicit. Consider the element of the
symplectic group $Sp(4,\zed)$ given by
\beq
g\=\begin{pmatrix}0&0&-1&0\\0&1&0&0\\1&-1&0&0\\0&0&1&1
\end{pmatrix}~=:~\begin{pmatrix}A&B\\C&D \end{pmatrix} \ .
\label{Prymg}\eeq
It induces the change in basis of $H_1(\hat\Sigma,\zed)$ represented
by
\beq
{\sf M}\={\sf M}'\,\begin{pmatrix}D^\top&B^\top\\
C^\top&A^\top\end{pmatrix} \qquad \mbox{with} \quad
{\sf M}'\=\begin{pmatrix}1&-1&0&0\\0&0&1&-1\end{pmatrix} \ ,
\label{Mprimedef}\eeq
and the genus two modular transformation
\beq
\tau_0\big(\tau^\bullet\,,\,\tau^\#\big)=\big(A\,\tau_0'(\tau^\bullet,
\tau^\#)+B\big)\,\big(C\,\tau_0'(\tau^\bullet,\tau^\#)+D\big)^{-1}
\label{tau0primemodtransf}\eeq
with
\beq
\tau_0'\big(\tau^\bullet\,,\,\tau^\#\big)=
\frac12\,\begin{pmatrix}\Pi+\tau^\bullet&\Pi-\tau^\bullet\\
\Pi-\tau^\bullet&\Pi+\tau^\bullet\end{pmatrix}
\label{tau0primedef}\eeq
where we have used (\ref{Pitaurel}) with $r=2$. From (\ref{fpushM}) it
follows that
\beq
f_*\big(\hat\alpha_1\big)\=-f_*\big(\hat\alpha_2\big)\=\alpha
\qquad \mbox{and}
\qquad f_*\big(\hat\beta_1\big)\=-f_*\big(\hat\beta_2\big)\=\beta \ .
\label{abpushprime}\eeq
Integrating both sides of (\ref{holdiffsrel}) in this basis thus gives
$h_1'=-h_2'=1$, and hence
\beq
f^*(\dd z)=\omega_1-\omega_2 \ .
\label{fpulldzprime}\eeq

What makes the instance of a double cover $f:\hat\Sigma\to\torus$
special is that it has a canonical conformal automorphism
$\iota:\hat{\Sigma}\to\hat{\Sigma}$, satisfying $f\circ\iota=f$, which
is the involution permuting the sheets of the cover. It uniquely
determines the covering with $\torus=\hat\Sigma/\iota$. From
(\ref{abpushprime}) it follows that
\beq
\iota\big(\hat\alpha_1\big)\=-\iota\big(\hat\alpha_2\big)
\qquad \mbox{and}
\qquad \iota\big(\hat\beta_1\big)\=-\iota\big(\hat\beta_2\big) \ ,
\label{iotaabs}\eeq
and hence that
\beq
\iota^*(\omega_1)=-\omega_2 \ .
\label{iotaomega}\eeq
The holomorphic one-form
\beq
\nu=\omega_1+\omega_2
\label{Prym00def}
\eeq
is called the Prym differential and it is the unique holomorphic
differential on the two-sheeted cover $\hat\Sigma$ which is odd under
the defining involution with $\iota^*(\nu)=-\nu$. It follows from
(\ref{fpulldzprime})--(\ref{Prym00def}) and the form
(\ref{tau0primedef}) of the period matrix in this basis that the Prym
period is determined by
\beq
\Pi=\oint_{\hat\beta_1}\,\nu \ .
\label{Piperiod}\eeq
The Prym differential $\nu$ is normalized with respect to the
$\hat\alpha_1$ cycle, while it has vanishing periods around
$\hat\alpha_1-\hat\alpha_2$ and $\hat\beta_1-\hat\beta_2$. 
At the level of Jacobian
varieties, the Prym variety $\ker(\Omega_f)$ is isomorphic to the
subvariety of ${\rm Jac}(\hat\Sigma)$ consisting of degree zero
divisor classes which are odd under the involution $\iota$. Note that
from (\ref{fpulldzprime}) it follows that the embedding
$f^*:\torus\hookrightarrow{\rm Jac}(\hat\Sigma)$ is isomorphic to the
subvariety invariant under $\iota$.

Similarly to the even holomorphic one-form (\ref{fpulldzprime}), the
Prym differential (\ref{Prym00def}) may be given explicitly as the
pull-back $\nu=f^*(\prym(w_1,w_2))$ of a multiplicative differential
$\prym(w_1,w_2)=\prym(z;w_1,w_2)~\dd z$ on the base elliptic curve
$\torus$ with modulus $\tau^\bullet$. It is required to have a square
root cut singularity about each of the branch points
$w_1,w_2\in\torus$ of the cover and to have global periodicity under
$z\to z+m+n\,\tau^\bullet$ for any $m,n\in\zed$. This uniquely
determines the multiplicative differential on $\torus$ in terms of
Jacobi-Erd\'elyi elliptic functions as
\beq
\prym(z;w_1,w_2)=\frac{\theta_1\big(z-\frac{w_1+w_2}{2}\,\big|\,
\tau^\bullet\big)}{\sqrt{\theta_1\big(z-w_1\,\big|\,
\tau^\bullet\big)\,
\theta_1\big(z-w_2\,\big|\,\tau^\bullet\big)}} \ .
\label{Prymdiffbranch}\eeq
The Prym modulus (\ref{Piperiod}) may then be written as
\beq
\tau^\#\=\mbox{$\frac12$}\,\Pi\=\frac12~\frac{\displaystyle
\oint_\beta\,
\prym(w_1,w_2)}{\displaystyle\oint_\alpha\,\prym(w_1,w_2)} \ ,
\label{taunumbranch}\eeq
thereby determining the desired explicit dependence of the elliptic
modulus $\tau^\#$ on the branch point loci. As expected,
$\Pi\to\tau^\bullet$ in the unramified limit $w_1\to w_2$ wherein the
branch cut on $\torus$ closes up. It follows from (\ref{tau0primedef})
that this limit corresponds to approaching a separating boundary
component of moduli space, wherein the genus two Riemann surface
$\hat\Sigma$ degenerates into two copies of the base torus $\torus$.

Thus far we have not accounted for global monodromy $\Phi$ of the
covering map $f:\hat\Sigma\to\torus$, \emph{i.e.}, the above formulas
are written in the untwisted sector $(\varepsilon,\delta)=(0,0)$. For
each twisted sector $(\varepsilon,\delta)\in(\zed/2\zed)^2$ there is a
holomorphic Prym form $\nu_{\varepsilon,\delta}$ which is odd under
the involution $\iota$ and which has non-vanishing periods only around
the $(\hat\alpha_1,\hat\beta_1)$ cycles of the homology group
$H_1(\hat\Sigma,\zed)$. They project onto multiplicative differentials
$\prym_{\varepsilon,\delta}(w_1,w_2)$ on $\torus$ which have square
root cut singularities about the branch points $w_1,w_2\in\torus$. The
Prym form corresponding to the characteristic $(\varepsilon,\delta)$
can be gotten from the untwisted one via a crossing transformation of
the branch points
\beq
w_1~\longrightarrow~w_1+\delta+\varepsilon\,\tau^\bullet \qquad
\mbox{and} \qquad w_2~\longrightarrow~w_2
\label{crossingtransf}\eeq
to get
\beq
\prym_{\varepsilon,\delta}(z;w_1,w_2)=\prym\big(z\,;\,
w_1+\delta+\varepsilon\,\tau^\bullet,w_2\big)
\label{prym}\eeq
with $\prym_{0,0}(w_1,w_2)=\prym(w_1,w_2)$. The corresponding Prym
modulus is defined by
\beq
\Pi_{\varepsilon,\delta}=\frac{\displaystyle\oint_\beta\,
\prym_{\varepsilon,\delta}(w_1,w_2)}
{\displaystyle\oint_\alpha\, \prym_{\varepsilon,\delta}(w_1,w_2)}
\label{Pitwisted}\eeq
with $\Pi_{0,0}=\Pi$.

These constructions of Prym varieties and Prym differentials have
natural generalizations to double covers $\hat\Sigma$ of a genus $g$
surface $\Sigma$ with $k=2n$ branch points ($n=0,1$), with genus $\hat
g=2g+n-1$ determined by the Riemann-Hurwitz formula (\ref{rh}). In
this case the Prym variety is a complex torus of dimension $g+n-1$. By
the Riemann-Roch theorem, there are exactly $g+n-1$ independent
holomorphic one-forms which are odd under the automorphism $\iota$ and
which form a basis for the Prym differentials. The remaining $g$ even
ones on $\hat\Sigma$ are preimages of the holomorphic differentials on
the base space $\Sigma$. A further generalization exists to more
general abelian automorphism groups of a cover. The action of the
group on $H^{1,0}(\hat\Sigma,\complex)$ is then always diagonal on a
suitable basis of holomorphic differentials and the subspace
corresponding to a non-trivial set of eigenvalues are pull-backs of
multiplicative elliptic differentials, whose multiplicative factors
are given by these eigenvalues. This is exploited implicitly in the
computation of $\zed_N$~orbifold twist field amplitudes
in~\cite{Atick}.

\subsection{Correlation Functions of Twist Field Operators\label{Z2Twist}}

We now come to the computation of the two-point function
$\langle\sigma(z)\,\sigma(0)\rangle^{\zed_2}$ of $\zed_2$ twist fields
$\sigma(z)=\sigma_{12}(z)$ in the $\real^{24}\wr\zed_2$ permutation
orbifold. We begin by discussing some general aspects concerning
global monodromy in the covering surface construction of
Section~\ref{TwistFields}. Recall that the sum appearing in the
correlation function (\ref{correlator}) of interest (computed with the
amplitude (\ref{Z2bos})) is restricted to the set of admissible monodromy
homomorphisms $\Phi$ such that each connected component of the
corresponding cover $\hat\Sigma$ of the base torus $\torus$ is a
surface of genus two. This is ensured by the requirement that the
monodromy of the generators of $\pi_1(\torus_{\,\underline{w}\,})$
encircling the punctures be a simple transposition in each orbit
$\xi\in{\cal O}(\Phi)$. The period matrix $\tau^{\xi,\,\underline{w}}$
depends on the monodromy only via its stabilizer subgroups, which are
the finite index subgroups $H<\pi_1(\torus_{\,\underline{w}\,})$
obeying the admissibility criterion (\ref{gcov}). Consider the
stabilizer subgroup $H=H_a$ of a given sheet $a$ corresponding to a
transitive homomorphism $\Phi:\pi_1(\torus_{\,\underline{w}\,})\to
S_N$. Since it is isomorphic to
$\pi_1(\hat\Sigma_{\,\underline{\hat w}\,})$ and since there are $2N-2$
preimages of the two branch points of $\torus_{\,\underline{w}}$, it
is a group freely generated by $2N+1$ elements. The kernel of the
forgetful homomorphism
$\hat\imath_*:\pi_1(\hat\Sigma_{\,\underline{\hat w}\,})\to\pi_1(\hat\Sigma)$
is given by the normal closure
\beq
\widehat{N}_H\big(\hat\gamma_1,\dots,\hat\gamma_{2N-2}\big)=\,
<h\,\hat\gamma_1\,h^{-1},\dots,h\,\hat\gamma_{2N-2}\,h^{-1}~
\big|~h\in H>
\label{normclosuredef}\eeq
of the generators $\hat\gamma_i$ encircling the ramification
points.

When $N=2$ the generators $\hat\gamma_i$ are easily determined. Let us
use the presentation
$\pi_1(\torus_{\,\underline{w}\,})=\,<\alpha,\beta,\gamma>$. The
generators of $\pi_1(\hat\Sigma_{\,\underline{\hat w}\,})$ encircling the
ramification points are the (pullbacks of the) squares of the
generators of $\pi_1(\torus_{\,\underline{w}\,})$ which encircle the
punctures. For $N=2$, the preimages of the punctures are precisely the
ramification points, and hence one has
\beq
\ker\big(\hat\imath_*\big)=\widehat{N}_H\big(\gamma^2\,,\,
([\alpha,\beta]\,\gamma)^2\big) \ .
\label{kernormalN2}\eeq
There are four homomorphisms with the prescribed monodromy
representing the four twisted sectors
$(\varepsilon,\delta)\in(\zed/2\zed)^2$, and all of them are
transitive. There are correspondingly exactly four admissible
subgroups $H$ of index two. Since $\zed_2$ is an abelian group,
conjugacy classes of homomorphisms contain only one element. Their
precise forms and the corresponding stabilizers can be determined
explicitly.

The simplest example is provided by the admissible homomorphism
$\Phi_1$ which sends $\gamma$ to the transposition $(1~2)$ and
$\alpha,\beta$ both to the identity. Its stabilizer $H_1$ is freely
generated by the words
$\alpha,\beta,\alpha\,\gamma\,\alpha^{-1},\beta\,\gamma\,\beta^{-1},\gamma^2$.
We then seek a presentation of the generators
$\hat\alpha_1,\hat\alpha_2,\hat\beta_1,\hat\beta_2,\hat\gamma$ of
$\pi_1(\hat\Sigma_{\,\underline{\hat w}\,})$ such that the quotient by
the relations $\gamma^2=([\alpha,\beta]\,\gamma)^2=1$ yields the group
$\pi_1(\hat\Sigma)$ with
$[\hat\alpha_1,\hat\beta_1]\,[\hat\alpha_2,\hat\beta_2]
\in\widehat{N}_{H_1}(\gamma^2,([\alpha,\beta]\,\gamma)^2)$. For the
case at hand, one sees that the assignments
$\hat\alpha_1=\alpha,\hat\beta_1=\beta,\hat\alpha_2=
\alpha\,\gamma\,\alpha^{-1},\hat\beta_2=
\beta\,\gamma\,\beta^{-1},\hat\gamma=\gamma^2$ suffice. This determines the
homomorphism of fundamental groups $\hat\imath_*\circ\tilde f_*^{-1}$,
where $\tilde{f}$ is the restriction of the covering map to the marked
surfaces. Since the abelianization of $\pi_1(\torus_{\,\underline{w}\,})$
  factors through this map, the powers of $\alpha,\beta$ in the
  canonical homology generators
  $\hat\alpha_1,\hat\alpha_2,\hat\beta_1,\hat\beta_2$ gives the map
  (\ref{fpushM}). This yields the covering homology matrix
\beq
{\sf M}_1=\begin{pmatrix} 1&1&0&0\\0&0&1&1 \end{pmatrix}
\eeq
which obeys the Hopf condition. Reduction of this matrix via an
$Sp(4,\zed)$ modular transformation as in Section~\ref{Prym} above
yields the normal form (\ref{sfMints}) with $r=2,m=t=1,s=0$. The other
three admissible homomorphisms are similarly treated.

However, the above formalism is sensitive only to the induced
homomorphism $f_*$ between homology groups rather than homotopy
groups, and it is difficult to proceed further with the explicit
construction of the modular invariant amplitude (\ref{correlator}). We
will return to this issue in some more detail in the next
section. Here we shall compute the twist field correlation function
using results of~\cite{dvvc1} where the correlation functions are
computed for a free boson $X$ in the geometric orbifold
$\sphere^1/\zed_2$ using the covering space method explained in
Section~\ref{TwistFields}. The two-point correlation function on the
torus $\torus$ with twist field insertions may be computed from the
path integral over field configurations $\hat X$ on the double cover
$\hat\Sigma$ which are odd under the canonical involution with $\hat
X\circ\iota=-\hat X~{\rm mod}~2\pi\,R$. As in Section~\ref{TSvsPO}
above, in each twisted sector $(\varepsilon,\delta)$ the amplitude is
a product of a radius independent quantum piece and a classical
piece. The instanton configurations on the worldsheet $\hat\Sigma$
that contribute to the classical part of the correlation function are
analogous to the untwisted ones used in Section~\ref{TSvsPO} above. In
the homology basis specified by (\ref{Mprimedef}), the boundary
conditions of the boson $\hat X$ in the given twisted sector are
characterized by the Prym differential $\nu_{\varepsilon,\delta}$. The
classical contribution is then completely analogous to that in
(\ref{ycft}) with the period $\tau$ equal to the Prym
modulus~$\Pi_{\varepsilon,\delta}$.

The quantum contributions may be computed by equating the two-loop
orbifold amplitude with that of the circle theory at the self-dual
radius as before, with the additional observation that the twist
fields in this correspondence are equivalent to magnetic vertex
operators~\cite{dvvc1}. At this radius the momentum lattices appearing in
the classical partition sums can be built up from a finite number of
square sublattices. A term by term comparison of the chiral blocks
gives an expression for the ratio of a twisted determinant to the
untwisted determinant ${\mathfrak z}(\tau^\bullet)$ as the modulus
squared of a holomorphic function of the positions of the branch
points on $\torus$. In this way the normalized twist field two-point
function on $\torus$ with the twist characteristic
$(\varepsilon,\delta)$ in the $\zeds_2$ target space orbifold of the
compactified boson $X$ can be written as~\cite{dvvc1}
\beq
\big\langle\sigma(z)\,\sigma(0)
\big\rangle^{\varepsilon,\delta}_{\rm orb}
={\mathfrak z}\big(\tau^\bullet\big)~
\left|c \big({}^\varepsilon_\delta\big)\right|^{-2}~
{\mathfrak z}^{\rm cl}(\Pi_{\varepsilon,\delta},R)\ ,
\label{corr1bos}\eeq
where
\beq
c\big({}^\varepsilon_\delta\big)=E(z)^{{1}/{8}}~
\frac{\theta\big({}^a_b\big)(0|\Pi_{\varepsilon,\delta})}
{\sqrt{\theta\big({}^{a+\varepsilon}_{b+\delta}\big)
\big(\frac{z}{2}\,\big|\,\tau^\bullet\big)~
\theta\big({}^a_b\big)\big(0\,\big|\,\tau^\bullet\big)}} \ .
\label{ctwisted}\eeq
Here we have used translation invariance to fix one of the twist field
insertion points at the origin, and $(a,b)\neq (1,1)$ is a
fixed arbitrary characteristic. The quantity $E(z)$ is the prime form
of the elliptic curve $\torus$ given by
\beq E(z)=\frac{\theta_1\big(z\,\big|\,\tau^\bullet\big)}
{\theta'_1\big(0\,\big|\,\tau^\bullet\big)} \label{primeform}\eeq
with $\theta_1'(z|\tau):=\frac{\partial}{\partial z}\theta_1(z|\tau)$,
and it is the doubly periodic elementary solution of the Laplace
equation on the torus. The independence of the expression
(\ref{ctwisted}) on the choice of characteristic $(a,b)$ is
the mathematical statement of the Schottky relations~\cite{fay} (see
Section~\ref{DLCQDVV} below).

We can now write down the desired amplitude in the permutation orbifold 
$\real^{24}\wr\zeds_2$. For this, we redefine the independent bosons
$X_i^a$, $i=1,\dots,24$, $a=1,2$ to $X_i^\pm=X_i^1\pm X^2_i$ as in
Section~\ref{TSvsPO} above. Since the $\zed_2$ permutation group acts on
the~$24$ bosons simultaneously, both the global and local monodromy of
the fields $X_i^+$ are trivial, and the twist operators act as the
identity on these fields. The path integral over $X_i^+$ thus leads
simply to an overall factor ${\mathfrak z}(\tau^\bullet,R)^{24}$. On
the other hand, the twist operators act as a $\zed_2$ twist field
simultaneously on all sigma model fields $X_i^-$. It follows that the
correct prescription is to raise the geometric $\zed_2$ orbifold twist
field correlation function in each sector to the power~$24$, and then
sum over the twisted sectors. The $X_i^+$ contribution is cancelled in
the suitably normalized correlation function by the same factors
coming from the partition function (\ref{egy}). One should then take
the decompactification limit $R\to\infty$, wherein ${\mathfrak z}^{\rm
  cl}(\Pi_{\varepsilon,\delta},R=\infty)=1$ as before. This gives the
two-point function
\beq
\big\langle\sigma(z)\,\sigma(0)\big\rangle^{\zed_2}=
\lim_{R\to\infty}~\frac12~\sum_{(\varepsilon,\delta)\in(\zed/2\zed)^2}~
\Big(\big\langle\sigma(z)\,\sigma(0)\big
\rangle_{\rm orb}^{\varepsilon,\delta}\Big)^{24} \ .
\label{PO2ptdef}\eeq
Substituting (\ref{eq:evaldet}) and
(\ref{corr1bos})--(\ref{primeform}), and using the identity
\beq
\theta_1'(0|\tau)=-2\pi\,\eta(\tau)^3 \ ,
\label{thetapreta}\eeq
then leads to the explicit formula
\beq
\big\langle\sigma(z)\,\sigma(0)\big\rangle^{\zed_2}=\frac12\,\left(
\frac{4\,\sqrt2\,\pi^{5/2}\,\alpha'}{\tau_2^\bullet}\right)^{12}\,
\left|\frac{\theta\big({}^a_b\big)\big(0\,\big|\,\tau^\bullet
\big)^{4}}{\theta_1\big(z\,\big|\,\tau^\bullet\big)\,
\eta\big(\tau^\bullet\big)^{5}}\right|^6~
\sum_{(\varepsilon,\delta)\in(\zed/2\zed)^2}~\left|
\frac{\theta\big({}^{a+\varepsilon}_{b+\delta}\big)
\big(\frac z2\,\big|\,\tau^\bullet\big)}
{\theta\big({}^a_b\big)\big(0\,\big|\,
\Pi_{\varepsilon,\delta}\big)^{2}}\right|^{24} \ .
\label{PO2ptexpl}\eeq

\subsection{DLCQ Free Energy = DVV Correlator\label{DLCQDVV}}

We will now prove the main result of this section, establishing the
equivalence
\beq
{\cal F}_2\big(\tau^\bullet\big)=\frac{4\lambda^2}{\tau_2^\bullet\,
\mu(0)}\,\int_\torus\,\dd\mu(z)~\big\langle\sigma(z)\,\sigma(0)\big
\rangle^{\zed_2}
\label{DLCQequivDVV}\eeq
between the DLCQ free energy on the double cover $\hat\Sigma\to\torus$
given by (\ref{10th2}) and the translationally invariant correlator
(\ref{transinvcorr}) of the DVV vertex operator determined by the
twist field two-point function (\ref{PO2ptexpl}) on
$\real^{24}\wr\zed_2$. We begin by observing that the right-hand side
of the formula (\ref{DLCQequivDVV}) is independent of the twist
characteristic $(\varepsilon,\delta)$ in (\ref{PO2ptexpl}). This
follows from the fact that one can get any twisted sector from the
untwisted one $(\varepsilon,\delta)=(0,0)$ by a crossing
transformation~(\ref{crossingtransf}). Crossing symmetry of the
orbifold theory, along with modular invariance at genus one, is the
remnant of genus two modular invariance on the covering
space~\cite{dvvc1}. One can check this invariance explicitly by
showing that the $z$-dependent part of the correlation function
(\ref{corr1bos}) transforms under the crossing transformation
(\ref{crossingtransf}) precisely by changing
$(0,0)\to(\varepsilon,\delta)$, just like the Prym modulus according
to (\ref{prym}).

Next we examine the change of integration variables from the modulus
$\tau^\#$ in (\ref{10th2}) to the branch point location in
(\ref{DLCQequivDVV}). For this, we require the Jacobian
$|\dd\tau^\#/\dd z|^2$. The explicit dependence of the Prym modulus
$\Pi$ on the branch point loci is given by the formula
(\ref{taunumbranch}) with $w_1=z,w_2=0$, but this is not
convenient for computing the requisite derivative $\dd\Pi/\dd
z$. Instead, it is more useful to use the {\it implicit} dependence of
the Prym modulus on the branch point $z$ dictated by the Schottky
relations. For zero characteristics $(\varepsilon, \delta)=(0,0)$,
they are given by
\beq
\frac{\sqrt{\theta_i\big(\frac{z}{2}\,\big|\,\tau^\bullet\big)\,
\theta_i\big(0\,\big|\,\tau^\bullet\big)}}{\theta_i(0|\Pi)}=
\frac{\sqrt{\theta_j\big(\frac{z}{2}\,\big|\,\tau^\bullet\big)\,
\theta_j\big(0\,\big|\,\tau^\bullet\big)}}{\theta_j(0|\Pi)} \ .
\label{Schottkyij}\eeq

By separating the explicit $z$ and $\Pi$ dependences for $i=4$ and
$j=2$, we can write (\ref{Schottkyij}) as
\beq
\frac{\theta_2(0|\Pi)}{\theta_4(0|\Pi)}=\sqrt{
\frac{\theta_2\big(0\,\big|\,\tau^\bullet\big)\,
\theta_2\big(\frac{z}{2}\,\big|\,\tau^\bullet\big)}
{\theta_4\big(0\,\big|\,\tau^\bullet\big)\,
\theta_4\big(\frac{z}{2}\,\big|\,\tau^\bullet\big)}} \ .
\label{Schottky42}\eeq
Taking the total derivative of the relation (\ref{Schottky42}) with
respect to $z$ yields
\beq \frac\partial{\partial\Pi}
\left(\frac{\theta_2(0|\Pi)}{\theta_4(0|\Pi)}\right)
~\frac{\dd\Pi}{\dd z}=\frac{\dd}{\dd z}\sqrt{
\frac{\theta_2\big(0\,\big|\,\tau^\bullet\big)\,
\theta_2\big(\frac{z}{2}\,\big|\,\tau^\bullet\big)}
{\theta_4\big(0\,\big|\,\tau^\bullet\big)\,
\theta_4\big(\frac{z}{2}\,\big|\,\tau^\bullet\big)}} \ .
\label{schder}\eeq
We can transform the $\Pi$ derivative by using the heat equation 
\beq
\frac{\pr\theta_i(z|\Pi)}{\partial\Pi}+\frac{\ii}{4\pi}\,
\frac{\pr^2\theta_i(z|\Pi)}{\partial z^2}=0
\eeq
to get the form
\beq
\frac\pr{\partial\Pi}
\left(\frac{\theta_2(0|\Pi)}{\theta_4(0|\Pi)}\right)=
-\frac{\ii}{4\pi\,\theta_4(0|\Pi)^2}\,\left.
\frac\pr{\partial w}\left(\theta_4(w|\Pi)^2~
\frac\pr{\partial w}\frac{\theta_2(w|\Pi)}{\theta_4(w|\Pi)}\right)
\right|_{w=0} \ .
\eeq
We may then use the identity for the derivative of a ratio of theta
functions given by
\beq
\frac{\partial}{\partial w}\left(
\frac{\theta_2(w|\Pi)}{\theta_4(w|\Pi)}\right)=
-\pi\;\theta_3(0|\Pi)^2\,
\frac{\theta_1(w|\Pi)\,\theta_3(w|\Pi)}{\theta_4(w|\Pi)^2}
\eeq
to arrive at
\beq \frac\partial{\partial\Pi}
\left(\frac{\theta_2(0|\Pi)}{\theta_4(0|\Pi)}
\right)=\frac{\ii}{4}\,\frac{\theta_3(0|\Pi)^3\,\theta'_1(0|\Pi)}
{\theta_4(0|\Pi)^2} \label{tauder} \ . \eeq

The differentiation on the right-hand side of (\ref{schder}) is an
easy exercise. This calculation can be repeated starting from the
Schottky relation (\ref{Schottkyij}) with $i=4$ and $j=3$. The final
result is identical to that above with the replacements
$\theta_2\leftrightarrow\theta_3$ of theta functions everywhere. In
this way we can finally write
\bea
\left|\frac{\dd\Pi}{\dd z}\right|^2&=&\pi^2\,\left|\,
\frac{\sqrt{\theta_2\big(0\,\big|\,\tau^\bullet\big)\,
\theta_2\big(\frac{z}{2}\,\big|\,\tau^\bullet\big)}}{\theta_2(0|\Pi)}\,
\frac{\sqrt{\theta_3\big(0\,\big|\,\tau^\bullet\big)\,
\theta_3\big(\frac{z}{2}\,\big|\,\tau^\bullet\big)}}
{\theta_3(0|\Pi)}\,\right| \nonumber \\ && \times\,\left|\,
\frac{\theta_2\big(0\,\big|\,\tau^\bullet\big)^2\,
\theta_3\big(0\,\big|\,\tau^\bullet\big)^2}
{\theta_4\big(0\,\big|\,\tau^\bullet\big)}\,
\frac{\theta_1\big(\frac{z}{2}\,\big|\,\tau^\bullet\big)^2}
{\theta_4\big(\frac{z}{2}\,\big|\,\tau^\bullet\big)^3}\,
\frac{\theta_4(0|\Pi)^4}{\theta'_1(0|\Pi)\,\sqrt{\theta_2(0|\Pi)\,
\theta_3(0|\Pi)}}\,\right| \ .
\label{realjaco}\eea

To compare (\ref{realjaco}) with the elliptic functions appearing in
the expressions (\ref{10th2}) and (\ref{PO2ptexpl}) for
$(\varepsilon,\delta)=(0,0)$, we exploit the identity (\ref{id1}) and
the Schottky relations (\ref{Schottkyij}) again to write
\beq
\left|\frac{\dd\Pi}{\dd z}\right|^2=\pi^2\,
\left|\,\frac{\theta_1\big(\frac{z}{2}\,\big|\,\tau^\bullet\big)^3}
{\theta_1\big(z\,\big|\,\tau^\bullet\big)\,\theta_1'(0|\Pi)^2}~
\prod_{i=1,2}\,\frac{\sqrt{\theta{a_i \choose b_i}
\big(\frac{z}{2}\,\big|\,\tau^\bullet\big)\,
\theta{a_i\choose b_i}\big(0\,\big|\,\tau^\bullet\big)}}
{\theta{a_i\choose b_i}(0|\Pi)}\,
\right| \ ,
\label{jaco}\eeq
where $(a_i, b_i)\in\{(0,0)\,,\,(0,1)\,,\,(1,0)\}$ are
arbitrary characteristics which we will choose conveniently. We can
now use the identities (\ref{thetaeta3id}), (\ref{id1}) and
(\ref{thetapreta}) along with
\beq
\mbox{$\theta_3\big(\frac{z}{2}\,\big|\,\tau^\bullet\big)^2\,
\theta_4\big(0\,\big|\,\tau^\bullet\big)^2-
\theta_4\big(\frac{z}{2}\,\big|\,\tau^\bullet\big)^2\,
\theta_3\big(0\,\big|\,\tau^\bullet\big)^2=
-\theta_1\big(\frac{z}{2}\,\big|\,\tau^\bullet\big)^2\,
\theta_2\big(0\,\big|\,\tau^\bullet\big)^2$}
\eeq
to expand the expression (\ref{jaco}) into
\bea
\left|\frac{\dd\Pi}{\dd z}\right|^2&=&\frac1{2^{18}}\,\left|\,
\frac{\eta\big(\tau^\bullet\big)^{-42}}{\theta_1\big(z\,\big|\,
\tau^\bullet\big)^6}~\prod_{i=1}^8\,
\frac{\sqrt{\theta{a_i \choose  b_i}
\big(\frac{z}{2}\,\big|\,\tau^\bullet\big)\,
\theta{a_i\choose b_i}\big(0\,\big|\,\tau^\bullet\big)}}
{\theta{a_i\choose b_i}(0|\Pi)}\,\right|\nonumber\\ &&
\times\,\Big|\mbox{$\theta_2\big(\frac{z}{2}\,\big|\,\tau^\bullet\big)\,
\theta_3\big(\frac{z}{2}\,\big|\,\tau^\bullet\big)\,
\theta_4\big(\frac{z}{2}\,\big|\,\tau^\bullet\big)\,
\theta_3\big(0\,\big|\,\tau^\bullet\big)^2\,
\theta_4\big(0\,\big|\,\tau^\bullet\big)^2$} \nonumber\\ && \times\,
\left[\mbox{$\theta_3\big(\frac{z}{2}\,\big|\,\tau^\bullet\big)^2\,
\theta_4\big(0\,\big|\,\tau^\bullet\big)^2-
\theta_4\big(\frac{z}{2}\,\big|\,\tau^\bullet\big)^2\,
\theta_3\big(0\,\big|\,\tau^\bullet\big)^2$}\right]\Big|^4 \ .
\label{dPidz8}\eea
We have again used (\ref{Schottkyij}) to infer that every term of the
product in (\ref{dPidz8}) is independent of the chosen characteristic
$(a_i, b_i)$.

Let us now substitute (\ref{dPidz8}) into the integral (\ref{10th2}),
recalling that $\Pi=2\tau^\#$. We can again exploit the freedom in
choice of characteristics $(a_i, b_i)$ to combine the theta
functions in (\ref{dPidz8}) with the ones
$\theta_i(0|\Pi)=:\theta({}^{a_i}_{ b_i})(0|\Pi)$ and
$\theta_i(0|\tau^\bullet)=:\theta({}^{a_i}_{ b_i})(0|\tau^\bullet)$
appearing in (\ref{10th2}) by re-expressing Dedekind functions as
theta functions using (\ref{thetaeta3id}). The simplification
effectively amounts to replacing each factor $\theta_i(0|\Pi)$ with
$\sqrt{\theta_i(\frac z2|\tau^\bullet)\,\theta_i(0|\tau^\bullet)}$. We
can use this trick to cancel the difference of theta functions
appearing in the integrand of (\ref{10th2}) by simply doing this
replacement for every term, and remembering that there are in total
$40$ factors of $\theta_i(0|\Pi)$ in each term of the expansion of the
fourth power of the difference.

In this way, it is straightforward to see after some inspection that
the free energy (\ref{10th2}) may be written in terms of an integral
over the branch point location on the torus $\torus$ as
\beq
{\cal F}_2\big(\tau^\bullet\big)=\frac{g_s^2}{\big(32\pi^2\,
\alpha'\,\big)^{12}}\,\frac{\big|\eta(\tau^\bullet)\big|^{-30}}
{4\,\big|\tau^\bullet\big|^8}\,\int_\torus\,\frac{\dd^2z}{\big(
{\rm Im}\,\Pi(z)\big)^{12}}~\left|\,\frac1{\theta_1(z|\tau^\bullet)^6}\,
\prod_{i=1}^{48}\,\frac{\sqrt{\theta{a_i \choose  b_i}
\big(\frac{z}{2}\,\big|\,\tau^\bullet\big)\,
\theta{a_i\choose b_i}\big(0\,\big|\,\tau^\bullet\big)}}
{\theta{a_i\choose b_i}\big(0\,\big|\,\Pi(z)\big)}\,\right| \ .
\label{calF2brptint}\eeq
It is now clear that with (\ref{PO2ptexpl}) the DLCQ free energy
function (\ref{calF2brptint}) can be expressed in the form
(\ref{DLCQequivDVV}) if we choose the measure
\beq
\dd\mu(z)\=\frac{\dd^2z}{\mu(z)} \qquad \mbox{with} \quad
\mu(z)\=\Big(\,\frac{2\pi^2\,\alpha'}{\tau_2^\bullet}~
{\rm Im}\,\Pi(z)\,\Big)^{d/2}
\label{measurechoicebos}\eeq
where $d=24$ is the spacetime dimension of the permutation
orbifold. Using $\Pi(0)=\tau^\bullet$, the coupling
constant $\lambda$ is then given by
\beq
\lambda=\frac{4g_s}{\pi^3\,\big|512\,\tau^\bullet\big|^4}~
\sqrt{\tau_2^\bullet} \ .
\label{lambdachoicebos}\eeq
Note that the coupling (\ref{lambdachoicebos}) has the correct
infrared behaviour $\lambda\to0$ as $\tau^\bullet_2\to\infty$ to
ensure that the interacting sigma model approaches a conformal fixed
point in the infrared limit.

From the genus two perspective the origin of the measure
(\ref{measurechoicebos}) is clear. It arises from the $Sp(4,\zed)$
modular invariant integration over the moduli space of genus two
branched covering maps $f:\hat\Sigma\to\torus$. From the genus one
perspective it is a consequence of the conformal anomaly, implying
that the local twist field correlation functions depend on the
coordinatization chosen on the Riemann surface $\torus$. For the twist
field operators the natural choice is the coordinate $z$ of $\torus$,
but to induce the modular invariant interactions of strings in the
symmetric product a non-trivial integration measure
(\ref{measurechoicebos}) must be adapted. We will see this explicitly
in the next section when we study the action of the mapping class
group of the punctured torus $\torus_{\,\underline{w}\,}$.

\newsection{Nonabelian Orbifolds\label{NonabOrbs}}

In this section we address some issues surrounding the extensions of
the results of the previous section to $S_N$ orbifolds with $N>2$. At
this stage, however, we have not succeeded in making the construction
as explicit as for the $\zed_2$ orbifold. The main technical obstruction
is the combined noncommutativity of the twist group $S_N$ and the
fundamental group $\pi_1(\torus_{\,\underline{w}\,})$ of the punctured
torus. For twist group $\zed_2$ the image of the latter group under a
given monodromy homomorphism $\Phi$ is of course an abelian group,
enabling explicit constructions. But these constructions become
ambiguous and inconsistent in the nonabelian case, as one must deal
with the full nonabelian homotopy group and not just its
abelianization to the homology group. We are not
aware of any direct computation of the twist field correlation
functions in these specific instances. In the following we will
highlight some of the main technical issues surrounding these
calculations in the higher degree permutation orbifolds, and in
particular to what extent the DLCQ free energy (\ref{FDLCQ2}) can be
used to provide an explicit representative for the DVV correlator
(\ref{transinvcorr}) using the combinatorial formula
(\ref{correlator}). One of the outcomes of this analysis will be a
more precise, general description of the measure $\dd\mu(z)$ required
in the definition of the vertex operator (\ref{DVVint}).

\subsection{Uniformization Construction\label{UniConstr}}

Let us recall the general construction of Section~\ref{TwistFields}. A
correlation function involving twist fields alone in any permutation
orbifold is defined through the generalized partition function
(\ref{correlator}). It gives a twist field correlation function on a
worldsheet $\Sigma$ as a sum over twisted sectors, each characterized by a
conjugacy class of monodromy homomorphisms. One term is given by the
partition function of the covering space $\hat{\Sigma}$ determined by
Hurwitz data, comprising the monodromy, the complex structure of the
worldsheet $\Sigma$ and the insertion points of the twist field
operators. The issue is how to determine the covering space and its
complex structure in terms of the Hurwitz data. The monodromy in the case
of $k$ distinct insertion points on the worldsheet is a homomorphism
$\Phi:\pi_1(\Sigma_{\,\underline{w}\,})\to G<S_N$, and the general
Riemann-Hurwitz formula (\ref{rh}) for ramified coverings gives the
genus $\hat g$ of the covering space. Determining the topological type
of the cover is analogous to the unramified case. The fundamental
group of the marked cover $\hat{\Sigma}_{\,\underline{\hat w}}$ is given by
a stabilizer subgroup $H_a<\pi_1(\Sigma_{\,\underline{w}\,})$. The index
$a$ is the label of a sheet, which is permuted by the twist group
$G<S_N$, and different choices of $a$ result in conjugate subgroups of
$\pi_1(\Sigma_{\,\underline{w}\,})$ corresponding to different choices
of pre-image of the base point of
$\pi_1(\Sigma_{\,\underline{w}\,})$ as the base point of
$\pi_1(\hat{\Sigma}_{\,\underline{\hat w}\,})$.

However, it is much more difficult to determine the complex structure
of the cover. Recall that the prescription for the unramified case was
to choose a uniformizing homomorphism $u:\pi_1(\Sigma)\to U$ such that
$\Sigma_{\tau}=U/u(\pi_1(\Sigma))$. Then one needs to restrict $u$ to
the stabilizer subgroup of $\pi_1(\Sigma)$ corresponding to the
monodromy homomorphism $\Phi$. But the domain of the monodromy is
$\pi_1(\Sigma_{\,\underline{w}\,})$ for the ramified case, which is a
group distinct from $\pi_1(\Sigma)$. Hence it is not straightforward
to extend this uniformization method to the case of branched
coverings. Consider the commutative diagram
\beq
\xymatrix{ {\hat{\Sigma}_{\,\underline{\hat w}}}~\ar[d]_{\tilde f}
\ar[r]^{\hat\imath}&~{\hat{\Sigma}} \ar[d]^{f} \\
{\Sigma_{\,\underline{w}}}~\ar[r]_{\imath}& ~{\Sigma} }
\eeq
where the maps $\imath$ and $\hat\imath$ are the canonical inclusions
(filling in the deleted points), and $\tilde{f}$ is the restriction of
the covering map $f$ to the punctured surfaces. Passing to the
corresponding pushforwards, this diagram induces a commutative diagram
of fundamental groups given by
\beq
\xymatrix{ {\pi_1\big(\hat{\Sigma}_{\,\underline{\hat w}\,}\big)} ~
\ar[r]^{\hat\imath_*} \ar[d]_{\tilde{f}_*}&~
        {\pi_1\big(\hat{\Sigma}\big)} \ar[d]^{f_*} \\
                  {\pi_1(\Sigma_{\,\underline{w}\,})} 
\ar[r]_{\imath_*}~&~{\pi_1(\Sigma)}} \ .
\label{fundgpdiag}\eeq

Let ${\cal T}(k,g)$ denote the Teichm\"uller space of genus $g$
Riemann surfaces with $k$ punctures. Let ${\cal M}(k,g)$ be the
mapping class group of the  (marked) Riemann surface
$\Sigma_{\,\underline{w}\,}$ acting on ${\cal T}(k,g)$. One seeks maps
which fit into the commutative diagram
\beq
\xymatrix{ {{\cal T}\big(\hat k\,,\,\hat{g}\big)}~ \ar[d] \ar[r]&~
                  {{\cal T}(0,\hat{g})} \ar[d] \\
                  {{\cal T}(k,g)}~ \ar[r]&~{{\cal T}(0,g)} }
\label{Teichdiag}\eeq
associated to the covering and the inclusions such that the vertical
arrow on the left is given by the surjective map
$U/u(\pi_1(\hat\Sigma_{\,\underline{\hat w}\,}))\cong U/u(H_a)\to
U/u(\pi_1(\Sigma_{\,\underline{w}\,}))$, where $u$ is a uniformizing map
of punctured surfaces. In this way one can incorporate the
information from the monodromy contained in the admissible finite
index subgroup $H_a$. Note that the corresponding complex dimensions
of the spaces involved in (\ref{Teichdiag}) map as
\beq
\xymatrix{ {3\hat{g}-3+\hat k} \ar[d] \ar[r]~&~
                  {3\hat{g}-3} \ar[d] \\
                 {3g-3+k} \ar[r]~&~{3g-3} }
\label{dimdiag}
\eeq
for $g>0$ (except for $\dim_{\complex}{\cal T}(0,1)=1$).

The problem rests in the construction of the horizontal arrows of
(\ref{Teichdiag}). Since the pushforward $\imath_*$ is
a group homomorphism, the image of an element of a uniformizing group
$u(\pi_1(\Sigma_{\,\underline{w}\,}))<PSL(2,\real)$, which we identify
with the complex structure given by
$\halfplane/u(\pi_1(\Sigma_{\,\underline{w}\,}))\in {\cal T}(k,g)$, is
a coset and thus not an element in $PSL(2,\real)$. Thus even though
the quotient of the uniformizing group of the marked surface by the
normal closure of the parabolic generators is isomorphic to
$\pi_1(\Sigma)$ (by the admissibility constraint), it is not a
subgroup of $PSL(2,\real)$. The same remarks apply to the map
$\hat\imath$ inducing the top horizontal arrow
in~(\ref{Teichdiag}). Therefore it is not possible to apply the method
of uniformization which worked for the unramified case, and the
forgetful maps (\emph{i.e.}, the horizontal arrows in
(\ref{Teichdiag})) need to be constructed by hand.

Let us specialize to our main problem of interest, where the base space
is the torus $\Sigma=\torus$ with $k=2$ simple branch points. For an
$N$-sheeted cover of genus $\hat g=2$ there are $\hat k=2N-2$
preimages of these branch points, so that two of the $N$ preimages of
a generic point of the base coincide for a branch point. The main
obstacle in constructing the map $\imath_{\cal T}:{\cal
  T}(2,1)\to{\cal T}(0,1)$ rests in the fact that a flat torus admits
a complete euclidean metric, whereas a punctured torus admits a complete
hyperbolic metric. Thus in order to apply uniformization one needs to
construct a map between the space of flat tori and the space of
hyperbolic tori. Let us assume that the branch points are
distinguished points of the flat metric on $\torus$. Using the
automorphism group of the torus we may fix the location of one of the
branch points at the origin. Then one requires a bijection ${\cal
  T}(2,1)\to{\cal T}(0,1)\times\halfplane$, where the second branch
point $z$ varies in the complex upper half plane $\halfplane$. This
must be done in such a way that a lift of the mapping class group
${\cal M}(0,1)=SL(2,\zed)$ to ${\cal M}(2,1)$ acts equivariantly on ${\cal
  T}(2,1)$ with respect to this bijection.

An element of ${\cal T}(2,1)$ is a twice punctured hyperbolic
torus. Using the uniformizing homomorphism
$u:\pi_1(\torus_{\,\underline{w}\,})\to PSL(2,\real)$, it can be
characterized as a discrete Fuchsian group
\beq
u\big(\pi_1(\torus_{\,\underline{w}\,})\big)=\,<\alpha,\beta,\gamma
\in PSL(2,\real)~\big|~|\tr\alpha\,|>2~,~|\tr\beta\,|>2~,~
|\tr\gamma\,|=|\tr[\alpha,\beta]\,\gamma\,|=2> \ .
\label{hypertorusPSL2R}\eeq
The hyperbolic generators $\alpha,\beta$ correspond to translation
along a canonical homology basis of the unmarked torus, while $\gamma$
and $[\alpha,\beta]\,\gamma$ are the parabolic generators
corresponding to the punctures.\footnote{We could have equivalently
  used an independent parabolic generator $\gamma'$ with the relation
  $[\alpha,\beta]\,\gamma\,\gamma'=1$.} Then the complex structure is
given by $\halfplane/u(\pi_1(\torus))$. The subgroup
(\ref{hypertorusPSL2R}) contains three real parameters for each
generator, two trace relations for parabolicity and a conjugation
symmetry which eliminates three parameters, hence the real dimension
of ${\cal T}(2,1)$ is $3\cdot3-2-3=4$, as anticipated.

The space ${\cal T}(0,1)\times\halfplane$ is coordinatized by ordered
pairs $(\tau,z)$, where $\tau$ is a genus one modulus and $z$ is a
distinguished point on $\torus$. The mapping class group ${\cal
  M}(0,1)\cong SL(2,\zed)$ of the flat torus acts on these pairs
through the generators
\beq
T\,:\,(\tau,z)~\longmapsto~(\tau+1,z) \qquad \mbox{and} \qquad
S\,:\,(\tau,z)~\longmapsto~\big(\mbox{$-\frac1\tau\,,\,\frac
  z\tau$}\big)
\label{SL2Ztauz}\eeq
obeying $S^4=(T\,S)^3\,S^2=1$. The modular $S$-transformation here is
defined via analytic continuation along a clockwise oriented path
around the origin in the complex $z$-plane. A lift of these generators
to the mapping class group ${\cal M}(2,1)$ of the twice punctured
hyperbolic torus is presented in~\cite{Smyrnakis:1996ii} as an action
on the generators of (\ref{hypertorusPSL2R}) by
\beq
\tilde T\,:\,\begin{pmatrix}\alpha\\\beta\\\gamma\end{pmatrix}~
\longmapsto~\begin{pmatrix}\alpha\\\beta\,\alpha\\\gamma\end{pmatrix}
\qquad \mbox{and} \qquad \tilde S\,:\,\begin{pmatrix}\alpha\\\beta\\
\gamma\end{pmatrix}~\longmapsto~\begin{pmatrix}\beta^{-1}\\
\alpha\\\beta^{-1}\,\gamma\,\beta\end{pmatrix} \ .
\label{SL2ZliftM21}\eeq
This lift of $SL(2,\zed)$ is not unique. In fact, the modular group
${\cal M}(2,1)$ is an extension of $SL(2,\zed)$ by ${\cal
  B}(2,1)/\Gamma_{{\cal M}(2,1)}$, where $\Gamma_{{\cal M}(2,1)}$ is
the center of ${\cal M}(2,1)$ and ${\cal B}(2,1)$ denotes the
two-stranded braid group of the
torus~\cite{Smyrnakis:1996ii}. Equivariance of the bijection
$\imath_{\cal T}:{\cal T}(2,1)\to{\cal T}(0,1)$ with respect to these
actions is then the statement
\beq
\imath_{\cal T}\circ\tilde T\=T\circ\imath_{\cal T} \qquad \mbox{and}
\qquad \imath_{\cal T}\circ\tilde S\=S\circ\imath_{\cal T} \ .
\label{forgetequiv}\eeq

We have not succeeded in constructing explicitly the required modular
equivariant bijections, and it is not possible to write an algebraic
formula~\cite{FarkasKra}. One could try to surpass this problem by
working directly with the hyperbolic presentation of the tori, and the
known bijection between the Fenchel-Nielsen coordinates of
Teichm\"uller space and the Fuchsian coordinates parametrizing the
uniformizing group~\cite{FarkasKra}. But there is a great deal of
ambiguity in this procedure which prevents an explicit construction,
and there is no canonical way to identify the modular parameters of
the torus itself and those corresponding to the branch points.

\subsection{Homology Construction\label{HomConstr}}

Given the technical difficulties encountered above, we now turn to an
alternative approach to determining the complex structure of the cover
via the push-forward induced on homology groups 
$f_*:H_1(\hat{\Sigma},\zeds)\to H_1(\Sigma,\zeds)$, which is provided
by the abelianization of the diagram (\ref{fundgpdiag}) for the
fundamental groups. If a canonical basis is fixed both in the homology
group of the base and that of the cover, then this map is given by a
$2g\times2\hat{g}$ matrix ${\sf M}^\top$. This matrix can then be used
to determine the period matrix of the cover in terms of the period
matrix of the base and some additional parameters~\cite{csz}. For
sufficiently low genus, the period matrix $\tau$ uniquely characterizes the
complex structure. We will go through this construction in detail for
the relevant case of the genus two cover $f:\hat\Sigma\to \torus$ for
the two point function of twist fields corresponding to simple branch
points. In this case the complex structure on $\hat\Sigma$ is
determined by a canonical map $H^{1,0}(\hat\Sigma,\complex)\otimes
H_1(\hat\Sigma,\zed)\to\complex$.

Let us see first how the matrix representation $\sf M$ of $f_*$ can be
determined and compared to the construction of Section~\ref{DLCQ}. The
main difference from the $N=2$ case studied at the beginning of
Section~\ref{Z2Twist} is that for $N>2$ the preimages of the punctures
are no longer just the ramification points, since there are $2N-2>2$
preimages of the branch points. Let 
$\pi_1(\torus_{\,\underline{w}\,})=\,<\a,\b,\g>$, the free group on
three generators such that $\ker(\imath_*)$ is the normalizer
$N_{\pi_1(\torus_{\,\underline{w}\,})}(\g,[\a,\b]\,\g)$. In other
words, the image of $\a$ and $\b$ are the standard generators of
$\pi_1(\torus)$, whereas $\gamma$ and $[\a,\b]\,\g$ correspond to
simple closed curves which are contractible to the branch points. The
stabilizer $H_a<\pi_1(\torus_{\,\underline{w}\,})$ corresponding to a
monodromy homomorphism $\Phi$ is a subgroup of index $N$ in the case
of an $N$-sheeted cover. It can be presented in terms of $4+(2N-2)-1$
words from $\pi_1(\torus_{\,\underline{w}\,})$ which freely generate
the group $H_a$. By identifying $H_a$ with
$\pi_1(\hat{\Sigma}_{\,\underline{\hat w}\,})$, this presentation gives the
homomorphism $\tilde{f}_*$ explicitly. There are $N-2$ independent
elements from $H_a$ which are conjugate to $\g$ in
$\pi_1(\torus_{\,\underline{w}\,})$, and another $N-2$ elements which
are conjugate to $[\a,\b]\,\g$. There is one further element conjugate
to $\g^2$ and another one conjugate to $([\a,\b]\,\g)^2$. This is
because the $N-2$ generators of
$\pi_1(\hat{\Sigma}_{\,\underline{\hat w}\,})$ corresponding to simple
closed curves contractible to $N-2$ preimages of a branch point
project to the simple closed curve contractible to the branch point,
whereas the other two generators project to curves with winding number
two about each of the branch points.

The normalizer of these $2N-2$ generators in $H_a$ is the subgroup
$\ker(\hat\imath_*)$. One then seeks $4+2N-3$ generating elements such
that $2N-3$ are in $\ker(\hat\imath_*)$ and also the commutator product
$[\hat\alpha_1,\hat\beta_1]\,[\hat\alpha_2,\hat\beta_2]$ of a suitably
chosen remaining four. In other words, $\hat\alpha_i$, $\hat\beta_i$
are representatives of the cosets that project to a canonical homology
basis of $\pi_1(\hat{\Sigma})$ under the map $\hat\imath_*$. Due to
the commutativity of the diagram (\ref{fundgpdiag}) and the
abelianization, the entry ${\sf M}_{ij}$ of the $2\times 4$ homology
covering matrix is the sum of powers of the $i$-th generator of
$H_1(\torus,\zeds)$ ($\a$ or $\b$) appearing in the expression of the
$j$-th generator of $H_1(\hat{\Sigma},\zeds)$ ($\hat\alpha_1$,
$\hat\beta_1$, $\hat\alpha_2$ or $\hat\beta_2$). In this way, the
two-point function may be computed by summing over admissible finite
index subgroups $H_a<\Gamma=\pi_1(\torus_{\,\underline{w}\,})$.

Let us look at an explicit example of how this works. For $N=3$, there
are $16$ conjugacy classes of transitive monodromy homomorphisms, each
class containing $6$ homomorphisms. Accordingly, there are $16$
conjugacy classes of admissible index three subgroups of $\Gamma$,
each class having $[\Gamma:N_\Gamma(\Gamma_k)]=3$
representatives. Consider the admissible monodromy homomorphism
$\Phi_1$ given by
\beq
\Phi_1\,:\, \a~\longmapsto~(2~3) \ , \quad \b~\longmapsto~(1~2) \quad
\mbox{and} \quad \g~\longmapsto~(2~3) \ . \eeq
The corresponding three sheeted cover may be depicted schematically as
\begin{equation}\includegraphics[height=4cm]{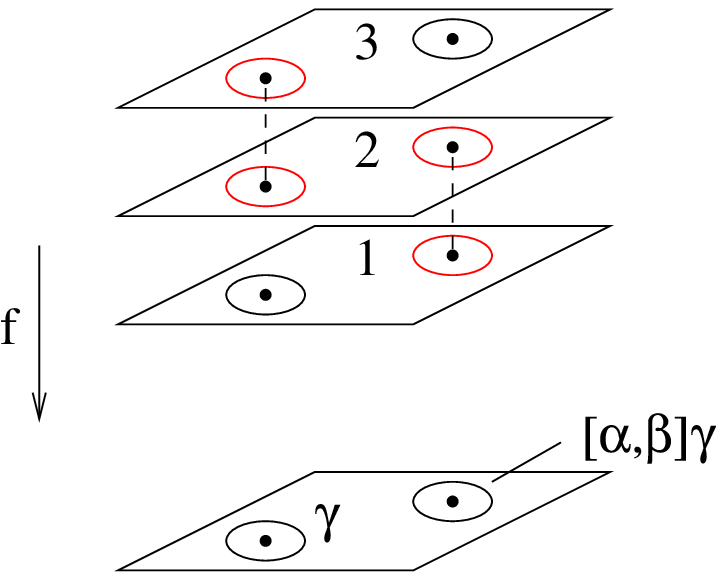}\end{equation}
with the parallelogram representing the base torus $\torus$.
The sheets $2$ and $3$ are ramified over the branch point corresponding
to $\g$, while the sheets $1$ and $2$ are ramified over the other
branch point corresponding to $[\a,\b]\,\g$ (since
$\Phi_1:[\a,\b]\,\g\mapsto (1~2)$).

The stabilizer subgroup of
$\Gamma=\pi_1(\torus_{\,\underline{w}\,})=\,<\a,\b,\g>$ can
be presented by\footnote{In practice it is easier to determine the
monodromy homomorphism corresponding to a given presentation of a
finite index subgroup.}
\beq H_1\=\,<g_1,\dots,g_7>~:=~
<\a,\, \b^2,\, \g,\, \b\,\a^2\,\b^{-1},\, \b\,\g\,\a^{-1}\,\b^{-1},\, 
\b\,\a\,\g\,\b^{-1},\, \b\,\a\,\b\,\a^{-1}\,\b^{-1}> \ . \eeq
The elements $g_1,\ldots,g_7$ generate the the group $H_1$ freely.
One can then determine the generators of $\ker(\hat\imath_*)$ as
\bea
\hat g_1&=&g_3\=\underline{\g} \ , \nonumber \\[4pt]
\hat g_2&=&g_6\,g_5\=\b\,\a\,\underline{\g^2}\,\a^{-1}\,\b^{-1} \ ,
\nonumber \\[4pt]
\hat g_3&=&g_4\,g_2\,g_1^{-1}\,g_2^{-1}\,g_5\=\b\,\a\,
\underline{[\a,\b]\,\g}\,\a^{-1}\,\b^{-1} \ , \nonumber\\[4pt]
\hat g_4&=&g_1\,g_4^{-1}\,g_7^{-1}\,g_6\,g_7\,g_3\=
\underline{\big([\a,\b]\,\g\big)^2} \ ,  \eea
where we have underlined the curves on the base that they are
conjugate to. Finally, it is possible to write down the generators
\beq
\hat\alpha_1\=g_7 \ , \quad \hat\beta_1\=g_1\,g_4^{-1} \ , \quad
\hat\alpha_2\=g_1\quad \mbox{and} \quad \hat\beta_2\=g_2
\label{quotientgens}\eeq
such that\footnote{One can check that the elements
  (\ref{quotientgens}) are independent representatives of the
  generators of the quotient modulo $\hat g_1$, $\hat g_2$ and $\hat
  g_4$, except for $\hat
  g_3=[\hat\alpha_1,\hat\beta_1]\,[\hat\alpha_2,\hat\beta_2]$.}
\beq
H_1/N_{H_1}(\hat g_1,\hat g_2,\hat g_3,\hat g_4)\=\,
<\hat\alpha_1,\hat\alpha_2,\hat\beta_1,\hat\beta_2\;\big|\;
[\hat\alpha_1,\hat\beta_1]\,[\hat\alpha_2,\hat\beta_2]=1>\,\=
\pi_1\big(\hat\Sigma\big) \ .
\eeq

We can now count the powers of $\a,\b$ appearing in
(\ref{quotientgens}) to determine the matrix representation ${\sf
  M}={\sf M}_1$ of $f_*$ in (\ref{fpushM}) with
\beq
{\sf M}_1=\begin{pmatrix}0&1&-1&0\\1&0&0&2\end{pmatrix} \ .
\eeq
The remaining $15$ admissible finite index subgroups are similarly
treated. All instances provide a matrix representation $\sf M$ which
satisfies the Hopf condition and which leads to the normal form
(\ref{sfMints}) after reduction using the symplectic group
$Sp(4,\zed)$. However, the map from the set of admissible finite index
subgroups to the set of normal forms (\ref{sfMints}) obeying the Hopf
condition is \emph{not} unique, and there is a large degree of
arbitrariness in this procedure. The reason is that the partial
reduction leading to (\ref{sfMints}) involves only $Sp(4,\zed)$
transformations, but not modular transformations of the base. It may
happen that an admissible finite index subgroup $H_a$ is invariant
under an $SL(2,\zed)$ transformation of the base (\emph{e.g.},
$\a\leftrightarrow\b$), in which case one may get matrices $\sf M$
leading to period matrices which are not related by a modular
transformation on the cover $\hat\Sigma$. Thus it is only onto the set
of fully reduced Poincar\'e normal forms of $\sf M$, which
incorporates a sum over all such $SL(2,\zed)$ transformations of the
base $\torus$, that this reduction map is unique. However, the reduced
moduli space for the Poincar\'e normal form is very complicated and
depends sensitively on number theoretic properties of the
degree~$N$~\cite{csz}.

\subsection{Equivariance of the DVV Correlator}

The construction of Section~\ref{HomConstr} above determines the
dependence of the $2\times2$ period matrix $\tau_H$ on a given
admissible monodromy homomorphism, or equivalently a given admissible
finite index subgroup $H<\Gamma$, with $\tau_H=\tau_{r,m,s,t}$ in
(\ref{Omegacover}). At this stage we are faced with the problem of
finding the dependence (either explicit or implicit) of the Prym
modulus $\Pi=r\,\tau^\#$ on the branch point location
$z\in\torus$. The construction of Prym differentials in
Section~\ref{Prym} does not carry through to the
higher degree branched covers, because for any genus two cover
$f:\hat\Sigma\to\torus$ of degree $N\geq3$ there are no non-trivial
automorphisms $\iota:\hat\Sigma\to\hat\Sigma$ such that
$f\circ\iota=f$~\cite{kani}. As any genus two Riemann surface is a
hyperelliptic curve, the cover $\hat\Sigma$ does have a canonical
hyperelliptic involution $\iota_{\hat\Sigma}$ and its hyperelliptic
divisor which is the effective divisor of degree six consisting of the
fixed points of $\iota_{\hat\Sigma}$. Then there is a unique
involution $\iota_\torus:\torus\to\torus$ of the base such that
$f\circ\iota_{\hat\Sigma}=\iota_\torus\circ f$~\cite{kani}. However,
given that the above construction is not invariant under $SL(2,\zed)$
transformations of the base, it is not clear how to exploit the
hyperelliptic representation of $\hat\Sigma$, and the corresponding
Schottky relations, to determine the branch point dependence as
before. This is further reflected in the fact that the standard
constructions of cut abelian differentials (such as
(\ref{Prymdiffbranch})) for cyclic orbifolds~\cite{Atick} become
ambiguous for nonabelian monodromy. We are not aware of any
constructions of Prym differentials or Prym moduli for higher degree
genus two covers $f:\hat\Sigma\to\torus$ in terms of branch point
loci.

On general grounds it follows that the complex structure on the
covering surface $\hat\Sigma$ is uniquely determined by the
holomorphic map $f:\hat\Sigma\to\torus$ in terms of the moduli
$\tau^\bullet$ and $z$, but not necessarily in an explicit
parametrization. We can use results of~\cite{Mason:2006dk} to
ascertain that the desired explicit branch point dependence
\emph{does} exist and can be used to give some insight into the
modular behaviour of the DVV correlator. One of the advantages of the
formalism of Section~\ref{HomConstr} over that of
Section~\ref{UniConstr} above is that one can study equivariance
properties in the genus two modular group $Sp(4,\zed)$, rather than in
the more complicated mapping class group ${\cal M}(2,1)$. For fixed
monodromy given by an admissible finite index subgroup $H<\Gamma$,
there is a holomorphic map
\beq
\tau_H\,:\,{\cal T}(0,1)\times\halfplane~\longrightarrow~
\halfplane^2 \ , \qquad (\tau^\bullet,z)~\longmapsto~
\tau_H(\tau^\bullet,z)
\label{tauHholmap}\eeq
which is determined generically in~\cite{Mason:2006dk} via a sewing
construction on twice-punctured tori in terms of Jacobi-Erd\'elyi
theta functions, Weierstrass functions and Eisenstein series on the
base torus $\torus$. The primary difference in our specific case is
that the modulus $\tau^\#$ has a square root cut singularity
at each of the branch points $w_1=z$ and $w_2=0$, rather than the
logarithmic cut singularity which arises in~\cite{Mason:2006dk}.

Consider the monomorphism $SL(2,\zed)\hookrightarrow Sp(4,\zed)$ given
by
\beq
\begin{pmatrix}a&b\\c&d\end{pmatrix}~\longmapsto~
\begin{pmatrix}a&0&b&0\\0&1&0&0\\c&0&d&0\\0&0&0&1\end{pmatrix} \ .
\label{SL2Zemb}\eeq
This lift of $SL(2,\zed)$ acts in the expected way on the domain of
the map (\ref{tauHholmap}) as
\beq
(\tau^\bullet,z)~\longmapsto~\mbox{$\big(\frac{a\,\tau^\bullet+b}
{c\,\tau^\bullet+d}\,,\,\frac z{c\,\tau^\bullet+d}\big)$} \ .
\label{SL2Zdomaction}\eeq
For each choice of branch for $\tau^\#$, the map $\tau_H$ is
equivariant with respect to this action of $SL(2,\zed)<
Sp(4,\zed)$~\cite{Mason:2006dk} and there is a commutative diagram
\beq
\xymatrix{ {\cal T}(0,1)\times\halfplane ~
\ar[r]^{~~~~~\tau_H} \ar[d]_{SL(2,\zed)}&~
        \halfplane^2 \ar[d]^{SL(2,\zed)} \\
                  {\cal T}(0,1)\times\halfplane
\ar[r]_{~~~~~\tau_H}~&~\halfplane^2} \ .
\label{tauHcommdiag}\eeq
This property determines the equivariance of the DVV correlator
(\ref{transinvcorr}), represented by the genus two DLCQ free energy
(\ref{FDLCQ2}) at a fixed value of the degree $N$. Since under
(\ref{SL2Zdomaction}) the flat area form on the torus transforms as
$\dd^2z\mapsto\dd^2z/|c\,\tau^\bullet+d|^2$, and since the local twist
field correlation functions
$\langle\sigma_{a_1b_1}(z)\,\sigma_{a_2b_2}(0)\rangle^{S_N}$ have
total scaling dimension $6$, the scaling properties of the measure
$\mu(z)$ under (\ref{SL2Zdomaction}) can be explicitly
determined.

Given the remarkable agreement of the $N=2$ free energy
with the twist field two-point function in the $\zed_2$ orbifold, it
is natural to extrapolate this correspondence and to take the
fixed~$N$ DLCQ free energy integrand in (\ref{FDLCQ2}) as the {\it
  definition} of the local twist field correlation function
$\langle\sigma_{a_1b_1}(z)\,\sigma_{a_2b_2}(0)\rangle^{S_N}$ on the
$\real^{24}\wr S_N$ permutation orbifold, according to the covering
surface principle of Section~\ref{TwistFields}. However, the explicit
form of the mapping (\ref{tauHholmap}) displayed
in~\cite[Proposition~6.2]{Mason:2006dk} is far too complicated for an
explicit determination of the required Jacobian $|\dd\tau^\#/\dd
z|^2$ (and furthermore one needs an $Sp(4,\zed)$ transformation
relating their period matrix to ours). Moreover, it is difficult to
arrive at explicit formulas which are illuminating, as the products
(\ref{Psi10Omega}) of theta functions (\ref{poi}) are rather involved
for $N\geq3$.

\newsection{Fermionic Orbifolds\label{FermOrb}}

In this final section we will study fermionic extensions of the
permutation orbifolds considered thus far, in particular those
orbifold sigma models arising in discrete light-cone quantization of
superstrings and heterotic strings in ten spacetime dimensions. We
will describe the modifications of the covering surface principle and
twist field operators of Section~\ref{CorrsPermOrbs} required in these
cases. The genus two DLCQ free energy amplitudes in these instances
are derived in~\cite{csz}. Given the success of the bosonic $\zed_2$
orbifold model of Section~\ref{Z2Orbifolds}, we will use the
appropriately modified versions of the generic covering
space principle of Section~\ref{TwistFields} to compute local one-loop
correlation functions of (spin) twist field operators in
supersymmetric and heterotic $\zed_2$ orbifolds. To the best of our
knowledge these correlation functions have not been previously
computed. The analysis of this section thus provides a powerful
application of DLCQ string theory to producing new explicit
expressions for correlation functions in orbifold superconformal field
theories on the one hand, and for the forms of the leading cubic
string interactions in the associated superstring field theories in
ten dimensions on the other hand. Throughout we work in the
Neveu-Schwarz-Ramond formalism.

\subsection{Spin Twist Fields\label{ExcitedTwist}}

Consider the superconformal sigma model on the torus with target space
$\real^8$ defined by the action
\beq
I(X,\psi)=\frac{1}{4\pi\,\alpha'}\,\int_\torus\,
\dd^2z~\frac{1}{2\ii\tau_2}\,\left(\partial X_i
(z)\,\overline{\partial}X_i(z)+\psi_i(z)\,\overline{\partial}
\psi_i(z)+\overline{\psi}{}_i(z)\,\partial\overline{\psi}{}_i
(z)\right) \ ,
\label{susysigmamodel}\eeq
where the real bosonic fields $X_i$, $i=1,\dots,8$ transform in the
eight-dimensional vector representation $\mbf 8_v$ of the R-symmetry
group $SO(8)$, while the components
$\psi_i,\overline{\psi}{}_i$, $i=1,\dots,8$ of the 16-component
Majorana-Weyl spinor field $\psi$ transform in the spinor $\mbf 8_s$
and conjugate spinor $\mbf 8_c$ representations of $SO(8)$,
respectively. The spinor fields are sections of the
twisted spin line bundle $S_\torus\otimes L_{\mbf\delta}$ over the
torus, where $L_{\mbf\delta}$ is a real line bundle
over $\torus$ with flat connection determined by one of the four spin
structures ${\mbf\delta}=\big({}^{\delta_\alpha}_{\delta_\beta}\big)\in
H^1(\torus,\zed/2\zed)=\zed^2/2\zed^2$ and $[S_\torus]\in{\rm
  Pic}^0(\torus)$ is chosen to correspond to the theta divisor in the
given homology basis $(\alpha,\beta)$. The ${\cal N}=8$ worldsheet
supersymmetry of the sigma-model is generated by the fermionic
supercurrents
\beq
{\sf G}^\ell(z)=-\mbox{$\frac12$}\,\gamma_{\ell'\ell}^i\,\left(
\psi_{\ell'}(z)\,\partial X_i(z)+
\overline{\psi}{}_{\ell'}(z)\,\overline{\partial}X_i(z)\right)
\label{supercurrent}\eeq
where $\gamma^i$ are the $Spin(8)$ Dirac matrices.

In the corresponding permutation orbifold, the monodromy conditions on
the bosonic fields $X$ in a given twisted sector $(P,Q)$ are as in
(\ref{mbond}), while the fermion monodromy is given by
\beq
\psi^a(z+1)\=(-1)^{\delta_\alpha}\,\psi^{P(a)}(z) \qquad \mbox{and}
\qquad \psi^a(z+\tau)\=(-1)^{\delta_\beta}\,\psi^{Q(a)}(z) \ ,
\label{fermmonodromy}\eeq
where for simplicity we have omitted a potential extra sign depending
on the reference spin structure $[S_\torus]$. This symmetry is
compatible with ${\cal N}=8$ worldsheet superconformal
invariance~\cite{Dixon:1988ac}, and it means that on the fermionic
fields the twist group $G$ is extended to
$G\times(\zed_2)^N$. The consistency
condition $P\,Q=Q\,P$ implies~\cite{Fuji:2001kt} that the spin
structure phases in (\ref{fermmonodromy}) are independent of the
coordinate label $a$ in the permutation orbifold, and hence that only
the diagonal subgroup of $(\zed_2)^N$ acts nontrivially on the
fermions. The asymmetry between the twistings of bosons and fermions
implies that the modular invariant sum over monodromy homomorphisms
breaks spacetime supersymmetry of the orbifold sigma model.

Generally, the sum over $(\zed_2)^N$ monodromy in the fermionic sector
is weighted by a consistent set of GSO phases
$\zeta[\mbf\delta;\Phi]$, generically dependent upon the twisted
sector $\Phi:\pi_1(\Sigma)\to G$, which are constrained by modular
covariance requirements. In the untwisted sector $\Phi(-)=e$, the
phase corresponding to a spin structure
$\mbf\delta=\big({}^{\mbf\delta_\alpha}_{\mbf\delta_\beta}\big)\in
\zed^{2g}/2\zed^{2g}$ is the mod~$2$ index of the Dirac operator on
$\Sigma$ twisted by the flat line bundle $L_{\mbf\delta}\to\Sigma$
given by~\cite{Mumford1}
\beq
\zeta[\mbf\delta;e]\=(-1)^{\dim H^0(\Sigma,S_\Sigma\otimes
  L_{\mbf\delta})} \= (-1)^{\mbf\delta_\alpha\cdot\mbf\delta_\beta} \ ,
\label{zetadeltae}\eeq
where $\dim H^0(\Sigma,S_\Sigma\otimes L_{\mbf\delta})$ is the number
of linearly independent holomorphic sections of the spin bundle
$S_\Sigma\otimes L_{\mbf\delta}$. Schematically then, the modification
of the formula (\ref{partf}) for the partition function of the
supersymmetric permutation orbifold is given by
\beq Z^{G\times(\zed_2)^N}(\tau)=\frac{1}{2^N\,|{G}|}\,
\sum_{\Phi:\pi_1(\Sigma)\to {G}}\;~\sum_{\mbf\delta\in
  H^1(\Sigma,\zed/2\zed)} \,\zeta[\mbf\delta;\Phi]~
\Big(\,\prod_{\xi\in {\cal O}(\Phi)}\,Z_{\mbf\delta}
\big(\tau^{\xi}\big)\,\Big)
\label{partfsusy}\eeq 
where $Z_{\mbf\delta}(\tau^\xi)$ is the partition function of the
parent superconformal field theory computed with the global fermionic
monodromy determined by the spin structure $\mbf\delta$.

For example, the partition function of the supersymmetric
$\real^8\wr(S_N\times(\zed_2)^N)$ permutation orbifold on
$\Sigma=\torus$ can be determined by first calculating the contribution
from a given spin structure (say the Ramond-Ramond sector) to the path
integral over the complex fermionic fields, and then summing over the
modular orbits using either of the two GSO projections of Type~II
string theory. Then the parent partition function appearing in the
formula (\ref{partf}) is given by~\cite{Fuji:2001kt}
\beq
Z(\tau)=\mbox{$\frac12$}\,\left|\mathfrak{z}\big({}^0_0\big)(\tau)^4-
\mathfrak{z}\big({}^0_1\big)(\tau)^4-
\mathfrak{z}\big({}^1_0\big)(\tau)^4\pm
\mathfrak{z}\big({}^1_1\big)(\tau)^4\right|^2 \ ,
\label{ZtauGSO}\eeq
where the $+/-$ sign corresponds to the Type~IIA/B string amplitude
and
\bea
\mathfrak{z}({\mbf\delta})(\tau)&=&
\left(\frac{4\pi^2\,\alpha'}{\tau_2}\right)^4~
\e^{\frac{\pi\ii}{12}\,(2 \delta_\alpha^2-1)\,\tau}~
\e^{\pi\ii\delta_\alpha\,\delta_\beta/4} \nonumber\\ && \times\,
\prod_{n=1}^\infty\,\left(1-(-1)^{\delta_\beta}~
\e^{\pi\ii\tau\,(2n-1+\delta_\alpha)}\right)\,
\left(1-(-1)^{\delta_\beta}~\e^{\pi\ii\tau\,(2n-1-\delta_\alpha)}
\right) \ .
\label{zdeltatau}\eea
The corresponding grand canonical partition function (\ref{combim})
matches the Type~II DLCQ free energy at finite temperature, with
(\ref{ZtauGSO}) producing the action of the (restricted) Hecke
operator on the partition function of the first quantized
Green-Schwarz superstring~\cite{GS1,Grignani:2000zm,csz}. A completely
analogous correspondence holds for the thermal partition function of
Type~IIB DLCQ superstrings on the maximally supersymmetric plane wave
background in ten dimensions~\cite{Sugawara:2002rs}.

The operators which create local monodromy in the superconformal sigma
model with respect to the action of $S_N$ are products
$\sigma_P(z)\,{\cal S}_P(z)$ of bosonic and fermionic twist
fields. Let us work in the sector of trivial global $\zed_2$ monodromy
for the spinor fields, \emph{i.e.}, with the Ramond-Ramond spin structure
${\mbf\delta}=\big({}^0_0\big)$. The other sectors are treated
similarly as in~\cite{Dijkgraaf:2003nw}. In
a $\zed_n$-twisted sector corresponding to a
cyclic permutation $P=(n)$, the vacuum state then carries an irreducible
representation of the Clifford algebra for $Spin(8)$. By using an
$SO(8)$ triality isomorphism, the representation space can be taken to
be the direct sum $\mbf 8_v\oplus\mbf 8_c$. The corresponding
components of the 16-dimensional ground state vector are created
respectively by the primary spin fields ${\cal S}^i_{(n)}(z)$ and
$\widetilde{\cal S}{}^{\,i}_{(n)}(z)$, $i=1,\dots,8$. They each have
conformal dimension~\cite{Arutyunov:1997gi}
\beq
\Delta_{(n)}^{\rm RR}=\frac n6+\frac1{3n} \ .
\label{DeltanRR}\eeq

To describe the supersymmetric version of the DVV interaction
vertex~\cite{Dijkgraaf:1997vv}, we need another kind of spin twist
field to ensure that the operators generating the basic joining and
splitting of superstrings yield an irrelevant deformation of the
superconformal sigma model. The bosonic twist field $\sigma_{ab}(z)$
transposing the fields $X^a$ and $X^b$ has conformal dimension
$\frac12$ when $d=8$ (see~(\ref{Deltand})), as does the fermionic
twist field ${\cal S}_{ab}(z)$ interchanging $\psi^a$ and $\psi^b$. To
increase the scaling dimension by $\frac12$ in a supersymmetric
fashion, we use the supersymmetric descendent of the primary twist
field operators $\sigma(z)\,\widetilde{\cal S}(z)$ given by
\beq
\big[{\sf Q}^\ell\,,\,\sigma(z)\,\widetilde{\cal
  S}{}^{\,\ell'}(z)\big]+
\big[\sigma(z)\,\widetilde{\cal
  S}{}^{\,\ell}(z)\,,\,{\sf Q}^{\ell'}\big]\=
\varrho^i(z)\,{\cal S}^i(z)~\delta^{\ell\ell'}~=:~\Lambda(z)~
\delta^{\ell\ell'}
\label{Lambdazdef}\eeq
where
\beq
{\sf Q}^\ell=\oint\,\frac{\dd z}{2\pi\ii}~{\sf G}^\ell(z)
\label{Qelldef}\eeq
are the ${\cal N}=8$ supercharges and the contour integral is taken
around the origin $z=0$. The descendent bosonic twist fields
$\varrho_{[P]}^i(z)$ create the first excited states in the twisted
sector $[P]$. Since the combination $\psi^a-\psi^b$ has Ramond
boundary conditions under transposition in $S_N$, the corresponding
spin field carries a representation of the Clifford algebra. The
twist field ${\cal S}_{ab}^i(z)$ transforms as a vector of $SO(8)$,
and it coincides with the standard spin field of the supersymmetric
$\real^8\wr\zed_2$ permutation orbifold which can be constructed
explicitly via bosonization of the fermion fields
$\psi_i$~\cite{Friedan:1984rv,Friedan:1985ge}.

The fermionic DVV vertex operator is now defined by
\beq
V_{\rm ferm}=-\frac{\lambda\,N}{{\rm vol}(\torus)}\,\int_\torus\,
\dd\mu(z)~\sum_{1\leq a<b\leq N}\,\Lambda_{ab}(z) \ .
\label{DVVintferm}\eeq
The descendent twist field $\Lambda_{ab}(z)$ is a primary field of
conformal weight $\frac32$. The interaction vertex (\ref{DVVintferm})
is spacetime supersymmetric, $SO(8)$ invariant and describes
elementary string interactions~\cite{Dijkgraaf:1997vv}.

The computation of the local twist field correlations functions
$\langle\Lambda_{a_1b_1}(z)\,\Lambda_{a_2b_2}(0)\rangle^{S_N\times(\zed_2)^N}$
requires a modification of the covering surface principle of
Section~\ref{TwistFields}. This is because
one should no longer simply close the punctures on the covering space
$\hat\Sigma$ corresponding to the branch points to get the identity
state at those points. Rather, one must insert the operator that
creates a Ramond vacuum at the insertion points in order to give the
fermions the correct local monodromy. Thus in the supersymmetric
orbifold theory one uses the same covering spaces $\hat\Sigma$ as in
the case of the bosonic orbifold, but instead of computing the
partition function on $\hat\Sigma$ one computes a correlation
function of spin fields on $\hat\Sigma$.\footnote{A similar statement
  is also true in the NS--NS sector. In the mixed R--NS and NS--R
  sectors, there are no combinations of $\psi^a$ which possess zero
  modes, so that these sectors have trivial local spin monodromy and
  the prescription instead follows that of Section~\ref{TwistFields}.}
Schematically, the modification of a generic, normalized bosonic twist
field correlation function (\ref{correlator}) is given by
\bea
&& \Big\langle\,\prod_{i=1}^k\,
\Lambda_{[P_i]}(w_i)\,\Big\rangle^{G\times(\zed_2)^N}
\label{fermcorrelator} \\ &&
\qquad\qquad ~=~ \frac{1}{2^N\,|G|}\,
\sum_{\Phi:\pi_1(\Sigma_{\,\underline{w}\,})\to G\times(\zed_2)^N}~
\frac1{Z^{G\times(\zed_2)^N}(\tau)}~
\prod_{\xi\in {\cal O}(\Phi)}\,\Big\langle\,\prod_{i=1}^{\hat k}\,
\hat{\cal S}_{[P_i]}(\hat w_i)\,\Big\rangle
\big(\tau^{\xi,\,\underline{w}\,}
\big) \ , \nonumber \eea
where the global $(\zed_2)^N$ monodromy acts trivially in the bosonic
sector and diagonally in the fermionic sector as in
(\ref{partfsusy}). A similar prescription for ${\cal N}=4$
supersymmetric orbifold sigma models is used
in~\cite{Lunin:2001pw}. When $\Sigma=\torus$, this will be provided by
the corresponding DLCQ free energy through the required modification
of the GSO projection at finite temperature which breaks supersymmetry
by making spacetime fermions antiperiodic around the thermal
cycle~\cite{csz}.

Similar considerations also apply to the heterotic sigma model on the
torus with target space $\real^8$, which is defined by the action
\beq
I(X,\psi,\chi)=\frac{1}{4\pi\,\alpha'}\,\int_\torus\,
\dd^2z~\frac{1}{2\ii\tau_2}\,\left(\partial X_i
(z)\,\overline{\partial}X_i(z)+\psi_i(z)\,\overline{\partial}
\psi_i(z)+\chi_A(z)\,\partial\chi_A(z)\right)
\label{hetsigmamodel}\eeq
where the Majorana-Weyl fermion fields $\chi_A$, $A=1,\dots,32$ are
$SO(8)$ singlets. The holomorphic sector of this worldsheet field
theory coincides with that of the supersymmetric sigma model
(\ref{susysigmamodel}), while after bosonization of $\chi_A$ the
antiholomorphic sector coincides with the bosonic sigma model
(\ref{bosPolyakov}) in $d=24$ with $16$ of the bosons compactified on
the Cartan torus of the heterotic gauge group ${\cal G}=SO(16)\times
SO(16)$. The heterotic sigma model (\ref{hetsigmamodel}) is a
superconformal field theory with chiral $(8,0)$ worldsheet
supersymmetry.

The corresponding $(8,0)$ supersymmetric permutation
orbifold~\cite{Rey:1997hj,Lowe:1997sx} is $(\real^8\times{\cal
  G})\wr(G\ltimes(\zed_2)^N)$. The twist subgroup $(\zed_2)^N$ acts
on the holomorphic sector exactly as in the supersymmetric case. In
the antiholomorphic sector, the gauge fermions $\chi_A$ are sections
of flat real line bundles $L_{\mbf \delta}\to\torus$ like $\psi_i$,
and so have global fermionic monodromy conditions as in
(\ref{fermmonodromy}). In contrast to the fields $\psi_i$, however,
the spin structure phases for $\chi_A$ in the permutation orbifold
generally depend on the coordinate label $a$. Perturbative string
interactions are now generated by the heterotic version of the DVV vertex
operator~\cite{Rey:1997hj,Lowe:1997sx}. For this, we must explicitly
write worldsheet fields as products of holomorphic and antiholomorphic
fields (which was implicitly understood in all previous formulae). As
the holomorphic sector consists of the usual supersymmetric orbifold
theory in eight dimensions, the holomorphic part of the vertex is
constructed using the dimension $\frac32$ spin twist operators
$\Lambda_{ab}(z)$ defined in (\ref{Lambdazdef}). On the other hand,
the antiholomorphic sector consists of $d=24$ bosons, and since the
local monodromies about branch points are insensitive to the
compactness of the $16$ bosons on the Cartan torus, the
antiholomorphic part of the vertex is built from the dimension
$\frac32$ bosonic twist fields $\overline{\sigma}{}_{ab}(z)$ of
Section~\ref{TwistFields}. It follows that the heterotic DVV vertex
operator is defined by
\beq
V_{\rm het}=-\frac{\lambda\,N}{{\rm vol}(\torus)}\,\int_\torus\,
\dd\mu(z)~\sum_{1\leq a<b\leq N}\,\big(\,\Lambda\otimes\overline{
\sigma}\,\big)_{ab}(z) \ .
\label{DVVinthet}\eeq
The computation of local twist field two-point functions proceeds by
using a formula analogous to~(\ref{fermcorrelator}).

\subsection{Supersymmetric DLCQ Strings\label{SUSYDLCQ}}

The genus two DLCQ free energy $F_{\rm
  ferm}^{(2)}(\tau^\bullet,\kappa)$ for Type~IIA superstrings at finite
temperature is computed in~\cite{csz}. To write the result, we require
some preliminary definitions. The ten even reduced, genus two integer
characteristics
$\bigl({\mbf{a}\atop\mbf{b}}\bigr)=\bigl({}^{a_1~b_1}_{a_2~b_2}\bigr)
\in H^1(\hat\Sigma,\zed/2\zed)=\zed^4/2\zed^4$ obey
$\mbf{a}\cdot\mbf{b}\equiv 0~\rm{mod}~2$ and are denoted by
\bea
&&\mbf\delta_1\=\left({}^{0~0}_{0~0}\right) \ , \quad
\mbf\delta_2\=\left({}^{0~0}_{0~1}\right) \ , \quad
\mbf\delta_3\=\left({}^{0~1}_{0~0}\right) \ , \quad
\mbf\delta_4\=\left({}^{0~1}_{0~1}\right) \ , \quad
\mbf\delta_5\=\left({}^{0~0}_{1~0}\right)  \ , \label{10evenchars}\\
&&\mbf\delta_6\=\left({}^{0~1}_{1~0}\right) \ , \quad
\mbf\delta_7\=\left({}^{1~0}_{0~0}\right) \ , \quad
\mbf\delta_8\=\left({}^{1~0}_{0~1}\right) \ , \quad
\mbf\delta_9\=\left({}^{1~0}_{1~0}\right) \quad \mbox{and} \quad
\mbf\delta_{0}\=\left({}^{1~1}_{1~1}\right) \ . \nonumber
\eea
We use the shorthand notation
$\vartheta_i:=\Theta(\mbf\delta_i)(\tau)^4$, where the genus two
period matrix $\tau=\tau_{r,m,s,t}(\tau^\bullet,\tau^\#)$ is given by
(\ref{Omegacover}). On the last four characteristics in
(\ref{10evenchars}) we define genus two functions
$\Xi_6(\mbf\delta_i)(\tau)$ of modular weight six by the formulae
\bea
\Xi_6(\mbf\delta_7)&=&
\vartheta_2\,\vartheta_3\,\vartheta_5+\vartheta_8\,
\vartheta_9\,\vartheta_0-\vartheta_1\,\vartheta_4\,\vartheta_6 \ ,
\nonumber\\[4pt]
\Xi_6(\mbf\delta_8)&=&\vartheta_7\,\vartheta_9\,\vartheta_0-
\vartheta_1\,\vartheta_4\,\vartheta_5+\vartheta_2\,\vartheta_3\,
\vartheta_6\ , \nonumber\\[4pt]
\Xi_6(\mbf\delta_9)&=&\vartheta_7\,\vartheta_8\,\vartheta_0-
\vartheta_1\,\vartheta_2\,\vartheta_6+\vartheta_3\,\vartheta_4\,
\vartheta_5\ , \nonumber\\[4pt]
\Xi_6(\mbf\delta_0)&=&\vartheta_7\,\vartheta_8\,\vartheta_9+
\vartheta_3\,\vartheta_4\,\vartheta_6-\vartheta_1\,\vartheta_2\,
\vartheta_5\ .
\eea

Then one has
\bea
F_{\rm ferm}^{(2)}\big(\tau^\bullet\,,\,\kappa\big)&=&
-\frac{g_s^2}4\,\left|\frac{\tau^\bullet}{64\pi^2\,\alpha'}
\right|^{4}\,\sum_{N=2}^\infty\,\frac{\kappa^N}{N}~
\sum_{\stackrel{\scriptstyle r\,m=N}{m~{\rm odd}}}\,\frac1{m^4}~
\sum_{\stackrel{\scriptstyle s,t\in\zed/r\,\zed}{\scriptstyle
    t\neq0}}~\int_\triangle\,\frac{\dd^2\tau^\#}{\big(\tau^\#_2
\big)^{4}}~\big|\Psi_{10}(\tau)\big|^{-2} \nonumber\\ && \times\,
\Bigl|\Xi_6(\mbf\delta_7)(\tau)\,\Theta(\mbf\delta_7)
(\tau)^4+\Xi_6(\mbf\delta_8)(\tau)\,\Theta(\mbf\delta_8)(\tau)^4
\Bigr.\nonumber\\ &&\qquad+\Bigl.\,
\Xi_6(\mbf\delta_9)(\tau)\,\Theta(\mbf\delta_9)(\tau)^4+
\Xi_6(\mbf\delta_0)(\tau)\,\Theta(\mbf\delta_0)(\tau)^4
\Bigr|^2 \ .
\label{Fferm2}\eea
Note that the fermionic contribution to (\ref{Fferm2}) consists of a
sum of four terms in the Weierstrass-Poincar\'e reduction. We may
identify these terms as resulting from the modular invariant sum over
genus one spin structures, as in (\ref{ZtauGSO}). The free
energy (\ref{Fferm2}) should now be equated to the translationally
invariant correlator $\langle\NO V_{\rm ferm}\,V_{\rm
  ferm}\NO\rangle^{S_N\times(\zed_2)^N}$. As in the bosonic case, one
is then faced with the problem of equating the two continuous
parametrizations of the partially discretized genus two moduli space,
one in terms of the elliptic Prym modulus $\Pi=r\,\tau^\#$ and the
other in terms of the branch point location $z\in\torus$. This can
again be done explicitly for the degree two contribution to
(\ref{Fferm2}), corresponding to double covers of the torus $\torus$,
and used to compute local spin twist field correlation functions
explicitly in each twisted sector of the $\real^8\wr(\zed_2)^3$
permutation orbifold.

The $N=2$ contribution to (\ref{Fferm2}) is given by
\beq
{\cal F}_2^{\rm ferm}\big(\tau^\bullet\big)=-\frac{g_s^2}4\,
\,\left|\frac{\tau^\bullet}{64\pi^2\,\alpha'}\right|^{4}\,
\int_\triangle\,\frac{\dd^2\tau^\#}{\big(\tau^\#_2
\big)^{4}}~\left|\frac{{\cal C}\big(\tau_0(\tau^\bullet,\tau^\#)
\big)}{\Psi_{10}\big(\tau_{0}(\tau^\bullet,\tau^\#)
\big)}\right|^{2}
\label{FZ2ferm}\eeq
where
\beq
{\cal C}=\Xi_6(\mbf\delta_7)\,\vartheta_7+\Xi_6(\mbf\delta_8)\,
\vartheta_8+\Xi_6(\mbf\delta_9)\,\vartheta_9+
\Xi_6(\mbf\delta_0)\,\vartheta_0 \ .
\label{calCdef}\eeq
We will begin by simplifying the elliptic function (\ref{calCdef})
using the decomposition (\ref{poi}) for $N=2$. We use the notation of
Section~\ref{DLCQDouble} throughout. To simplify the
formulae somewhat, we momentarily omit the overall factor of
$1/2\,\sqrt{-\ii\tau^\#}$ in (\ref{Theta2decomp}) and reinstate it at
the end of the calculation. For reference, let us tabulate the
ten reduced even genus two theta constants according to the spin
structures (\ref{10evenchars}) as
\beq
\begin{array}{c|c|c|c|c}
\mbf\delta_1&\mbf\delta_2&\mbf\delta_3&\mbf\delta_4&\mbf\delta_5 \\[5pt]
\tb_3\,\ts_3+\tb_4\,\ts_4&\tb_3\,\ts_3-\tb_4\,\ts_4&\tb_4\,
\ts_3+\tb_3\,\ts_4&
\tb_4\,\ts_3-\tb_3\,\ts_4&2\,\ttb_3\,\tts_3\\ \hline\mbf\delta_6&
\mbf\delta_7&\mbf\delta_8&\mbf\delta_9&\mbf\delta_0 \\[5pt]
2\,\ttb_3\,\tts_3&\ts_2\,\tb_2&\ts_2\,\tb_2&2\,\ttb_1\,
\tts_1&-2\ii\tts_1\,\ttb_1
\end{array} \ .
\label{spintable}\eeq
Since one has the equalities $\vartheta_5=\vartheta_6$,
$\vartheta_7=\vartheta_8$ and $\vartheta_9=\vartheta_0$ for the given
reduction, we immediately find that
$\Xi_6(\mbf\delta_7)=\Xi_6(\mbf\delta_8)$ and
$\Xi_6(\mbf\delta_9)=\Xi_6(\mbf\delta_0)$.

After some elementary manipulations we can bring (\ref{calCdef}) into
the form
\bea
{\cal C}&=&2^8~\tb_2\;^4\,\ts_2\,^4\,
\tb_3\,\tb_4\,\ts_3\,\ts_4\,\big(\tb_3\,^2-\tb_4\,^2\big)\,
\big(\ts_3\,^2-\ts_4\,^2\big)\,\ttb_3\,^4\,\tts_3\,^4
\nonumber \\ && \qquad \times\,
\left[\tb_3\,^2\,\tb_4\,^2\,\big(\ts_3\,^2-\ts_4\,^2\big)^2+
\ts_3\,^2\,\ts_4\,^2\,\big(\tb_3\,^2-\tb_4\,^2\big)^2\right]
\nonumber \\ && +\,2^{10}~\tb_2\,^8\,\ts_2\,^8\,\ttb_1\,^8\,
\tts_1\,^8+2^{11}~\tb_2\,^4\,\tb_3\,^2\,\tb_4\,^2\,
\ts_2\,^4\,\ts_3\,^2\,\ts_4\,^2\,\ttb_1\,^4\,\ttb_3\,^4\,
\tts_1\,^4\,\tts_3\,^4 \nonumber \\ && -\,2^9~\ttb_1\,^4\,\ttb_3\,^4\,
\tts_1\,^4\,\tts_3\,^4\,\tb_2\,^4\,\ts_2\,^4\,\big(\tb_3\,^4-\tb_4\,^4
\big)\,\big(\ts_3\,^4-\ts_4\,^4\big) \ .
\eea
We can now proceed as in Section~\ref{DLCQDouble} by doubling the
modulus of the theta functions. In addition to the identities
displayed in (\ref{doublingids}), we will also require the doubling
identities
\bea
\th_1(z|\t)\,\th_2(z|\t)&=&\th_1(2z|2\t)\,\th_4(0|2\t) \ ,
\nonumber \\[4pt]
\th_3(z|\t)\,\th_4(z|\t)&=&\th_4(2z|2\t)\,\th_4(0|2\t) \eea
with $z=\frac14$. We may then take into account that the theta
functions with argument $z=\frac14$ satisfy $\ttb_3=\ttb_4$ and
$\ttb_1=-\ttb_2$, and analogously for $\tts_i$. The calculation is
neither difficult nor illuminating, and the result is
\beq
{\cal C}=\frac{2}{\big(\t^\#\big)^8}~\bar{\th}_2^\bullet\,^8\,
\bar{\th}_3^\bullet\,^4\,\bar{\th}_4^\bullet\,^4\,\bar{\th}_2^\#\,^8\,
\bar{\th}_3^\#\,^4\,\bar{\th}_4^\#\,^4\ ,
\label{calCillum}\eeq 
where we have inserted back the factor
$\big(1/2\,\sqrt{-\ii\t^\#}\,\big)^{16}$ and the bar stands for
doubled modulus as in Section~\ref{DLCQDouble}.

Let us now perform a modular $S$ transformation (\ref{Stransfs}) on
the modulus of both types of theta functions in
(\ref{calCillum}). Then the final result for the numerator of the
integrand in (\ref{FZ2ferm}) reads
\beq
\Big|{\cal C}\big(\tau_0(\tau^\bullet,\tau^\#)
\big)\Big|^2=2^{34}\,\Big|\big(\t^\bullet\big)^{16}\,\eta
\big(2\t^\#\big)^{24}\,\eta\big(\t^\bullet\big)^{24}\,
\th_4\big(2\t^\#\big)^8\,\th_4\big(\t^\bullet\big)^8\Big|\ .
\label{modcalCfinal}\eeq
Substituting (\ref{modcalCfinal}) along with (\ref{Psi10final}) into
(\ref{FZ2ferm}), and using the abstruse identity (\ref{id2}), we
arrive at the final form for the supersymmetric two-loop DLCQ free
energy given by
\beq
{\cal F}_2^{\rm ferm}\big(\tau^\bullet\big)=-\,\frac{16\,g_s^2}
{\big(\pi^2\,\alpha'\,\big)^4}\,\left|\theta_4\big(
\tau^\bullet\big)\right|^8\,
\int_\triangle\,\frac{\dd^2\tau^\#}{\big(\tau^\#_2\big)^4}~
\left|\frac{\theta_4\big(2\tau^\#\big)^2}{
\theta_3\big(\tau^\bullet\big)^4\,
\theta_4\big(2\tau^\#\big)^4-
\theta_4\big(\tau^\bullet\big)^4\,\theta_3\big(2\tau^\#\big)^4}
\right|^{4} \ .
\label{FZ2fermfinal}\eeq
Analogously to the bosonic case of Section~\ref{DLCQDVV}, this
integral should be matched to the worldsheet averaged two-point
correlation function of spin twist field operators
$\Lambda(z)=\Lambda_{12}(z)$ in the $\real^8\wr(\zed_2)^3$ permutation
orbifold given by
\beq
{\cal F}_2^{\rm ferm}
\big(\tau^\bullet\big)=\frac{4\lambda^2}{\tau_2^\bullet\,
\mu(0)}\,\int_\torus\,\dd\mu(z)~\big\langle\Lambda(z)\,\Lambda(0)\big
\rangle^{(\zed_2)^3} \ .
\label{DLCQequivDVVf}\eeq

We recall that, by modular invariance at genus two, the branch point
integration in (\ref{DLCQequivDVVf}) projects all contributions to the
correlation function onto the trivial twist sector
$(\varepsilon,\delta)=(0,0)$, so that the local integrand that we can
read off from (\ref{DLCQequivDVVf}) is
$4\cdot\frac1{2^3}\,\langle\Lambda(z)\,\Lambda(0)
\rangle_{0,0}^{(\zed_2)^3}$. We substitute (\ref{measurechoicebos})
with $d=8$ and (\ref{lambdachoicebos}), and recall that the Prym
modulus is given by $\Pi=\Pi_{0,0}=2\tau^\#$. The crucial observation
is that the bosonic contribution to (\ref{FZ2fermfinal}) involving the
difference of theta functions is identical to that of the purely bosonic case
(\ref{10th2}), due to the universal dimension independent contribution
of the modular form $\Psi_{10}(\tau)$ to the bosonic genus two partition
function (\ref{Z2bos}). We can therefore use the same calculation of
the Jacobian $|\dd\tau^\#/\dd z|^2$ that was carried out in
Section~\ref{DLCQDVV}, wherein it was shown that
\bea
&& \left|\frac{\dd\tau^\#}{\dd z}\right|^2\,
\Big|\theta_3\big(\tau^\bullet\big)^4\,
\theta_4\big(2\tau^\#\big)^4-
\theta_4\big(\tau^\bullet\big)^4\,\theta_3\big(2\tau^\#\big)^4
\Big|^{-4} \label{tauJac}\\ && \qquad\qquad ~=~
\frac1{2^{21}}\,\left|\frac{\eta(\Pi)^4\,\eta\big(\tau^\bullet
\big)^{-1}}{\theta_1\big(z\,\big|\,\tau^\bullet\big)}\right|^6\,
\left|\frac{\theta\big({}^{a}_{b}\big)
\big(\frac z2\,\big|\,\tau^\bullet\big)\,
\theta\big({}^a_b\big)\big(0\,\big|\,\tau^\bullet\big)}
{\theta\big({}^a_b\big)\big(0\,\big|\,
\Pi\big)^{2}}\right|^{24} \nonumber
\eea
for an arbitrary fixed characteristic $(a,b)\neq(1,1)$. Using the
identity (\ref{thetapreta}) and recalling the definition of the prime
form (\ref{primeform}), after a little algebra we can use
(\ref{FZ2fermfinal})--(\ref{tauJac}) to compute
\beq
\big\langle\Lambda(z)\,\Lambda(0)\big
\rangle^{(\zed_2)^3}_{0,0}=\hat{\mathfrak{z}}\big(\tau^\bullet
\big)^8\,\big|E(z)\big|^{-6}\,\big|64\tau^\bullet\,
\eta(\Pi)^3\,\theta_4(\Pi)\big|^8\,
\left|\frac{\theta\big({}^{a}_{b}\big)
\big(\frac z2\,\big|\,\tau^\bullet\big)^3\,
\theta\big({}^a_b\big)\big(0\,\big|\,\tau^\bullet\big)^3}
{\theta\big({}^a_b\big)\big(0\,\big|\,
\Pi\big)^{6}}\right|^{8}
\label{fermtwist00}\eeq
where
\beq
\hat{\mathfrak{z}}(\tau)=\sqrt{\frac{4\pi^2\,\alpha'}{\tau_2}}~
\frac1{\big|\eta(\tau)\big|^2}\,
\left|\frac{\theta_4(\tau)}{\eta(\tau)}\right|
\label{GSpartfnR}\eeq
is the one-loop, first quantized partition function of the
Green-Schwarz superstring in~$\real$ evaluated with the genus one spin
structure $\big({}^0_1\big)$.

We can generate from (\ref{fermtwist00}) the contribution of a generic
twisted sector $(\varepsilon,\delta)\in(\zed/2\zed)^2$ to the spin twist
field correlation function by using a crossing transformation $z\mapsto
z+\delta+\varepsilon\,\tau^\bullet$ and the corresponding twisted Prym
modulus (\ref{Pitwisted}), along with the transformation
formula for Jacobi elliptic functions given by
\beq
\theta\big({}^a_b\big)\big(z+\delta+\varepsilon\,\tau^\bullet\,
\big|\,\tau^\bullet\big)=\exp\left(\mbox{$-\frac{\pi\ii}4\,
\varepsilon^2\,\tau^\bullet-\pi\ii \varepsilon\,z-\frac{\pi\ii}2\,
(b+\delta)\,\varepsilon$}\right)~\theta\big(
{}^{a+\varepsilon}_{b+\delta}\big)\big(z\,\big|\,\tau^\bullet\big)
\label{Jacobicrosstransf}\eeq
which is valid for arbitrary $a,b\in\rat$ and
$\varepsilon,\delta\in\rat$. In fact, the $z$-dependence of the
correlation function (\ref{fermtwist00}) is identical to that of
Section~\ref{Z2Twist} (up to an overall power), and hence the twisted
sector two-point function is an appropriate supersymmetric completion
of the bosonic correlation function (\ref{corr1bos}) (with $R=\infty$
and $d=8$). The final result
is
\beq
\big\langle\Lambda(z)\,\Lambda(0)\big
\rangle^{(\zed_2)^3}_{\varepsilon,\delta}=\hat{\mathfrak{z}}
\big(\tau^\bullet\big)^8~\big|\hat c\big({}^\varepsilon_\delta\big)
\big|^{-16} \ ,
\label{corr1ferm}\eeq
where
\beq
\hat c\big({}^\varepsilon_\delta\big)=
\frac{c\big({}^\varepsilon_\delta\big)^3}{8\,\sqrt{\tau^\bullet\,\eta(
\Pi_{\varepsilon,\delta})^3\,\theta_4(\Pi_{\varepsilon,\delta})}}
\label{hatctwisted}\eeq
and the twisted bosonic determinant $c\big({}^\varepsilon_\delta\big)$
is given by (\ref{ctwisted}). The cubic power in the supersymmetric
twisted determinant (\ref{hatctwisted}) reflects the fact that
the effective twist group of the supersymmetric permutation orbifold
is $(\zed_2)^3$.

\subsection{Heterotic DLCQ Strings}

Finally, we come to the thermodynamic, genus two DLCQ free energy
$F_{\rm het}^{(2)}(\tau^\bullet,\kappa)$ for
heterotic strings with heterotic gauge group $\hat{\cal
  G}=Spin(32)/\zed_2$ or $\hat{\cal G}=E_8\times E_8$. The holomorphic
sector consists of the usual chiral superstring contribution at genus
two. In the antiholomorphic sector, the non-compact bosons produce the
usual antichiral bosonic contribution, while the compactified bosonic
fields produce an instanton sum over the root lattice of $\hat{\cal
  G}$. The latter contribution yields a theta function of the root
lattice which is the unique genus two modular form of weight eight
given by
\beq
\Psi_8(\tau)=\sum_{i=0}^9\,\Theta(\mbf\delta_i)(\tau)^{16} \ .
\label{Psi8tau}\eeq
In the notation of Section~\ref{SUSYDLCQ} above, one then
has~\cite{csz}
\bea
F_{\rm het}^{(2)}\big(\tau^\bullet\,,\,\kappa\big)&=&
\frac{g_s^2}8\,\left|\frac{\tau^\bullet}{2048\pi^4\,(\alpha'\,)^2}
\right|^{4}\,\sum_{N=2}^\infty\,\frac{\kappa^N}{N}~
\sum_{\stackrel{\scriptstyle r\,m=N}{m~{\rm odd}}}\,\frac1{m^4}~
\sum_{\stackrel{\scriptstyle s,t\in\zed/r\,\zed}{\scriptstyle
    t\neq0}}~\int_\triangle\,\frac{\dd^2\tau^\#}{\big(\tau^\#_2
\big)^{4}}~\frac{\overline{\Psi_8(\tau)}}{
\big|\Psi_{10}(\tau)\big|^{2}} \nonumber\\ && \times\,
\Bigl(\Xi_6(\mbf\delta_7)(\tau)\,\Theta(\mbf\delta_7)
(\tau)^4+\Xi_6(\mbf\delta_8)(\tau)\,\Theta(\mbf\delta_8)(\tau)^4
\Bigr.\nonumber\\ &&\qquad+\Bigl.\,
\Xi_6(\mbf\delta_9)(\tau)\,\Theta(\mbf\delta_9)(\tau)^4+
\Xi_6(\mbf\delta_0)(\tau)\,\Theta(\mbf\delta_0)(\tau)^4
\Bigr) \ .
\label{Fhet2}\eea

Again we deal explicitly only with the contribution of double covers
to the formula (\ref{Fhet2}), which is given by
\beq
{\cal F}_2^{\rm het}\big(\tau^\bullet\big)=\frac{g_s^2}8\,
\left|\frac{\tau^\bullet}{2048\pi^4\,(\alpha'\,)^2}
\right|^{4}\,\int_\triangle\,\frac{\dd^2\tau^\#}{\big(\tau^\#_2
\big)^{4}}~\frac{\overline{\Psi_8\big(\tau_0(\tau^\bullet,\tau^\#)
\big)}~{\cal C}\big(\tau_0(\tau^\bullet,\tau^\#)
\big)}{\big|\Psi_{10}\big(\tau_{0}(\tau^\bullet,\tau^\#)
\big)\big|^{2}}
\label{FZ2ferm}\eeq
where from (\ref{modcalCfinal}) one has
\beq
{\cal C}\big(\tau_0(\tau^\bullet,\tau^\#)
\big)=2^{17}\,\big(\t^\bullet\big)^{8}\,\eta
\big(2\t^\#\big)^{12}\,\eta\big(\t^\bullet\big)^{12}\,
\th_4\big(2\t^\#\big)^4\,\th_4\big(\t^\bullet\big)^4 \ .
\label{calChet}\eeq
To simplify the combination of elliptic functions arising in the genus
two modular form (\ref{Psi8tau}), we follow the same steps as in the
bosonic and supersymmetric calculations. Namely, we expand the terms
in the sum over even genus two spin structures in (\ref{Psi8tau})
using the table (\ref{spintable}), transform it to a form that is
suitable for doubling the moduli of the Jacobi theta functions, write
the doubling identities, and then make an elliptic $S$
transformation. The final result is again conveniently written in
terms of theta functions of moduli $2\tau^\#$ and $\tau^\bullet$ as
\beq
\Psi_8\big(\tau_0(\tau^\bullet,\tau^\#)\big)=2^{10}\,\big(
\t^\bullet\big)^8~\theta_4\big(\t^\bullet\big)^{16}\,
\theta_4\big(2\t^\#\big)^{16}~P_{\hat{\cal G}}\Big(\mbox{$\frac{
\theta_3(\t^\bullet)^4}{\theta_4(\t^\bullet)^4}\,,\,\frac{\theta_3(2\t^\#)^4}
{\theta_4(2\t^\#)^4}$}\Big) \ ,
\label{Psi8simpl}\eeq 
where $P_{\hat{\cal G}}(x,y)$ is the symmetric polynomial defined by
\begin{eqnarray}
P_{\hat{\cal G}}(x,y) &=& 256~\big(x^4\,y^4+1\big)-512~
\big(x^4\,y^3+x^3\,y^4+x+y\big)+1984~\big(x^3\,y^3+x\,y\big)
\nonumber\\ &&+\,
288~\big(x^4\,x^2+x^2\,y^4+x^2+y^2\big)-2016~
\big(x^3\,y^2+x^2\,y^3+x^2\,y+x\,y^2\big) \nonumber \\ &&+\,
x^4+y^4+604~\big(x^3\,y+x\,y^3\big)+3654~x^2\,y^2 \ .
\label{Psisympoly}\end{eqnarray}

Substituting (\ref{calChet}) and (\ref{Psi8simpl}), along with
(\ref{Psi10final}) and the abstruse identity (\ref{id2}), we find that
the heterotic DLCQ free energy is given by
\bea
{\cal F}_2^{\rm het}\big(\tau^\bullet\big)&=&\frac{g_s^2}{64}\,
\left|\frac{\theta_4\big(\tau^\bullet\big)}
{4\pi^2\,\alpha'}\right|^8\,\left(\frac{\theta_4\big(-
\overline{\tau^\bullet}\,\big)}{\eta\big(-\overline{\tau^\bullet}\,
\big)}\right)^{12} \nonumber\\ && \times\,
\int_\triangle\,\frac{\dd^2\tau^\#}{\big(\tau^\#_2\big)^4}~
\left|\frac{\theta_4\big(2\tau^\#\big)^2}{
\theta_3\big(\tau^\bullet\big)^4\,\theta_4\big(2\tau^\#\big)^4-
\theta_4\big(\tau^\bullet\big)^4\,\theta_3\big(2\tau^\#\big)^4}
\right|^{4} \nonumber\\ && \qquad\qquad\qquad \times\,
\left(\frac{\theta_4\big(-2\,\overline{\tau^\#}\,\big)}
{\eta\big(-2\,\overline{\tau^\#}\,\big)}\right)^{12}\,P_{\hat{\cal G}}
\Big(\mbox{$\frac{\theta_3(-\overline{\t^\bullet}\,)^4}
{\theta_4(-\overline{\t^\bullet}\,)^4}\,,\,\frac{\theta_3(-2\,
\overline{\t^\#}\,)^4}{\theta_4(-2\,\overline{\t^\#}\,)^4}$}\Big)
\label{FZ2hetfinal}\eea
where we have used the complex conjugation properties
$\overline{\theta_i(\tau)^4}=\theta_i(-\overline{\tau}\,)^4$ and
$\overline{\eta(\tau)^{12}}=\eta(-\overline{\tau}\,)^{12}$. We equate
(\ref{FZ2hetfinal}) to the integrated two-point correlation function
in the heterotic $(\real^8\times{\cal G})\wr(\zed_2\ltimes(\zed_2)^2)$
permutation orbifold given by
\beq
{\cal F}_2^{\rm het}
\big(\tau^\bullet\big)=\frac{4\lambda^2}{\tau_2^\bullet\,
\mu(0)}\,\int_\torus\,\dd\mu(z)~\big\langle(\,\Lambda\otimes
\overline{\sigma}\,)(z)\,(\,\Lambda\otimes\overline{\sigma}\,)(0)\big
\rangle^{\zed_2\ltimes(\zed_2)^2} \ .
\label{DLCQequivDVVh}\eeq
Using the identities (\ref{primeform}), (\ref{thetapreta}) and
(\ref{tauJac}) we then arrive at the two-point function of twist
fields in the untwisted sector given by
\bea
&& \big\langle(\,\Lambda\otimes
\overline{\sigma}\,)(z)\,(\,\Lambda\otimes\overline{\sigma}\,)(0)\big
\rangle^{\zed_2\ltimes(\zed_2)^2}_{0,0} \nonumber\\ &&
\qquad\qquad ~=~ 32\,\left(\frac{\hat{\mathfrak{z}}\big(\tau^\bullet
\big)}{\sqrt{4\pi^2\,\alpha'}}\right)^8\,\left(\frac{\theta_4\big(-
\overline{\tau^\bullet}\,\big)}{\eta\big(-\overline{\tau^\bullet}\,
\big)}\right)^{12}\,\big|E(z)\big|^{-6}\,\big|8\tau^\bullet\,
\eta(\Pi)^3\,\theta_4(\Pi)\big|^8 \label{hettwistcorr00} \\ &&
\qquad\qquad\qquad \times\,\left(\frac{\theta_4\big(-
\overline{\Pi}\,\big)}{\eta\big(-\overline{\Pi}\,
\big)}\right)^{12}\,P_{\hat{\cal G}}
\Big(\mbox{$\frac{\theta_3(-\overline{\t^\bullet}\,)^4}
{\theta_4(-\overline{\t^\bullet}\,)^4}\,,\,\frac{\theta_3(-
\overline{\Pi}\,)^4}{\theta_4(-\overline{\Pi}\,)^4}$}\Big)\,
\left|\frac{\theta\big({}^{a}_{b}\big)
\big(\frac z2\,\big|\,\tau^\bullet\big)^3\,
\theta\big({}^a_b\big)\big(0\,\big|\,\tau^\bullet\big)^3}
{\theta\big({}^a_b\big)\big(0\,\big|\,
\Pi\big)^{6}}\right|^{8} \nonumber
\eea
with $(a,b)\neq(1,1)$, where $\hat{\mathfrak{z}}(\tau)$ is the
supersymmetric partition function (\ref{GSpartfnR}).

The structure of the formula (\ref{hettwistcorr00}) can be understood
as follows. Generally, the separating degeneration limit $\tau_{12}\to0$
of the genus two modular form (\ref{Psi8tau}) factorizes into the
unique elliptic modular form of weight eight under $SL(2,\zed)$ as
\beq
\Psi_8(\tau)=\big(\theta_2(\tau_{11})^{16}+\theta_3(\tau_{11})^{16}+
\theta_4(\tau_{11})^{16}\big)\,\big(\theta_2(\tau_{22})^{16}+
\theta_3(\tau_{22})^{16}+\theta_4(\tau_{22})^{16}\big)+
{O}\big(\tau_{12}^2\big) \ .
\label{Psi9deg}\eeq
For the covering surface $\hat\Sigma$, in the homology basis wherein
the period matrix is given by (\ref{tau0primedef}) this degeneration
limit corresponds to $\Pi\to\tau^\bullet$, or equivalently
$z\to0$. Since the $x\to y$ limit of the symmetric polynomial
(\ref{Psisympoly}) factorizes as
\beq 
P_{\hat{\cal G}}(x,x)=64\,\big((x-1)^4+x^4+1\big)^2 \ ,
\label{Psisympolyfact}\eeq
we see that the $z\to0$ limit of the two-point function
(\ref{hettwistcorr00}) factors into the one-loop heterotic string
partition function on $\real^8$ evaluated with the spin structure
$\big({}^0_1\big)$ which is given by
\beq
\hat{\mathfrak{z}}_{\rm
  het}(\tau)=\left(\frac{4\pi^2\,\alpha'}{\tau_2}\right)^4~
\frac1{\big|\eta(\tau)\big|^{16}}\,
\left(\frac{\theta_4(\tau)}{\eta(\tau)}\right)^4~\left(
\frac{\theta_2\big(-\overline{\tau}\,\big)^{16}+\theta_3
\big(-\overline{\tau}\,\big)^{16}+\theta_4\big(-\overline{
\tau}\,\big)^{16}}{2\,\eta\big(-\overline{\tau}\,\big)^{16}}
\right) \ .
\label{hetpartfnR8}\eeq

However, in contrast to the bosonic and supersymmetric twist field
correlation functions, for distinct branch points the
two-point function (\ref{hettwistcorr00}) does not neatly factor out a
component corresponding to the untwisted fluctuation determinant of
the heterotic orbifold sigma model. The reason generally is that the
effective twist group is now a semi-direct product
$S_N\ltimes(\zed_2)^N$ acting on the gauge fermions $\chi^a$. This
means that the discrete $(\zed_2)^N$ gauge symmetry acts in the gauge
sector together with the monodromy conditions of the permutation
orbifold, and a disentanglement of the twisted and untwisted
determinants arising from integration over the fermion fields $\chi$
in terms of branch point data as previously is not possible.

For example, by applying a crossing transformation to
(\ref{hettwistcorr00}) as before one arrives at the twisted sector
two-point functions
\bea
\big\langle(\,\Lambda\otimes
\overline{\sigma}\,)(z)\,(\,\Lambda\otimes\overline{\sigma}\,)(0)\big
\rangle^{\zed_2\ltimes(\zed_2)^2}_{\varepsilon,\delta}&=&
32\,\left(\frac{\hat{\mathfrak{z}}\big(\tau^\bullet
\big)}{\sqrt{4\pi^2\,\alpha'}}\right)^8~
\big|\hat c\big({}^\varepsilon_\delta\big)\big|^{-16}~
\left(\frac{\theta_4\big(-
\overline{\tau^\bullet}\,\big)}{2\eta\big(-\overline{\tau^\bullet}\,
\big)}\right)^{12} \label{hettwistcorred} \\ && \times\,
\left(\frac{\theta_4\big(-
\overline{\Pi_{\varepsilon,\delta}}\,\big)}
{2\eta\big(-\overline{\Pi_{\varepsilon,\delta}}\,
\big)}\right)^{12}\,P_{\hat{\cal G}}
\Big(\mbox{$\frac{\theta_3(-\overline{\t^\bullet}\,)^4}
{\theta_4(-\overline{\t^\bullet}\,)^4}\,,\,\frac{\theta_3(-
\overline{\Pi_{\varepsilon,\delta}}\,)^4}
{\theta_4(-\overline{\Pi_{\varepsilon,\delta}}\,)^4}$}\Big) \nonumber
\eea
with the supersymmetric twisted determinant $\hat
c\big({}^\varepsilon_\delta\big)$ given by (\ref{hatctwisted}). The
extra gauge symmetry is implemented by $O(N)$ vector reflections of
$\chi^a$ and holonomies of the corresponding flat real line bundles
$L_{\mbf\delta}\to\torus$. The latter phases correspond to
$\zed_2$-valued Wilson lines which break the spacetime heterotic
gauge group $\hat{\cal G}$ to ${\cal G}=SO(16)\times SO(16)$. They
yield the extra GSO projection required to match to the spectrum of
the free $E_8\times E_8$ heterotic
string~\cite{Rey:1997hj,Lowe:1997sx,Banks:1997it} and to light-cone
heterotic string field theory.

\subsection*{Acknowledgments}

We thank R.~Accola, P.~B\'antay, H.~Braden, J.C.~Eilbeck, H.~Farkas,
J.~Howie, A.~Konechny, S.~Ramgoolam, A.~Recknagel, R.~Reis and
G.~Semenoff for helpful discussions and correspondence. This work was
supported in part by the Marie Curie Research Training Network Grants
{\sl ForcesUniverse} (contract no.~MRTN-CT-2004-005104) and {\sl
  ENRAGE} (contract no.~MRTN-CT-2004-005616) from the European
Community's Sixth Framework Programme. The work of H.C. was supported
in part by a Postgraduate Studentship from the Engineering and
Physical Sciences Research Council~(U.K.).

\end{document}